\documentclass[amssymb,
amsmath,
amsfonts,
reprint,
superscriptaddress,
longbibliography,
physrev, 
nofootinbib,
aps]{revtex4-2}

\usepackage[utf8]{inputenc}
\usepackage{graphicx}
\usepackage{dcolumn}
\usepackage[T1]{fontenc}
\usepackage{tabularx}
\usepackage{enumitem}
\usepackage{filecontents}
\usepackage{soul}
\usepackage{booktabs}
\usepackage{multirow}

\let\svthefootnote\thefootnote
\newcommand\freefootnote[1]{%
  \let\thefootnote\relax%
  \footnotetext{#1}%
  \let\thefootnote\svthefootnote%
}

\newif\iffastcompile
\fastcompilefalse

\usepackage[mathlines]{lineno}
\usepackage{graphicx}
\usepackage[margin=1in]{geometry}
\usepackage{graphicx}
\usepackage[table]{xcolor}

\usepackage[hidelinks, hypertexnames=false]{hyperref}
\definecolor{color1}{rgb}{0,0.25,0.70}
\hypersetup{colorlinks=true,
    linkcolor={color1},
    citecolor={color1},
    urlcolor={color1}
}

\newcommand{\closedm}[1]{\left[ #1 \right]}
\newcommand{\closeds}[1]{\left( #1 \right)}

\newcommand{\ket}[1]{| #1 \rangle}
\newcommand{\bra}[1]{\langle #1 |}

\newcommand{\px}[1]{\hspace{#1 pt}}

\begin{document}

\title{Molecular Quantum Control Algorithm Design by Reinforcement Learning}
\author{Anastasia Pipi}
\thanks{These two authors contributed equally.}
\affiliation{Department of Physics and Astronomy, University of California, Los Angeles (UCLA), California 90095, USA}
\author{Xuecheng Tao$^\dagger$}
\thanks{These two authors contributed equally.}
\affiliation{Division of Physical Sciences, College of Letters and Science, University of California, Los Angeles (UCLA), California 90095, USA}
\author{Arianna Wu}
\affiliation{Department of Chemistry and Biochemistry, University of California Los Angeles (UCLA), Los Angeles, California 90095, USA}
\author{Prineha Narang$^\ddagger$}
\affiliation{Division of Physical Sciences, College of Letters and Science, University of California, Los Angeles (UCLA), California 90095, USA}
\affiliation{Electrical and Computer Engineering Department, University of California, Los Angeles (UCLA), California, 90095, USA}
\author{David R. Leibrandt$^\S$}
\affiliation{Department of Physics and Astronomy, University of California, Los Angeles (UCLA), California 90095, USA}

\freefootnote{
$^\dagger$xuechengtao@gmail.com, \\
$^\ddagger$prineha@ucla.edu, \\
$^\S$leibrandt@ucla.edu. 
}

\date{\today}
\begin{abstract}
Precision measurements of molecules offer an unparalleled paradigm to probe physics beyond the Standard Model.
The rich internal structure within these molecules makes them exquisite sensors for detecting fundamental symmetry violations, local position invariance, and dark matter.
While trapping and control of diatomic and a few very simple polyatomic molecules have been experimentally demonstrated, leveraging the complex rovibrational structure of more general polyatomics demands the development of robust and efficient quantum control algorithms.
In this study, we present reinforcement-learning quantum-logic spectroscopy (RL-QLS), a general, reinforcement-learning-designed, quantum logic approach to prepare molecular ions in single, pure quantum states.
The reinforcement learning agent optimizes the pulse sequence, each followed by a projective measurement, and probabilistically manipulates the collapse of the quantum system to a single state.
The performance of the RL-QLS control algorithm is numerically demonstrated 
for the polyatomic molecule H$_3$O$^+$ with 130 thermally populated eigenstates and degenerate transitions within inversion doublets, as well as for CaH$^+$ under the disturbance of environmental thermal radiation.
We expect that the developed theoretical framework can be readily implemented for quantum control of polyatomic molecular ions with densely populated structures, thereby facilitating experimental tests of fundamental theories. \\
\end{abstract}

\maketitle
\clearpage


\section{Introduction}
Low-energy, high-precision measurements provide a powerful tool to explore fundamental physics beyond the Standard Model (BSM) \cite{safronova_search_2018, demille2024quantum}.
The rich internal energy-level structure of molecules,
particularly polyatomic molecules, presents sensitive probes to test BSM hypotheses.
For example, the frequencies of the inversion transitions of hydronium are used by astronomers to search for violation of local position invariance and would be sensitive to potential dark energy carriers \cite{kozlov2010sensitivity};
a minuscule energy shift is predicted in molecular enantiomers as a result of parity violation \cite{letokhov1975difference} and awaits experimental observation \cite{quack2022perspectives, landau2023chiral}.
However, high-fidelity control of those molecules remains challenging, because many rovibrational states are populated by thermal radiation and the transition frequencies between those states commonly overlap each other.
In fact, preparation of molecules into single, pure states is a central yet non-trivial quantum control task \cite{mitra_qcontrolFP_2022, patterson2018method}.

Several methods have been developed for state preparation,
including sympathetic cooling \cite{hudson2016sympathetic, BGcooling_DeMille}, photoassociation of cold atoms \cite{Fesh_Jin}, optical cycling \cite{augenbraun_molecular_2020, Ocycling_Nicholas, dickerson2023single}, and quantum-logic spectroscopy \cite{schmidt2005spectroscopy} (QLS).
Among these, QLS stands out as a unique control scheme \cite{leibfried2012quantum, ding2012quantum},
requiring no specific restrictions on the internal structure of the molecular ion and enabling non-destructive detection of molecular ion states.
Prominent experiments \cite{chou2017preparation, lin2020quantum, chou2020frequency, liu2023quantum, QLS_Kienzler, QLS_Willitsch, QLS_schmidt} have demonstrated the ability to measure and manipulate the quantum states of simple diatomic ions with QLS.
Increasing complexity in the molecular Hilbert space, as in polyatomic species,
demands robust and scalable state preparation techniques beyond current capabilities.
More broadly, the rich internal degrees of freedom of polyatomics serve as unique sensors for BSM physics, which in turn necessitates the development of {\it molecular quantum control} approaches tailored to their inherent complexity.

\begin{figure*}[!htb]
    \includegraphics[width=0.95\textwidth]{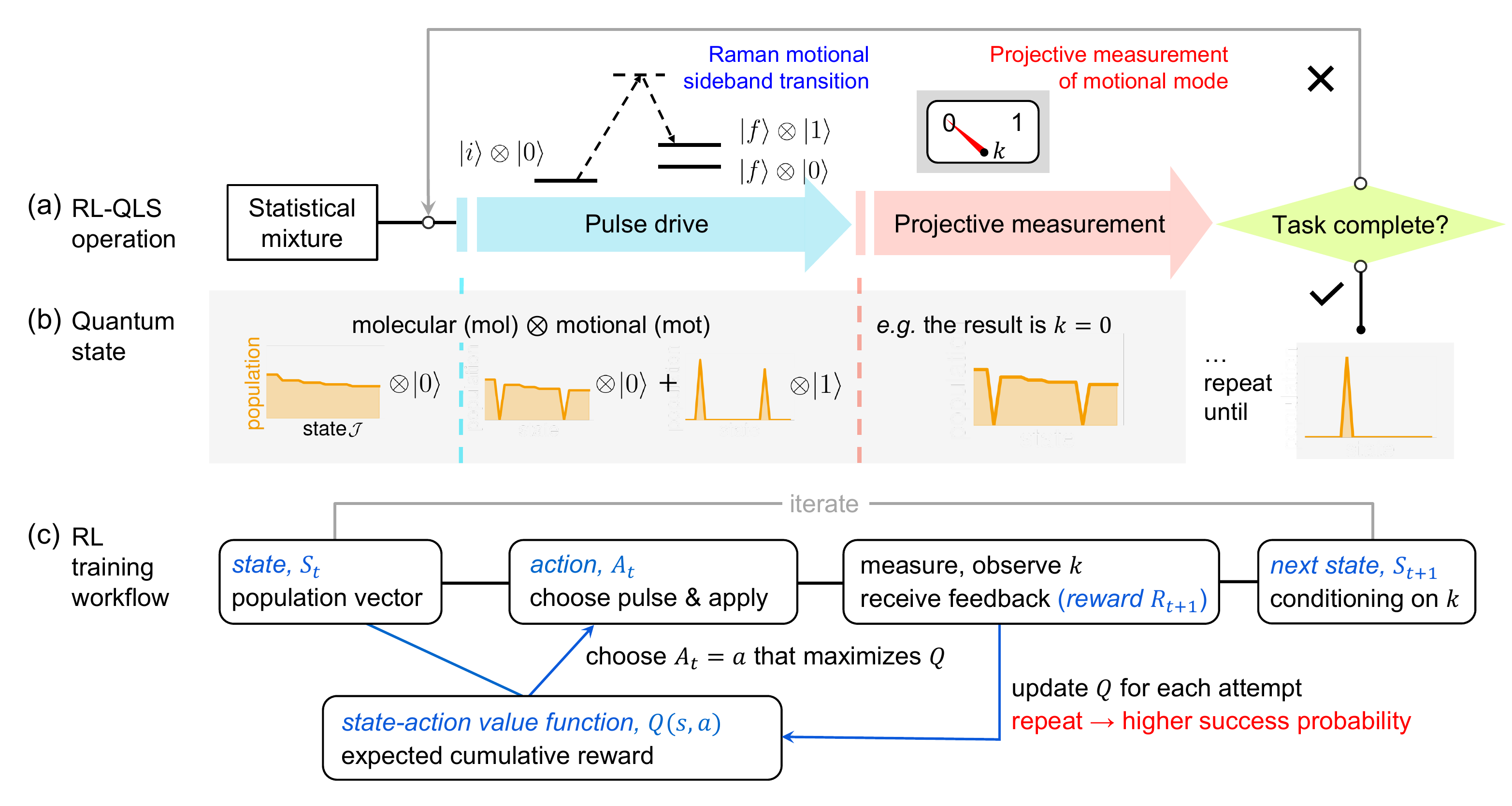}
    \caption{RL-QLS framework---state preparation via projective measurements. 
    \textbf{(a)} A single quantum state is obtained by taking repetitive steps each consisting of two parts; ({\it i}) 
    a laser pulse driving the blue-sideband of a molecular state transition, followed by ({\it ii}) a projective measurement of the motional state.
    \textbf{(b)} Time evolution of the state in terms of density matrices. 
    $u_{\mathcal{J}\mathcal{J}'}$ and $v_{\mathcal{J}\mathcal{J}'}$ are obtained by solving the time-dependent Schr\"odinger equation. 
    In the illustration, the $k=0$ measurement result is more probable than $k=1$ (areas of the shadow, second column). 
    \textbf{(c)} 
    Training of the RL-QLS agent. The agent explores the pulse library and updates the state–action value function $Q$ with the accumulating experience.
    As training proceeds, the learned $Q$-function guides pulse selection and increases the success probability of the molecular control task.}
    \label{fig:fig1}
\end{figure*}

In this article, we establish and demonstrate RL-QLS, 
a theoretical framework that unifies tools from quantum chemistry, AMO physics, and artificial intelligence for molecular quantum control.
Initially, the molecular level structure is obtained from experiments or calculated and the evolution of the molecular state under control pulses is numerically simulated.
For each step of the experiment, a control pulse is selected by the reinforcement learning (RL) agent and applied to the molecule,
followed by a QLS-based projective measurement that probabilistically collapses part of the molecular quantum state. 
Through repetitive measurements, a single, pure state is prepared.
We show that, by leveraging the complete history of control pulses and measurement results,
RL-QLS enables single-state preparation for molecular ions with increasingly complex energy-level structures and subject to environmental disruption both faster and with higher fidelity than current protocols.
Building on the success of learning-based control in simple quantum systems, tasks such as state engineering \cite{zhang2019does, an2021quantum, mackeprang2020reinforcement, paparelle2020digitally} and gate optimization \cite{niu2019universal, preti2024hybrid},
RL-QLS directly addresses the distinct challenge of molecular complexity in molecular quantum control.
Based on the following numerical demonstrations, we anticipate that RL-QLS will facilitate precision measurement experiments that were previously inaccessible.

\section{Method}
We begin by introducing the QLS framework to prepare a single molecular state with projective measurements.
For a simple molecular ion where a set of signature transitions are resolved in frequency, the protocol was proposed \cite{leibfried2012quantum, ding2012quantum} and experimentally demonstrated \cite{chou2017preparation, lin2020quantum, chou2020frequency, liu2023quantum}.
Starting from a Boltzmann mixture of energetically accessible states ($\ket{\mathcal{J}}$),
the protocol repeatedly drives blue-sideband transitions and performs projective measurements (realized with quantum logic gates with a co-trapped auxiliary ion) of the motional state (Fig.~\ref{fig:fig1}a).

We consider a molecular spectroscopy ion that occupies the ground electronic and vibrational states, 
while a substantial number of its rotational and hyperfine states are populated due to thermal radiation. 
The density matrix of the rotational manifold is 
$\rho_{\rm mol} = \sum_{\mathcal{J}=1}^{N_S} 
P_\mathcal{J} \ket{\mathcal{J}} \bra{\mathcal{J}} $, with $N_S$ the number of states.
$P_\mathcal{J}$ denotes the occupation probability of the molecular ion state $\ket{\mathcal{J}}$, 
and $P_\mathcal{J} = e^{-\beta E_\mathcal{J}} / \sum_\mathcal{J}  e^{-\beta E_\mathcal{J}} $ follows a Boltzmann distribution.
The spectroscopy ion and the co-trapped logic ion share a common motional mode $\ket{k}$, initialized to $\ket{k=0}$.
Thus $\rho = \rho_{\rm mol} \otimes \ket{0} \bra{0}$.
In the Lamb-Dicke regime (thus $k \in \{0, 1\}$ in the model), the applied laser pulse drives a molecular transition and results in a quantum state with a density matrix of 
\begin{equation}
\begin{aligned} \label{eq:density_matrix_on_pulse}
    \rho = \sum_{\mathcal{J}=1}^N P_\mathcal{J} 
    \Bigg{[} &\closeds{\sum_{\mathcal{J}'} u_{\mathcal{J} \mathcal{J}'} \ket{\mathcal{J}', 0} 
            + \sum_{\mathcal{J}'} v_{\mathcal{J} \mathcal{J}'} \ket{\mathcal{J}', 1} } \\
          &\closeds{\sum_{\mathcal{J}'} u^*_{\mathcal{J} \mathcal{J}'} \bra{\mathcal{J}', 0} 
            + \sum_{\mathcal{J}'} v^*_{\mathcal{J} \mathcal{J}'} \bra{\mathcal{J}', 1} }
    \Bigg{]}. 
\end{aligned}
\end{equation}
In Eq.~\ref{eq:density_matrix_on_pulse}, 
$u_{\mathcal{J} \mathcal{J}'}$ and $v_{\mathcal{J} \mathcal{J}'}$ describes the time evolution of a pure state $|\mathcal{J}, k=0\rangle$ under the influence of the applied pulse.
As illustrated in Fig.~\ref{fig:fig1}b, the blue-sideband pulses partition the molecular Hilbert space into two subspaces, each associated with one motional quantum number, $k$.

A projective measurement of the motional state is subsequently performed with a bright detection on the quantum state of the logic ion.
The projective measurement collapses the state to either the ground or excited motional state manifold according to the result.
As illustrated in Fig.~\ref{fig:fig1}b, the $k=1$ outcome concentrates the population in a small subspace that consists of the ending states of the driven transition while $k=0$ outcome eliminates population in that subspace.
The probability of each measurement outcome is given by
\begin{equation}
\begin{aligned} \label{eq:qevo_2}
p_0 = \sum_\mathcal{J,J'} P_\mathcal{J} |u_{\mathcal{J} \mathcal{J}'}|^2, \px{4}
p_1 = \sum_\mathcal{J,J'} P_\mathcal{J} |v_{\mathcal{J} \mathcal{J}'}|^2.
\end{aligned}
\end{equation}
Furthermore, after the measurement and the subsequent motional state cooling, the state of the molecular ion,
$\rho_{\rm mol}$, is 
\begin{equation}
\begin{aligned} \label{qevo_3}
    \begin{cases}
     (1/p_0) \sum_\mathcal{J'} \closeds{\sum_{\mathcal{J}} P_\mathcal{J} |u_{\mathcal{J} \mathcal{J}'}|^2} \ket{\mathcal{J}'} \bra{\mathcal{J}'}, &{\rm if \ } k=0, \\  
     (1/p_1) \sum_\mathcal{J'} \closeds{\sum_{\mathcal{J}} P_\mathcal{J} |v_{\mathcal{J} \mathcal{J}'}|^2} \ket{\mathcal{J}'} \bra{\mathcal{J}'}, &{\rm if \ } k=1. \\
    \end{cases}
\end{aligned}
\end{equation}
Note that we have assumed that coherence is destroyed in the motional cooling, leading to a population-based description for computational efficiency.
Laser pulses and projective measurements are then repeated many times until a pure state has been prepared.
As such, the population distribution, $P(\mathcal{J})= {\rm tr} (\rho \ket{\mathcal{J}} \bra{\mathcal{J}})$, is controlled, in a probabilistic manner, to collapse to a single, pure molecular state 
with confidence above the purity threshold.
A more detailed discussion of the state evolution is presented in Sec.~SA \cite{supplemental}.

This state preparation framework is quite general, and there is a lot of flexibility in the selection of the molecular sideband pulse for each step.
In previous work with CaH$^+$ \cite{chou2017preparation, lin2020quantum, chou2020frequency, liu2023quantum},
signature transitions with unique frequencies (under an external magnetic field) were identified, and the pulse sequence was designed to sweep the possible transitions sequentially.
This simple `sweeping' strategy was experimentally demonstrated
for up to 48 hyperfine states, however, it encounters difficulties in more complex molecular ions, where hundreds of states are thermally accessible and transition frequencies often overlap.
More importantly, the sweeping protocol used in the pioneering experiments does not take advantage of the historical measurement data, thus the number of pulses and measurements (i.e., {\it steps}) needed for state preparation can be significantly optimized.

Reinforcement learning (RL) is a promising approach for optimizing the state preparation task, by leveraging historical information to decide on the next action.
The physical state preparation process straightforwardly maps onto a sequential decision-making task, formalized as a Markov decision process (MDP) in Fig.~\ref{fig:fig1}c.
In the RL framework \cite{sutton2018reinforcement},
the agent explores how a pulse choice may drive the population dynamics and exploits the information from past attempts to guide current control decisions.
The {\it state} at time $t$ is tracked as an $N_S$-dimensional population vector $S_t \in [0, 1]^{N_S}$ in the eigenstate space.
The agent selects a pulse $A_t=a$ from the {\it action} library ($N_A$ choices) to apply each step.
The quantum-state evolution resulting from the selected pulse is then calculated and inputted into the MDP as transition matrices, $\mathcal{A}^{(a)}$ (Fig.~S1 and Sec.~SB-SC \cite{supplemental}).
To account for the motional mode measurement,
a different $\mathcal{A}^{(a)}$ matrix is needed for each possible motional state measurement outcome.
Taking both the coherent state evolution driven by laser pulses and the probabilistic wavefunction collapse during measurement together, the state-action dynamics are specified by the following two equations.
The conditional probability of the measurement outcome, $k$, is 
\begin{subequations}
\begin{equation}
p(k|S_t, a) = ||\mathcal{A}_{k}^{(a)} S_t||_1, \px{4} k\in\{0,1\}
\end{equation}
with $||\cdot||_1$ the vector 1-norm.
The post-measurement state is then
\begin{equation}
S_{t+1} = \mathcal{A}_{k}^{(a)} S_t. 
\end{equation}
\end{subequations}
Specifically, we do not distinguish $k \ge 1$ results, and perfect measurement is assumed for now; infidelity will be addressed later.
The reward function $R$ is set to a negative number, e.g., $R=-1$ for each step, regardless of the action, to encourage fast task completion.
Overall, we expect the RL agent to learn the state-action value function, $Q(s,a)$, or the effectiveness of the actions for state preparation given the current state, 
\begin{equation}
Q(s, a) = \mathbb{E} \closedm{\sum_{\tau=0}^T R_{t+1+\tau} | S_t=s, A_t=a},
\end{equation}
with $T$ the terminal step of task completion and $\mathbb{E}$ the expectation value.
In this study, we focus on the deep $Q$-learning algorithm \cite{mnih2013playing, mnih2015human, watkins1989learning} for its exploration efficiency in the discrete action space (Sec.~SD \cite{supplemental}).
The algorithm works by finding the current action $a$ that 
maximizes the estimated expected cumulative reward, i.e., $a = \arg \max_{a} Q(s,a)$
with $Q(s, a)$ expressed on a simple, fully-connected neural network so that the operation time (on a CPU embedded FPGA) to evaluate the optimal pulse choice can be shorter than the wall-clock pulse duration.
We use a three-layer neural network with 128 nodes per layer (we also tested four-layer and wider networks and the same qualitative conclusion holds).
The computation workflow is implemented using PyTorch \cite{paszke2019pytorch}.

\begin{figure}[!tb]
    \centering
    \includegraphics[width=\columnwidth]{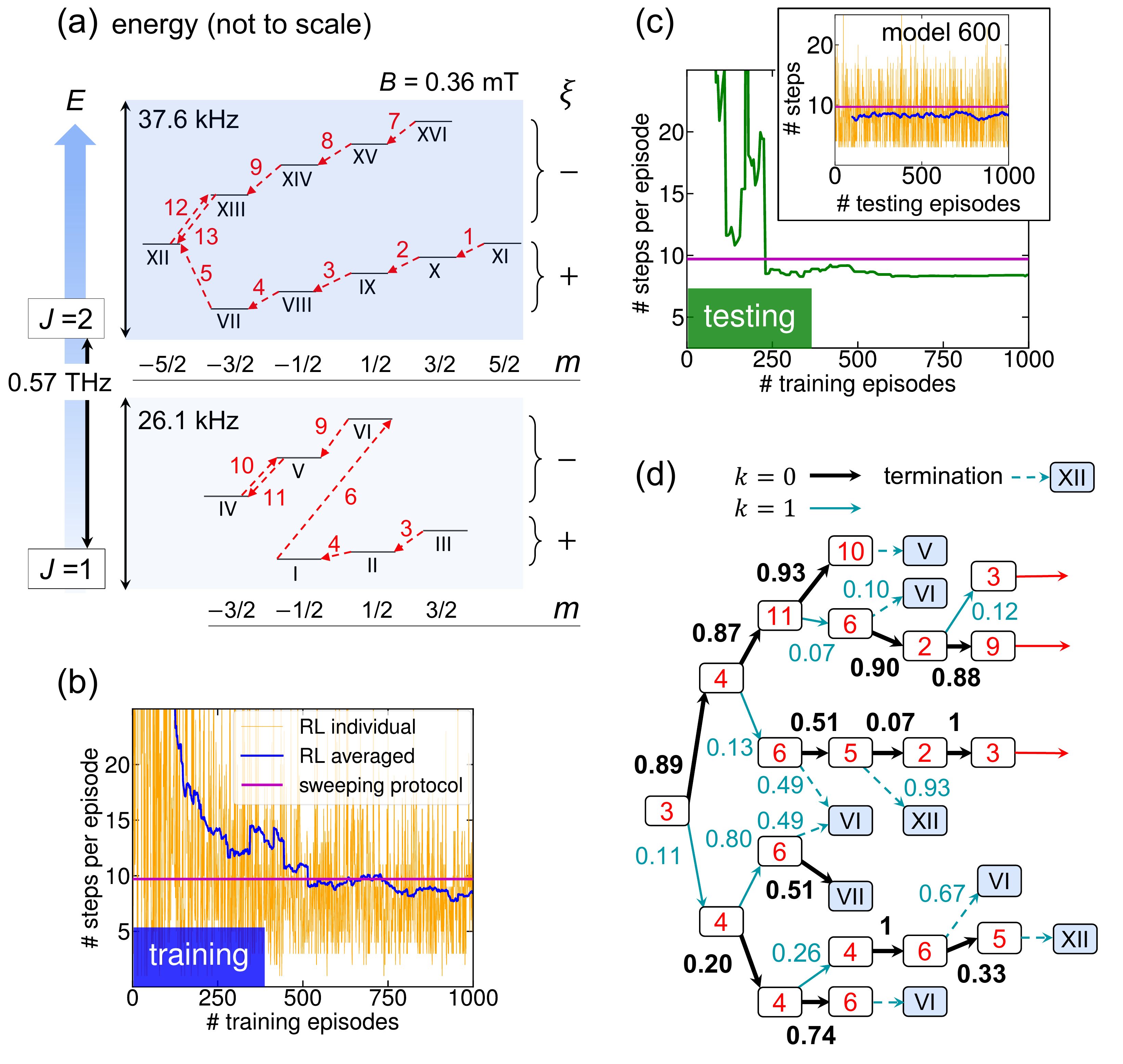}
    \caption{\label{fig:fig2}
    \textbf{(a)} Energy level diagram of CaH$^+$ featuring the thermally occupied, low-lying rotational $J \in \{1,2\}$ manifolds with states labeled I through XVI. 
    A set of uni-directional, blue-sideband $\pi$-pulses are used to concentrate the population (red arrows, 1--9) or to drive the population between the $\ket{J,-J+1/2,-}$ and $\ket{J,-J-1/2,-}$ states (10--13).
    \textbf{(b-c)} The number of steps, or length, to prepare a single molecular state with the RL-designed protocol.
    {(b)} Training process; the length (blue) is obtained by moving averages over the most recent 100 individual episodes (orange). For instance, the blue point at episode 600 is averaged over the range $(500, 600]$ in orange.
    The trained models are then tested in (c), e.g. the inset presents the testing process for the model trained with 600 episodes.
    The main curve (green) is then obtained by averaging the length of 1000 testing episodes for each model.
    \textbf{(d)} A truncated decision tree of the RL-QLS protocol (complete version in Fig.~S4 \cite{supplemental}). 
    Pulse choices and the terminal states (blue boxes) are reported in red numbers and Roman numerals, respectively.
    The branching probabilities are color-coded to match the legend of the measurement outcomes.}
\end{figure}

\section{Results and Discussions \label{sec:results}}
Fig.~\ref{fig:fig2} presents the usage of the RL-QLS approach for state preparation.
For illustration purposes, initially we consider only the $J \le 2$ manifolds of CaH$^+$ at a magnetic field of 0.36~mT to match the NIST experiments \cite{chou2017preparation, lin2020quantum, liu2023quantum} (Fig.~\ref{fig:fig2}a).
Laser pulses driving two-photon stimulated Raman transitions form the action library (Fig.~S2 \cite{supplemental}) for RL simulations.  Previously, a similar pulse library was used \cite{chou2017preparation} in the `sweeping' protocol for state preparation;
pulses were sequentially and periodically applied to concentrate the population, followed by a final projective measurement to obtain a single state.
In contrast, here we choose to perform measurements after every blue-sideband pulse to obtain feedback on the instantaneous populations (Sec.~SC \cite{supplemental}), 
although other choices are possible. For example, within the RL-QLS framework, the number of measurements during state preparation can be reduced by tailoring the reward function to the specific application.

A sweeping protocol attempt is simulated in Fig.~\ref{fig:fig2}b; 
population dynamics of a few representative episodes are shown in Fig.~S3 \cite{supplemental}.
Typically, a single state is prepared (i.e., the {\it episode} terminates) in 1--2 sweeping cycles.
Episodes sometimes require >1 sweeping cycle to terminate since certain pulses from the library (Tab.~S2 \cite{supplemental}) drive the population into multiple destinations, 
a consequence of degenerate transitions.
The average number of steps (or episode length, 9.7) needed to prepare a pure state is slightly lower than the number in one sweeping cycle (13), indicating probable terminations from projective measurements collapsing onto low-population states.
To ensure a fair comparison with the RL-QLS results below, 
the sweeping protocol implementation uses the entire history of actions and measurement results to determine the molecular state and terminate the trial (in contrast with the experimental implementation of the sweeping protocol at NIST \cite{chou2017preparation}).
Overall, the sweeping protocol is most effective for molecules with simple level structures and thus frequency-resolved transitions.

Now, the same state preparation task is assigned to the RL agent.
During the training, the RL agent progressively learns an increasingly effective pulse-selection policy for each instantaneous molecular state population, as reflected by the decrease in the moving-average episode length (blue in Fig.~\ref{fig:fig2}b).
Episodes with longer lengths are also observed throughout the training (frequent orange spikes), particularly early in the training, due to the intentionally suboptimal choices.
Such deliberate applications of suboptimal pulses allow the RL agent to explore the pulse choices that are not locally optimal but may yield faster state preparation eventually.
The trained models from Fig.~\ref{fig:fig2}b are then evaluated in the panel Fig.~\ref{fig:fig2}c (specifically, one evaluation example presented in the inset). 
In the testing sessions, exploration is disabled and the actions are selected deterministically according to the state-action value function. As a consequence, the averaged episode length in Fig.~\ref{fig:fig2}c (green) is generally lower than that in panel b (blue).
The episode length approaches consistent near-optimal behavior after $\sim$250  episodes, and the training converges to the optimal policy after $\sim$550  episodes. 
In practice, the evaluations can be performed on the fly, and the training is finished once such consistent behavior is achieved.

The success of RL-QLS molecular control is straightforward to observe, as the average number of required pulses and measurement steps (i.e., 8.3) per preparation episode (green in Fig.~\ref{fig:fig2}c, main figure) outperforms that achieved by the sweeping protocol (purple).
The end product of the RL training is the learned pulse-selection strategy.
One example of the resulting decision tree is presented in Fig.~\ref{fig:fig2}d.
The cumulative probability of the successful state preparation episodes with RL-QLS outperforms that of the sweeping protocol when the same number of pulses is applied (Tab.~\ref{tbl:success}).
The RL-designed protocol applies available pulses non-repetitively at the beginning, which resembles the sweeping protocol, while
the repetitive application of one pulse is more common as the state preparation progresses.
Among different training results, typically $\sim$60\% of the episodes end on the $\ket{J, m=-J\pm1/2, -}$ states (Fig.~S5 \cite{supplemental}, top).
We note that the reported decision tree is not unique, and different decision trees with similar success probabilities can be obtained with independently trained models due to stochastic initialization. 
However, as shown in the action histogram (Fig.~S5, bottom), smart utilization of the pulses that drive multiple transitions is common in those decision trees. Computational details are reported in Sec.~SD \cite{supplemental}.

\begin{table}[t]
\caption{\label{tbl:success} The percentage of successfully finished episodes versus the number of pulses applied.}
\begin{ruledtabular}
\begin{tabular}{lccccccccc}
\shortstack[l]{\# pulses\\applied} & 2 & 3 & 4 & 5 & 6 & 7 & 8 & \ldots & 18 \\
RL       & 0\% & 15\% & 35\% & 35\% & 35\% & 45\% & 56\% &        & 99\% \\
sweeping & 0\% & 0\%  & 9\%  & 34\% & 47\% & 47\% & 47\% &        & 94\% \\
\end{tabular}
\end{ruledtabular}
\end{table}

\begin{figure}[!tb]
    \centering
    \includegraphics[width=0.48\textwidth]{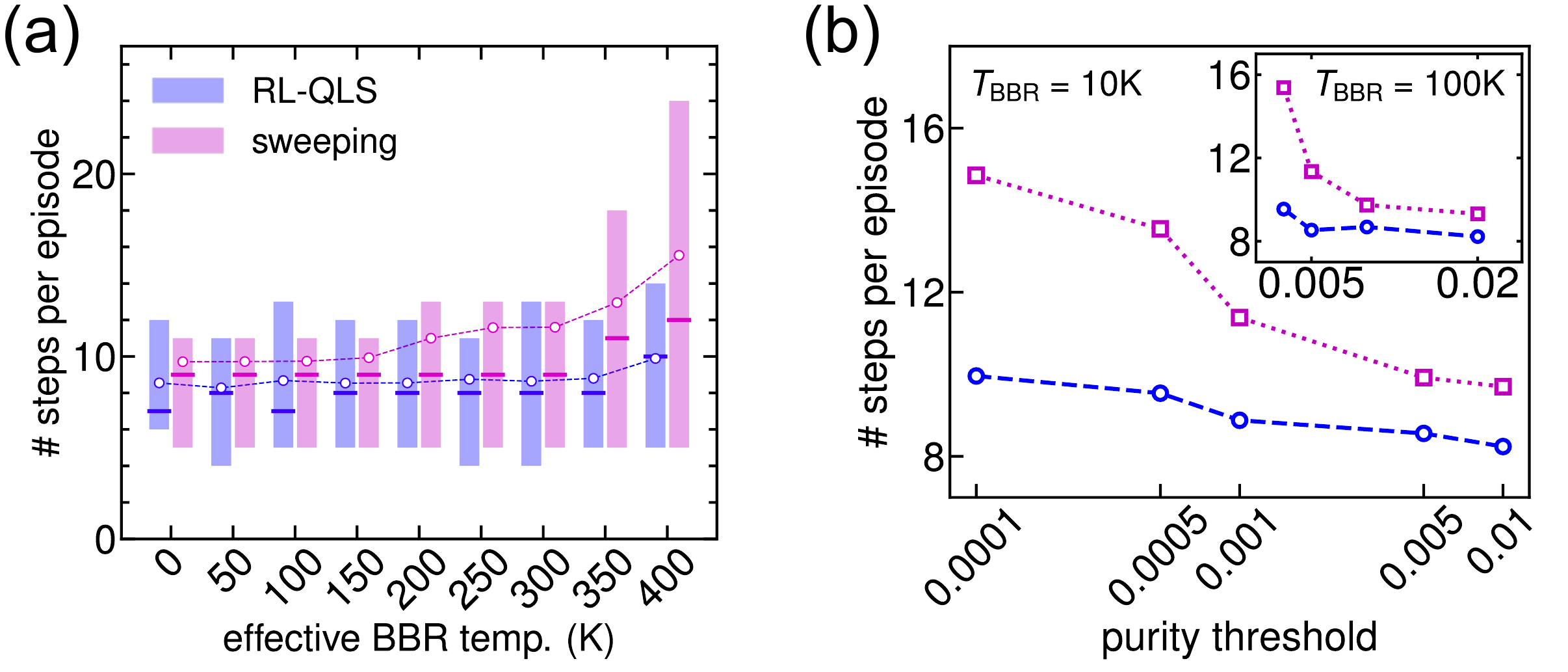}
    \caption{ \label{fig:fig3} 
    \textbf{(a)} 
    The mean number and distribution of steps (i.e., episode lengths) to prepare a pure molecular state under different magnitudes of thermal radiation with RL-QLS and the sweeping protocols.
    The strength of the thermal radiation is quantified by effective BBR temperatures.
    Circular markers denote the mean numbers of steps and are connected by dashed lines.
    Short horizontal lines indicate the medians, and boxes denote the interquartile ranges (25\%–75\% of the distributions).
    The initial population of the molecular states follows a Boltzmann distribution at 300 K.
    The purity threshold is 0.01, the same as that used in Fig.~\ref{fig:fig2}.
    \textbf{(b)} Mean number of steps to prepare a pure molecular state for different purity thresholds, with the effective BBR temperatures at 10 and 100 K. 
    The number of energy levels included in the simulation is constant across data points with different effective BBR temperatures. 
    The distribution of the episode lengths is reported in Fig.~S7 \cite{supplemental}.}
\end{figure}

Fig.~\ref{fig:fig3} examines the performance of RL-QLS subject to environmental thermal radiation (TR), one major source of noise in molecular control \cite{liu2023quantum}.
The strength of TR is quantified by effective black body radiation (BBR) temperature, $T_{\rm BBR}$. 
TR drives the system towards thermal equilibrium and thus hinders the state preparation progress (e.g. purple in Fig.~\ref{fig:fig3}a, $T_{\rm BBR}=400$ K v.s. 0 K).
An environment with stronger thermal noise requires more steps to prepare a pure state, 
and the RL agent is able to complete the task with nearly the same small number of pulses under moderate TR noises, a clear advantage (blue v.s. purple).
Fig.~\ref{fig:fig3}b further examines the degree to which a pure state can be prepared.
The TR noise limits the achievable purity of the prepared state (Fig.~S6 \cite{supplemental}), and increased episode lengths are needed when the threshold tightens.
Consistent with previous results, the RL-QLS protocol outperforms the referenced one to prepare a pure state up to a purity of 0.9999 at $T_{\rm BBR} = 10$~K.
In contrast, a recent experimental demonstration of CaH$^+$ state preparation and measurement using a Bayesian state tracking scheme achieved 0.998 fidelity at $T_{\rm BBR} = 16$~K \cite{chaffee2025high}.
Overall, Fig.~\ref{fig:fig3} demonstrates that RL-QLS can be readily adapted to mitigate realistic experimental disturbance.

\begin{figure}[!tb]
    \centering
    \includegraphics[width=0.48 \textwidth]{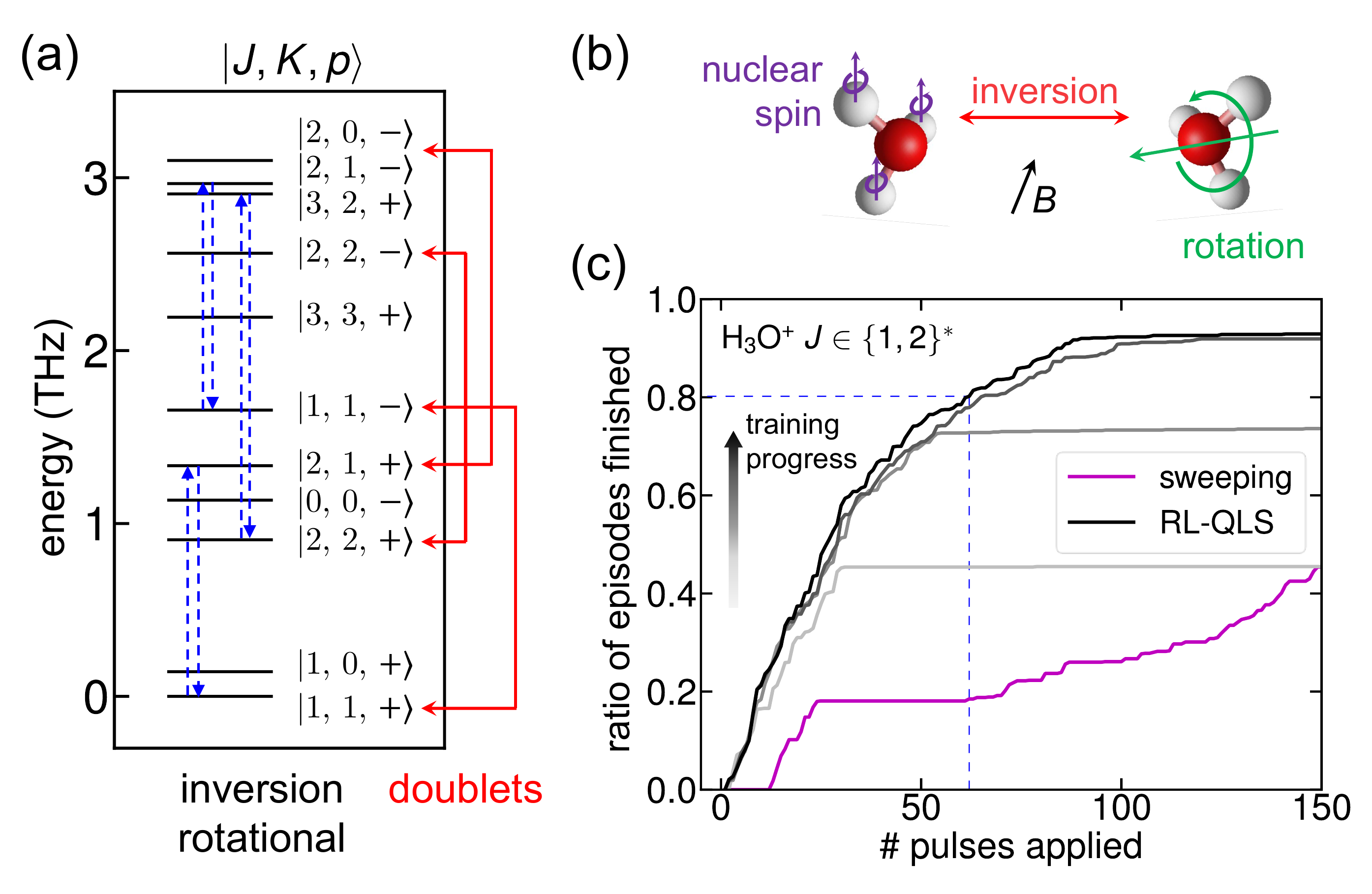}
    \caption{\label{fig:fig4}
    (\textbf{a}) Energy level diagram of H$_3$O$^+$ featuring the low-lying rotational $J \in \{1,2\}$* manifolds (energies and Rabi rates in Tabs. S2--3, pulse library in Fig.~S10 \cite{supplemental}). 
    The asterisk indicates that $J=3$ states with energies below the highest $J=2$ state are included in the simulations (99.8\% of the populations at 20~K).
    A $\ket{J, K}$ rotational manifold splits into doublets for two parities of the inversion mode (connected with red lines), and two-photon cross-$J$ pulses (blue arrows) are necessary to address this transition degeneracy between the inversion doublets. 
    (\textbf{b}) The complexity in the level structure of the
    hydronium ion arises from its multiple coupled, internal degrees of freedom.
    (\textbf{c}) Percentage of finished episodes v.s.~the number of pulses applied for H$_3$O$^+$ ion control, with 130 states and a library of 218 pulses to choose from.
    The purity threshold is set to 0.01.
    The dashed lines indicate that 80\% of episodes terminate within 62 pulses with RL-QLS state preparation.
    The sequence of RL-QLS curves represents the training progression, with darker curves (gray to black) corresponding to later training steps. Details of the hyperparameter tuning underlying these results are reported in Sec.~SD \cite{supplemental}.}
\end{figure}

Fig.~\ref{fig:fig4} addresses the scalability challenge, applying RL-QLS to a polyatomic ion. We aim to enable precision measurements of the inversion transition frequencies of hydronium (H$_3$O$^+$) in an ion trap with controlled systematics.
Here we focus on state preparation;  we report experimental considerations and the computation of the energy levels and the coupling rates in a subsequent article \cite{hydronium_level}.
The control task becomes more complex as many more states are thermally accessible.
In addition, originating from the two parity states of the inversion mode, the rotational manifolds split into doublets with similar level structures, leading to degenerate transitions within each doublet (Fig.~\ref{fig:fig4}a-b).
As a result, pulses that drive $\Delta J= \pm 1$ transitions (blue arrows) are required to separate the population in doublets.
Those pulses are not necessary in CaH$^+$ control---even when higher rotational manifolds are included (Fig.~S8 \cite{supplemental}, the high success ratios for CaH$^+$ ion control, $J \in \{1,2,3,4\}$ and $J \in \{1,2,3,4, 5, 6\}$ for the two panels, with 48 and 96 populated states, respectively), as the Zeeman and hyperfine split monotonically increases as $J$ increases.

To address these challenges, we introduce two core components into RL-QLS: quantum MDP modeling and a physics-informed reward function to promote exploration in learning (Sec.~SD \cite{supplemental}).
The quantum MDP \cite{barry2014quantum} modeling explicitly incorporates the measurement process in the $Q$-value estimate update and improves the learning efficiency (Fig.~S9 \cite{supplemental}, the loss function obtained with qMDP modeling is three magnitudes smaller than that with MDP modeling); 
the reward function is modified to discourage applying a pulse if the resulting state closely resembles the previous one.
As shown in Fig.~\ref{fig:fig4}c,
RL-QLS operates effectively in the molecular Hilbert space of H$_3$O$^+$
to prepare a pure molecular state from a Boltzmann mixture at $20$ K. 
At the target purity threshold of 0.01, RL-QLS yields more successful terminations with fewer pulses and reaches a high success ratio plateau (93.4\%) much more rapidly than the reference protocol. 
Exploration of this high-dimensional state-action space still relies heavily on the choice of hyperparameters, which lack direct interpretability in the control simulations.

We finally note that quantum MDP modeling, in addition, introduces a simple, approximate solution to the measurement errors. 
Since only the measurement outcomes, not populations, are directly available in QLS experiments, the control process is intrinsically a partially observable MDP.
When we assume an ideal noise-free environment, the population is fully determined by the measurement outcome sequence.
Similarly, in a realistic environment with measurement infidelities, 
measurement outcomes (together with the initial state) yield a {\it belief state}, $b(s)$, a probabilistic distribution of the current {\it state}.
The state-action value can then be approximated \cite{littman1995learning} using that of the corresponding fully observable environment, 
$Q(b(s),a) \approx \sum_s b(s) Q(s,a)$.
That is, without additional training, the $Q(s,a)$ value from this work provides a first-order solution to the partial observability arising from measurement infidelity.

\section{Conclusions}
In summary, the RL-QLS theoretical framework integrates quantum chemistry, AMO physics, and AI approaches to control the quantum state of a trapped molecular ion.
Combined with projective measurements realized with quantum logic spectroscopy (QLS),
the reinforcement learning (RL) agent leverages historical information of pulse choices and measurement outcomes to perform efficient and robust single state preparation.
RL-QLS is especially powerful for polyatomic molecular control where complex rovibrational structures emerge and an abundance of occupied states are of interest.
RL-QLS decision trees (Fig.~\ref{fig:fig2}d) can be directly implemented in experiments with minimal real-time computational cost.
We also note that the RL-QLS framework can be broadly applied to other state preparation problems where projective measurements are not realized by QLS, for example, in large-scale quantum computing architectures that utilize indirect ancilla qubit measurements for error correction \cite{nielsen2010quantum}.

This work may spark future developments at the intersection of physical science and AI.
Naturally, the utilization of other RL algorithms and neural network architectures 
that can effectively explore the immense state-action space would be beneficial for the control with even more molecular complexity.
In addition, the present method assumes prior knowledge of molecular energy level structure with sufficient precision to resolve the transitions used in the control protocol. 
Designing RL-based protocols that are resilient to perturbative shifts in the energy levels and coupling rates would further improve the applicability of the method.
At the same time, RL could also offer a complementary tool to understand the uncertainties in the molecular energy levels and coupling rates from a bottom-up perspective.
In conclusion, the RL-QLS framework opens possibilities at the nexus of AI-enabled precision control, quantum information science, and AMO physics, catalyzing future advancements in precision measurements and quantum metrology.

\section*{Acknowledgements}
The authors acknowledge Kristian D. Barajas, Dr. Muhammad M. Khan, Byoungwoo Kang, Dr. Zhong Zhuang, Dr. Tingrei Tan, Dr. Yu Liu and Dr. Hannah Knaack for helpful discussions of various aspects of this work.
This research used resources of the National Energy Research
Scientific Computing Center, a DOE Office of Science User Facility supported by the Office of Science of the U.S. Department of Energy under Contract No. DE-AC02-05CH11231.
This work was supported by NSF CAREER Award under grant number ECCS 2246394, NSF QuSeC-TAQS 2326840, NSF ExpandQISE 2231387, and NSF Physics 2309315. P.N. gratefully acknowledges support from the Gordon and Betty Moore Foundation grant No. GMBF 8048 and from the John Simon Guggenheim Memorial Foundation (Guggenheim Fellowship). D.R.L. acknowledges support from the Gordon and Betty Moore Foundation under grant No. GMBF 12252.

%


\end{document}


\title{Molecular Quantum Control Algorithm Design by Reinforcement Learning}
\author{Anastasia Pipi}
\thanks{These two authors contributed equally.} 
\affiliation{Department of Physics and Astronomy, University of California, Los Angeles (UCLA), California 90095, USA}
\author{Xuecheng Tao$^\dagger$}
\thanks{These two authors contributed equally.} 
\affiliation{Division of Physical Sciences, College of Letters and Science, University of California, Los Angeles (UCLA), California 90095, USA}
\author{Arianna Wu}
\affiliation{Department of Chemistry and Biochemistry, University of California Los Angeles (UCLA), Los Angeles, California 90095, USA}
\author{Prineha Narang$^\ddagger$}
\affiliation{Division of Physical Sciences, College of Letters and Science, University of California, Los Angeles (UCLA), California 90095, USA}
\affiliation{Electrical and Computer Engineering Department, University of California, Los Angeles (UCLA), California, 90095, USA}
\author{David R. Leibrandt$^\S$}
\affiliation{Department of Physics and Astronomy, University of California, Los Angeles (UCLA), California 90095, USA}

\date{\today}

\clearpage
\noindent \textbf{Supplementary Material for  
``Molecular Quantum Control Algorithm Design by Reinforcement Learning''}\\ 
{Anastasia Pipi$^*$, Xuecheng Tao$^{*, \dagger}$, Arianna Wu, Prineha Narang$^\ddagger$, 
and David R. Leibrandt$^\S$} \\

\noindent Contact author: $^\dagger$xuechengtao@gmail.com. \\
Contact author: $^\ddagger$prineha@ucla.edu. \\
Contact author: $^\S$leibrandt@ucla.edu.

\freefootnote{* These two authors contributed equally.}

\setcounter{figure}{0}
\setcounter{table}{0}
\setcounter{section}{0}
\setcounter{equation}{0}
\makeatletter
\renewcommand{\thefigure}{S\arabic{figure}}
\renewcommand{\thetable}{S\arabic{table}}
\renewcommand{\thesection}{Sec.~S\Roman{section}}
\renewcommand{\thesubsection}{Sec.~S\Alph{subsection}}
\renewcommand{\theequation}{S\arabic{equation}}

\subsection{State evolution in the preparation process} \label{sec_supp:purification_evolution}
We consider a molecular spectroscopy ion that occupies the ground electronic and vibrational states, 
while a substantial number of its rotational states are populated due to thermal radiation. 
The resulting density matrix of the rotational manifold is 
$\rho = \sum_{\mathcal{J}=1}^{N_S} 
P_\mathcal{J} \ket{\mathcal{J}} \bra{\mathcal{J}} $, with $N_S$ the number of states, 
and $P_\mathcal{J} = e^{-\beta E_\mathcal{J}} / \sum_\mathcal{J}  e^{-\beta E_\mathcal{J}} $ following a Boltzmann distribution.
A co-trapped Ca$^+$ logic ion is prepared in the 
$\ket{D_{5/2}}$ electronic
state, and a motional mode $\ket{k}$ is shared between the spectroscopy ion and the logic ion.
In this section, we present the equations for the case where one excited motional state is accessible (applicable in the Lamb-Dicke regime) for illustration purposes, 
approximating that the projective measurement gives identical signals for $k=1$ and higher excited motional states $k>1$. 
The applied laser pulse drives a molecular transition and results in a quantum state with a density matrix of 
\begin{align} \label{eq:density_matrix_on_pulse}
    \rho = \sum_{\mathcal{J}=1}^N P_\mathcal{J} 
    \Bigg{[} &\closeds{\sum_{\mathcal{J}'} u_{\mathcal{J} \mathcal{J}'} \ket{\mathcal{J}', 0} 
            + \sum_{\mathcal{J}'} v_{\mathcal{J} \mathcal{J}'} \ket{\mathcal{J}', 1} } \nonumber \\
          &\closeds{\sum_{\mathcal{J}'} u^*_{\mathcal{J} \mathcal{J}'} \bra{\mathcal{J}', 0} 
            + \sum_{\mathcal{J}'} v^*_{\mathcal{J} \mathcal{J}'} \bra{\mathcal{J}', 1} }
    \Bigg{]}.
\end{align}
In Eq.~\ref{eq:density_matrix_on_pulse}, the molecular ion has a probability of $P_\mathcal{J}$ 
to occupy the state $\ket{\mathcal{J}}$ and interacts with the laser pulse, 
and $u_{\mathcal{J} \mathcal{J}'}$ and $v_{\mathcal{J} \mathcal{J}'}$ describes the time evolution of a pure state $|\mathcal{J}, 0\rangle$ under the influence of the applied pulse.

Subsequently, a motional sideband pulse is applied on the logic ion in order to map the motional state onto the Ca$^+$ ion internal state, and a projective measurement of the motional state can now be performed 
with a fluorescence observation on the quantum state of the logic ion (i.e. to identify whether the Ca$^+$ ion is in state $\ket{D_{5/2}}$ or $\ket{S_{1/2}}$).
The projective measurement collapses the quantum state probabilistically according to the outcome and can be formalized as positive operator-valued measure (POVM).
The POVM is defined with two orthogonal projections on the motional state, 
$O_0=\ket{0}_{\rm mot} \bra{0}_{\rm mot}$ and $O_1=\ket{1}_{\rm mot} \bra{1}_{\rm mot}$, and 
$O_0 + O_1 = \mathcal{I}$.
When the motional state in Eq.~\ref{eq:density_matrix_on_pulse} is measured, the probability of each measurement is given by $p_k = {\rm tr} \closeds{\rho O_k}$, i.e.
\begin{align} \label{qevo_2}
p_0 = \sum_\mathcal{J,J'} P_\mathcal{J} |u_{\mathcal{J} \mathcal{J}'}|^2, \px{4}
p_1 = \sum_\mathcal{J,J'} P_\mathcal{J} |v_{\mathcal{J} \mathcal{J}'}|^2
\end{align}
with $k$ denoting the measured outcome of the motional state. 
After the measurement and the subsequent motional state cooling, the state of the molecular spectroscopy ion is 
\begin{align} \label{qevo_3}
    \rho=
    \begin{cases}
     (1/p_0) \sum_\mathcal{J'} \closeds{\sum_{\mathcal{J}} P_\mathcal{J} |u_{\mathcal{J} \mathcal{J}'}|^2} \ket{\mathcal{J}'} \bra{\mathcal{J}'}, \px{4} &{\rm if \ } k=0, \\  
     (1/p_1) \sum_\mathcal{J'} \closeds{\sum_{\mathcal{J}} P_\mathcal{J} |v_{\mathcal{J} \mathcal{J}'}|^2} \ket{\mathcal{J}'} \bra{\mathcal{J}'}, \px{4} &{\rm if \ } k=1. \\
    \end{cases}
\end{align}
Laser pulses and projective measurements are then repeated many times until a pure state has been prepared.
Note that we consider the molecular state to be pure when the probability $P_\mathcal{J}$ for the molecule to be in an arbitrary but known state $\ket{\mathcal{J}}$ is greater than $1-\eta$, where $\eta \ll 1$ is the state preparation infidelity threshold.

\subsection{Time evolution with the adiabatically-eliminated Hamiltonian} \label{sec:time_evolution}
Accurate construction of the transition matrices (TMs, i.e. $\mathcal{A}$s) in the Markov decision process is important so that the model faithfully reflects the physical process. 
Those TMs describe the effects of the pulse operations on the molecular quantum state 
and allow for a compact description of the system's time evolution.
Since motional state cooling is performed after every pulse/measurement in the current state preparation scheme and coherences are destroyed by the cooling as indicated in Eq.~\ref{qevo_3}, only the population vectors are tracked during the time evolution (Fig.~1c). 
This way, the description of population dynamics is condensed to a compact set of $2N_A$ TMs (with the size of $N_S \times N_S$) for input into the RL calculations.

\begin{figure}[hb]
    \centering 
    \begin{minipage}[b]{0.49\textwidth}
     \centering
     \includegraphics[width=\textwidth]{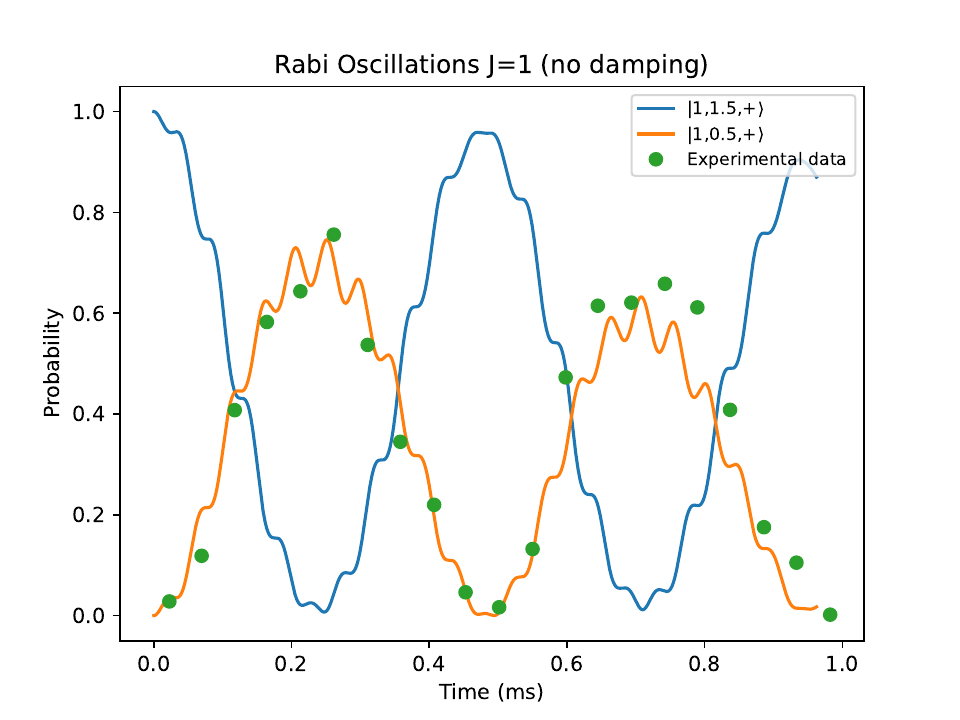}
    \end{minipage}
    \hfill
    \begin{minipage}[b]{0.49\textwidth}
     \centering
     \includegraphics[width=\textwidth]{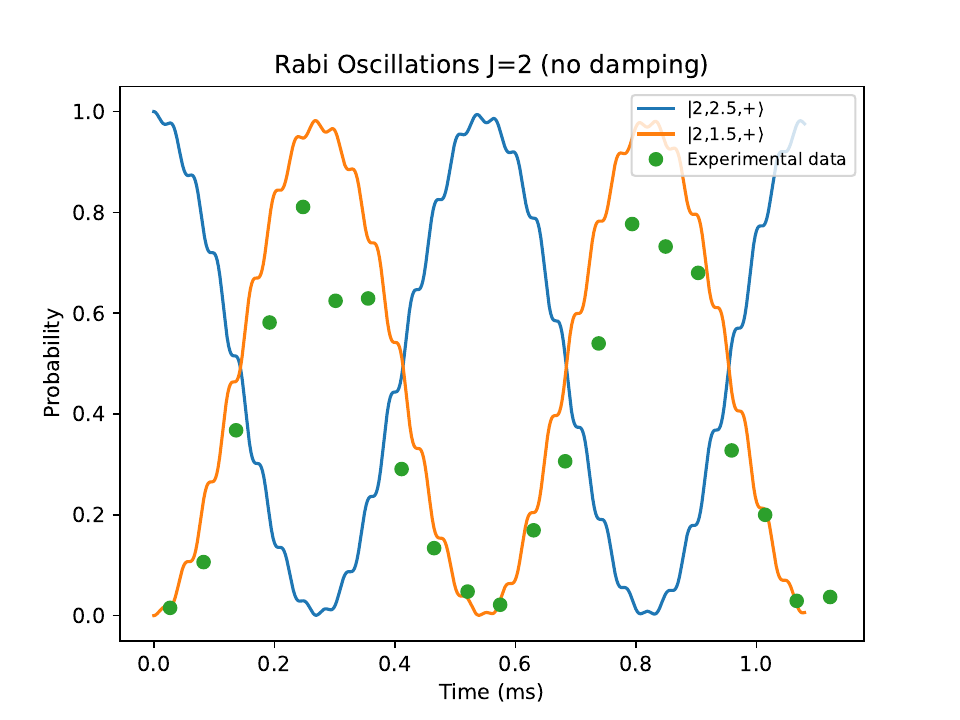}
    \end{minipage}
    \caption{Rabi oscillations between states $\ket{J,-J-1/2,-}\leftrightarrow \ket{J,-J+1/2,-}$ for the $J=1,2$ manifolds are simulated without noise, and compared to experimental data \cite{chou2017preparation}.
    In ~\ref{fig:ex_fig1}, the plots are obtained by dynamically evolving the system using the \textit{mesolve} function in QuTip. The initial state vector is set to have a concentrated population on state $\ket{J,-J+1/2,+}$. 
    }
    \label{fig:ex_fig1}
\end{figure}

We evaluate the TMs by numerically solving the time-dependent Schr\"{o}dinger Equation with the realistic pulse characteristics (frequencies, amplitudes, and durations). 
The total Hamiltonian of the system consists of two components, the time-independent Hamiltonian, $H_0$---molecular hyperfine Hamiltonian, Eq.~\ref{eq:cah_hamiltonian} or Eq.~\ref{eq:h3o_hamiltonian}, together with the harmonic trapping Hamiltonian, $\hbar \omega_{\rm mot} (a_{\rm mot}^{\dagger} a_{\rm mot} + 1/2 )$---and the time-dependent pulse-molecule interactions, 
\begin{align} \label{eq:int_hamiltonian}
    H_{\text{int}}(t) = \sum_{\ket{\mathcal{J}} \to \ket{\mathcal{J}'}} \frac{\Omega_{\mathcal{J}, \mathcal{J}'}}{2} 
    \left[ e^{i \left[\lambda_{\rm LD}(a_{\rm mot}+a_{\rm mot}^{\dagger}) - \omega t)\right]} 
    |\mathcal{J}' \rangle \langle \mathcal{J}| + h.c. \right]
\end{align}
where $\Omega_{\mathcal{J},\mathcal{J}'}$ is the Rabi frequency 
for the two-photon stimulated Raman transition $\ket{\mathcal{J}} \to \ket{\mathcal{J}'}$ and is obtained by adiabatically eliminating the intermediate states \cite{chou2017preparation} (see Eq.~\ref{eq:raman-rabi}).
$\lambda_{\rm LD}$ is Lamb-Dicke parameter.
$a_{\rm mot}$ and $a_{\rm mot}^\dagger$ are the annihilation and creation operators for the motional mode, and $\omega$ is the laser frequency.
For numerical stability, only the first-order terms in the Lamb-Dicke parameter are kept in the simulations,  
i.e. we assume that $\exp[i\lambda_{\rm LD}(a_{\rm mot}+a_{\rm mot}^{\dagger})] = \mathbb{I} + i\lambda_{\rm LD}(a_{\rm mot}+a_{\rm mot}^{\dagger}$) for $\lambda_{\rm LD} = 0.09$.
The Hamiltonian can also be expressed in the interaction picture with respect to $H_0$ for numerical efficiency, where
\begin{align} \label{eq:int_hamiltonian_rf}
    H_{\text{int}}'(t) & := e^{i H_0 t} H_{\text{int}}(t) e^{-i H_0 t} \nonumber \\
    &= \sum_{\ket{\mathcal{J}} \to \ket{\mathcal{J}'}} \frac{\Omega_{\mathcal{J}, \mathcal{J}'}}{2} 
    \left[ 
    \left[1 + i \lambda_{\rm LD}(a_{\rm mot}e^{-i \omega_{\rm mot}t} +a_{\rm mot}^{\dagger}e^{i \omega_{\rm mot}t} 
    )\right] e^{(E_{\mathcal{J}'} - E_\mathcal{J})/\hbar - i\omega  t} 
    |\mathcal{J}' \rangle \langle \mathcal{J}| + h.c. \right].
\end{align}
The numerical calculations are performed with the QuTiP software package \cite{johansson2012qutip, johansson2013qutip}. 
As plotted in Fig.~\ref{fig:ex_fig1}, a comparison of the Rabi oscillations between the simulation and the experiment observations \cite{chou2017preparation} demonstrates the effectiveness of the simulation protocol.

We also consider the effects of thermal radiation (TR), a major source of experimental noise. 
TR is modeled as black body radiation (BBR) at an effective temperature $T_{\rm BBR}$.
Competing with the aforementioned efforts to concentrate the population to a specific state, BBR drives the state populations back to their thermal equilibrium. 
Time evolution of the probability distribution of each state follows the coupled rate equations under the influence of BBR,
\begin{equation} \label{eq:bbr_main}
    \frac{dP_\mathcal{J}(t)}{dt} = 
    -\sum_{\mathcal{J}' \neq \mathcal{J}} \mathcal{R}_{\mathcal{J} \rightarrow \mathcal{J}'} \ P_\mathcal{J} 
    + \sum_{\mathcal{J}' \neq \mathcal{J}} \mathcal{R}_{\mathcal{J} \rightarrow \mathcal{J}'} \ P_\mathcal{J'},
\end{equation}
where $P_\mathcal{J}$ is the statistical population of occupying state $\ket{\mathcal{J}}$, 
$\mathcal{R}_{\mathcal{J} \rightarrow \mathcal{J}'}$ is the rate at which population from state $\mathcal{J}$ transfers to state $\mathcal{J}'$.
The rates are calculated using Einstein's A and B coefficients \cite{liu2023quantum} for spontaneous and stimulated transitions (Eq.~\ref{eq:bbr_rate_coefficients}). 
In the simulations, we discretize the time into small intervals and approximate the BBR influence as a first-order expansion with respect to the discretized timestep to propagate Eq.~\ref{eq:bbr_main}.  

Considering a mixed state that is under the influence of BBR, after a small time step $\delta t$, the initial probability distribution $\textbf{P}=\{P_1, P_2, ..., P_{N_S}\}$ evolves into 
\begin{equation}
    \textbf{P}(t+\delta t) = T \textbf{P}(t), 
\end{equation}
where 
\[
T = \mathbb{I} + 
\begin{bmatrix}
-\sum_{k=1}^{N}\mathcal{R}_{1\rightarrow k} & \mathcal{R}_{2\rightarrow 1} & \dots & \mathcal{R}_{N\rightarrow 1} \\
\mathcal{R}_{1\rightarrow 2} & -\sum_{k=1}^{N}\mathcal{R}_{2\rightarrow k} & \dots & \mathcal{R}_{N\rightarrow 2} \\
\vdots & \vdots & \ddots & \vdots \\
\mathcal{R}_{1\rightarrow N} & \mathcal{R}_{2\rightarrow N} & \dots & -\sum_{k=1}^{N}\mathcal{R}_{N\rightarrow k}
\end{bmatrix} \delta t
\]
under a first-order expansion to $\delta t$.
Therefore, for a duration of $\Delta t$, the BBR evolves the state population as
\begin{equation} \label{eq:bbr_evo}
    \textbf{P}(t+\Delta t) = 
    T^{\Delta t/\delta t} \textbf{P}(t).   
\end{equation}
We propagate the system dynamics under the influence of laser pulses and BBR effects sequentially, which is a good approximation to the actual dynamics because the laser pulses are much shorter than the time constant for BBR-driven transitions.

\subsection{Computational details: sweeping protocol, Hamiltonian, pulses } \label{sec_supp:cd_evo}
\textbf{Sweeping protocol ---} We note that the sweeping protocol presented in the article differs from the experimental protocol introduced by NIST.
Specifically, in NIST experiments, a number of pulses are applied between the projective measurements \cite{chou2017preparation, liu2023quantum}, while in our modeling, one projective measurement follows one applied pulse in every preparation step.
We choose to perform measurements between the pulses to receive feedback on the instantaneous populations.
Nevertheless, we use the sweeping protocol as a reference to report the reinforcement learning results and keep the name `sweeping protocol' to
credit the original authors for the development of the idea.

\textbf{Molecular Hamiltonian: CaH$^+$ ---} The time-independent molecular Hamiltonian for CaH$^+$, under the influence of an external magnetic field,  is \cite{chou2017preparation}
\begin{equation} \label{eq:cah_hamiltonian}
H_{\rm mol} =\frac{1}{\hbar}( 2\pi R \hat{\mathbf{J}}^{2}-g\mu_N\hat{\mathbf{J}} \cdot \mathbf{B} - g_I \mu_N \hat{\mathbf{I}} \cdot \mathbf{B}
- 2\pi c_{IJ}\hat{\mathbf{I}} \cdot \hat{\mathbf{J}}),
\end{equation}
where $\hat{\mathbf{J}}$ is the rotational angular momentum of the molecule, $\hat{\mathbf{I}}$ is the nuclear spin operator, $\mathbf{B}$ is the magnetic field. $R$ is the rotational constant, $\mu_N$ is the nuclear magneton, $g$ and $g_I$ are the rotational and nuclear $g$-factors, respectively, and $c_{IJ}$ is the spin-rotation constant. 
The eigenstates are denoted as $\ket{J,m,\xi}$, where $J$ is the total rotational quantum number, $m$ is the total magnetic quantum number and $\xi$ indicates the relative sign in the eigenstate coefficients.

\textbf{Raman Rabi rates: CaH$^+$ ---} For a given set of two pump/Stokes pulses (with known amplitudes, polarization, frequencies, and duration), the Raman-Rabi frequency is given by
\begin{align} \label{eq:raman-rabi}
    \Omega_{if} = \frac{1}{4\hbar^2}\sum_M \left( \frac{\bra{f} \mathbf{d} \cdot \mathbf{E}_2 \ket{M} \bra{M}  \mathbf{d} \cdot \mathbf{E}_1 \ket{i}}{\omega_{iM} - \omega_1} + \frac{\bra{f} \mathbf{d} \cdot \mathbf{E}_1 \ket{M} \bra{M} \mathbf{d} \cdot \mathbf{E}_2 \ket{i}}{\omega_{iM} + \omega_2} \right)
\end{align}
where $\mathbf{E}_1,\mathbf{E}_2$ are the two driving fields with respective frequencies $\omega_1,\omega_2$, $\mathbf{d}$ is the dipole operator, $\omega_{iM} = ({E_M-E_i})/{\hbar}$ is the frequency difference of the initial, $\ket{i}$, and intermediate $\ket{M}$, states.
The absorption pulse produces a $\pi$-polarized field and the stimulated emission pulse produces a $\sigma^+$/$\sigma^-$-polarized field. More details on the above expression can be found in \cite{chou2017preparation, collopy2023effects}, and pg.~22--23 of \cite{townes_microwave}.
It is worth mentioning that we do not apply the rotating wave approximation in Eq.~\ref{eq:raman-rabi}, 
because two-photon Raman transitions can utilize a pump/Stokes laser that is far detuned from the intermediate states (i.e. $|\omega_{iM}-\omega_1|$ is comparable to $|\omega_{iM}+\omega_2|$). 
Faster Raman-Rabi rates indicate that the population transition takes less time to drive.
The amplitudes of the laser pulses are set the same as in the experiment \cite{chou2017preparation} such that the Rabi rate for transition $\ket{1,-3/2,-} \to \ket{1,-1/2,-}$ is $2\pi \times 2.087$~kHz.

\textbf{Molecular Hamiltonian and Rabi rates: H$_3$O$^+$ ---} Computational details of the energy levels and the coupling rates of H$_3$O$^+$ ion are reported in a subsequent article \cite{hydronium_level}. Briefly, the H$_3$O$^+$ energy levels used in the current molecular control simulations are the eigenstates of the Hamiltonian
\begin{equation} \label{eq:h3o_hamiltonian}
H = \hat{H}_{\text{inv-rot}} + \hat{H}_\text{Zeeman} + \hat{H}_\text{s-r},
\end{equation}
which consists of three contributions, inversion-rotational Hamiltonian, nuclear and rotational Zeeman interaction, and nuclear spin-rotation coupling. 
A magnetic field of 0.36 mT is applied to lift the $m$ degeneracy. 
More specifically, $\hat{H}_\text{Zeeman}$ is comparable to $\hat{H}_\text{s-r}$ in energy scale at this magnetic field strength. 
The energy levels are reported in Table~\ref{table:h3o_energy}. Rabi rates of the two-photon transitions are then calculated with adiabatic elimination (Eq.~\ref{eq:raman-rabi}) and are reported in Table~\ref{table:h3o_rabi_rate}.

\textbf{Black Body Radiation ---} The BBR rates are evaluated as  
\begin{align}   \label{eq:bbr_rate_coefficients}
    R_{i \to f} = \rho^{\rm BBR}_{i,f}(\omega)  B_{i,f} + A_{i,f}
\end{align}
where $R_{i \to f}$ is the transition probability from state $i$ to state $f$, $A_{i,f}$ is stimulated emission probability, $B_{i,f}$ is the  Einstein coefficient for stimulated emission and $\rho^{\rm BBR}_{i,f}(\omega)$ is the energy density of black body radiation per unit bandwidth at angular frequency $\omega$. Details on calculating these coefficients can be found in Chapter 9 of Ref.~\onlinecite{corney_book}. 

\begin{figure}[tb]
    \centering
    \includegraphics[width=0.5\textwidth]{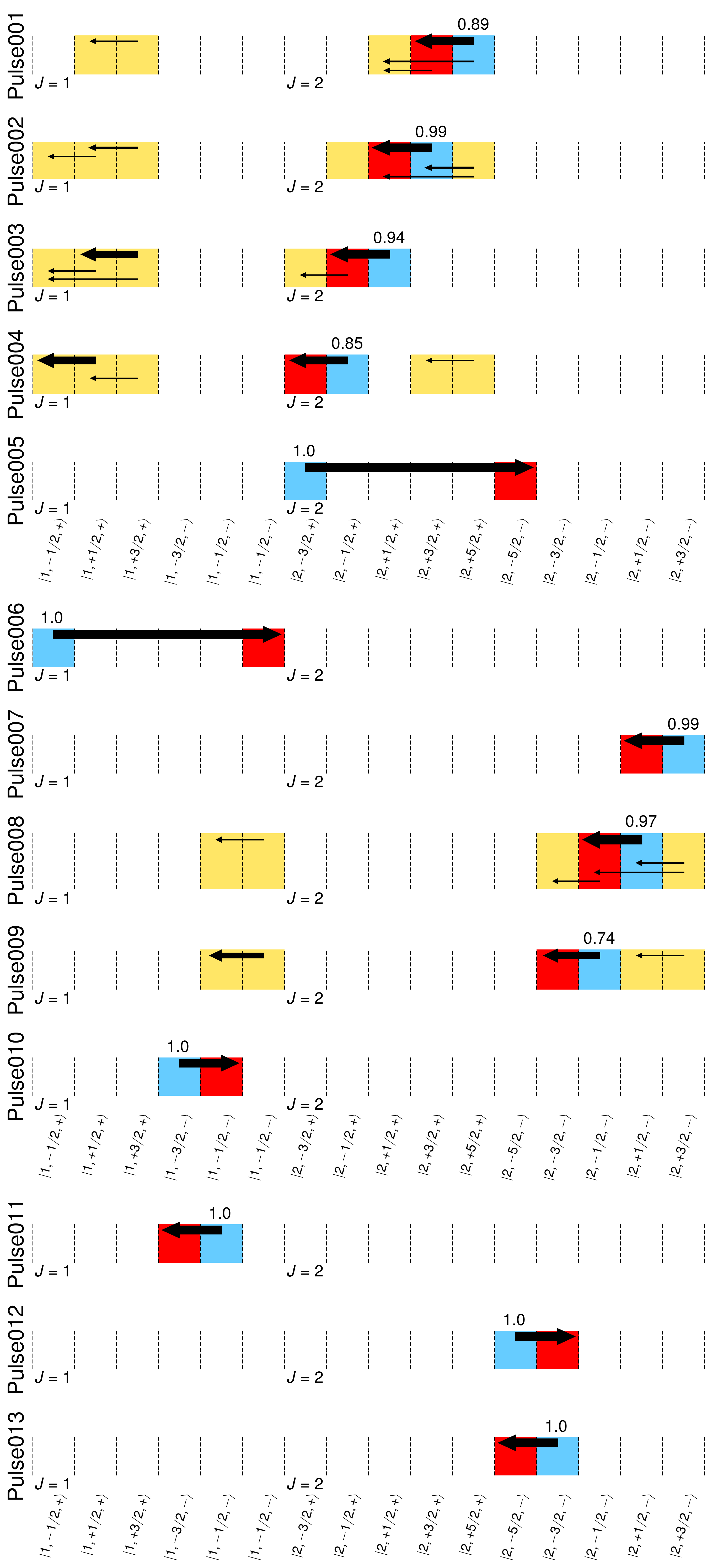}
    \caption{State population transfer driven by the 13 pulses described in Fig.~2 for $J=\{1, 2\}$ rotational manifold. 
    The main transition is color-coded as arrows from blue to red boxes, and the amount of the population transition is listed above the arrow. 
    The width of the arrows indicates the amount of the population transition and for each pulse, the most significant five transitions are plotted.
    By driving the blue-sideband transitions and controlling the pulse polarization, those pulses drive the population transfer in one direction.
    }
    \label{fig:ex_fig2}
\end{figure}

\textbf{Pulses that drive multiple transitions in \boldmath$J=1,2$ ---} In Fig.~2 (see also Fig.~\ref{fig:ex_fig2}), pulses 3, 4, 9 are shown to drive multiple transitions in different $J$-manifolds. The pulse frequencies and duration are chosen as the average of the two transition frequencies and $\pi$-pulse durations, respectively, except for pulse 3. For pulse 3, the Rabi frequency for one of the transitions is almost three times faster than the other, thus the pulse duration is set to the $\pi$-pulse duration of the slower transition.

\textbf{Computational Details: calculating the evolution transition matrices ---} We perform the simulation of state evolution with QuTip \cite{johansson2012qutip, johansson2013qutip} version 4.7.
We simulate the time-dependent Schr\"{o}dinger Equation with \textit{qutip.sesolve} and \textit{qutip.propagator} functions with the aforementioned Hamiltonians (Eq.~\ref{eq:cah_hamiltonian} or Eq.~\ref{eq:h3o_hamiltonian} with Eq.~\ref{eq:int_hamiltonian}, or Eq.~\ref{eq:int_hamiltonian_rf}), either in the laboratory frame or in the interaction picture.
We use the \textit{zvode} ODE integrator as implemented in the SciPy library \cite{2020SciPy-NMeth} and a timestep of $\sim$1 $\mu$s is used.
For the simulation of Raman transitions that are within the same rotational manifold, the tolerance thresholds {\it atol} and {\it rtol} are set to $10^{-8}$ and $10^{-6}$; for transitions that across rotational manifolds, the thresholds are $10^{-6}$ and $10^{-4}$. 
Despite the state dynamics being described in terms of population vectors (as in Fig.~1c), we keep the coherence in the evolution of the quantum state, and only make this approximation, i.e. take the diagonal term in the end to resolve the transition matrices $\mathcal{A}$s.
The time evolution of the statistical mixture with the total Hamiltonian, $H_{\rm tot}$, is obtained by separately simulating the time evolution of each eigenstate of the $H_{\rm mol}$, i.e. for 
$\rho(0) = \sum_\mathcal{J} P_\mathcal{J} \ket{\mathcal{J}} \bra{\mathcal{J}}$,
\begin{align}
\rho(t) = e^{-i H_{\rm tot}t} \rho(0) e^{+ i H_{\rm tot}t} = \sum_\mathcal{J} P_\mathcal{J} \ket{\mathcal{J}'(t)} \bra{\mathcal{J}'(t)},
\end{align}
with $\ket{\mathcal{J}'(t)} = e^{-i H_{\rm tot}t} \ket{\mathcal{J}}$.

\textbf{Weak pulses ---} It is recognized that transitions with weaker Rabi rates (for a given amplitude of Raman pulses) are harder to drive in the experiments. In addition, Fig.~\ref{fig:ex_weakpulse} presents the difficulty in the numerical procedure to simulate the time evolution for weak pulses. 
Numerical difficulties decrease if the pulses are intended to drive the stronger transition, quantified by larger Rabi rates (e.g. the results for 1', 2', 3', 4' under column `mot2').
We note that the numerical difficulties are associated with the use of {\it zvode} solver, and can be ameliorated by including more motional states in the simulations when using QuTip software.
In this work, we construct the pulse library with pulses with Rabi rates $\gtrsim  2\pi \times 0.1$~kHz.

\subsection{Computational details: Reinforcement learning agents \label{sec_supp:rl_agent}}
\textbf{RL agents ---} We prefer the use of the off-policy $Q$-learning algorithm for the state preparation task because off-policy exploration can lead to better sample efficiency in the discrete action space. Nevertheless, we also examined the performance of another widely recognized RL algorithm, proximal policy optimization (PPO) \cite{schulman2017proximal}. It turns out that for CaH$^+$ $J$~$\in\{1,2\}$ systems, the PPO agent has a profoundly smaller success ratio of episode termination compared to the DQN agent, for tested batch size and learning rate hyperparameters across 3 magnitudes.

\textbf{$Q$-learning algorithm ---} The $Q$-learning works by finding the 
current action $a$ that maximizes the estimated expected cumulative reward
\begin{align}
a &= \arg \max_{a} Q(s,a)
\end{align}
with the state-action value function
\begin{align}
Q(s, a) = \mathbb{E} \closedm{\sum_{\tau=0}^T R_{t+1+\tau} | S_t=s, A_t=a},
\end{align}
and $T$ is the terminal step when the state with a purity of $1$ (with a small tolerance of $\eta$) is prepared.
$Q(s,a)$ are updated as a temporal difference learning through the process of the agent interacting with the environment.
In this work, we focus on the deep $Q$-learning algorithm \cite{mnih2013playing, mnih2015human, watkins1989learning} with a simple, fully-connected neural network, and the RL training and testing are performed with PyTorch software\cite{paszke2019pytorch}.
In this work, the mean squared error (squared L2) and smooth L1 loss are used in the state-action value network optimization.

\textbf{Computational Details: training the RL agents ---} We implement the reinforcement learning agents to propose future actions according to the agent-environment interaction in the Markov decision process (Fig.~1c). 
In this work, the generalized policy iteration of the task is completed with the model-free, temporal difference learning agent---the deep $Q$-learning (DQN) agent \cite{mnih2013playing, mnih2015human}.
We use experience replay and double-$Q$ networks for robust and efficient training.
We perform the hyperparameter tuning for the neural network update rate ($\tau$) and the learning rate ($r_l$) by the analysis of state-action trajectories and the resulting decision tree obtained with training under different hyperparameters.
The soft action selection is implemented with the $\varepsilon-$greedy algorithm and the exploration parameter,
\begin{align} \label{eq:rl_epsilon}
\varepsilon = \varepsilon_{\rm end} + (\varepsilon_{\rm end}- \varepsilon_{\rm start}) e^{- n_{\rm training} /\tau_\varepsilon}
\end{align}
decreases with the progress of the training. In Eq.~\ref{eq:rl_epsilon}, we choose a $\varepsilon_{\rm start}=1$ and $\tau_\varepsilon = 0.3 N_{\rm training}$ with $n_{\rm training}$ and $N_{\rm training}$ the current and the total number of training episodes, respectively.
We performed the training with $\varepsilon_{\rm end}=$0.005, 0.025, and 0.125.

\textbf{Hyperparameter tuning for CaH$^+$ }{\boldmath$J=1,2$}\textbf{ system} --- Since the agent-environment interaction in the example system is relatively simple, we find that training with a range of hyperparameters ($\tau=0.0005, 0.001, 0.0025$, $r_l=0.0005, 0.001$) can all lead to a RL agent with good performance. We report in Fig.~2 the reinforcement learning results obtained with $\tau=0.001$, $r_l=0.0005, \varepsilon_{\rm end} =0.005$.

\textbf{Hyperparameter tuning for CaH$^+$ }{\boldmath$J=1,2$}\textbf{ system with BBR} --- We scanned a range of hyperparameters ($\tau=0.00025$, $0.0005$, $0.001$, $0.0025$, $0.005$; $r_l=0.00025$, $0.0005$, $0.001$, $0.0025$, $0.005$, $\varepsilon_{\rm end} = 0.0125,0.025, 0.05$) to ensure the RL agent's performance for this noisy system. The calculation procedure is described as follows and is done for each effective BBR temperature separately.
(\emph{i}) During training, we calculated the mean number of steps (length) of the testing episode with the model from the last training episode. The model (referred to as model A) with the lowest mean was identified as having the optimal hyperparameters. 
(\emph{ii}) To confirm our best model choice, we repeated the training using the optimal hyperparameters identified previously. We saved episodes with a mean equal to or lower than the average mean of model A. 
(\emph{iii}) We then tested all these saved models and selected the one with the lowest mean as the final optimal model from this search.
The lowest total mean for each BBR temperature is plotted in Fig.~3a. 
The purity of the state under BBR disturbance (i.e. $\textbf{P}(t+\delta t)$ from Eq.~\ref{eq:bbr_evo}) is examined to determine whether the termination condition has been met.
The optimal hyperparameters for each BBR temperature are outlined in Table \ref{tbl:rl_bbr_hyperparas}. 
We used $\varepsilon_{\rm end} = 0.025$ for all temperatures, as this setting yielded the best results.
The same hyperparameters are used in the calculation of Fig.~3b.
\begin{table}[h]
    \centering
    \begin{tabular}{|c|c|c|c|c|c|}
    \hline
    \textbf{BBR Temp. (K)} & \textbf{$\tau$} & \textbf{$r_l$} &
    \textbf{BBR Temp. (K)} & \textbf{$\tau$} & \textbf{$r_l$} \\ \hline
    0   & 0.005 & 0.0005 & 50  & 0.0005 & 0.0005 \\ \hline
    100 & 0.0005  & 0.0005 & 150 & 0.0005  & 0.001  \\ \hline
    200 & 0.001  & 0.0005 & 250 & 0.0025  & 0.0005 \\ \hline
    300 & 0.0005  & 0.0005 & 350 & 0.001  & 0.0005  \\ \hline
    400 & 0.0005  & 0.0005 & & & \\ \hline
    \end{tabular}
    \caption{Optimal hyperparameters for each BBR temperature.}
    \label{tbl:rl_bbr_hyperparas}
\end{table}

\textbf{Physics-informed reward function} --- As the number of occupied states and required control pulses increases, the state-value function to learn, $Q(s,a)$, spans a larger space. In fact, the same procedure that works well for $J \in \{1, 2\}$ system leads to unsatisfactory learning performance when applied to $J \in \{1,2,3,4\}$ and $J \in \{1,2,3,4,5,6\}$ systems due to under-exploration. To this end, we leverage our experience with the smaller system to perform physics-informed learning.
Specifically, we set the reward function to additionally discourage the application of a pulse on the state $S_t$, if the resulting state and the previous one largely overlap each other,
i.e. if $o(S_t, S_{t+1}) > 1-1/N_S$ with $o(S_t, S_{t+1}) := S_{t}\cdot S_{t+1} / (|S_{t}||S_{t+1}|)$. The extent to which the overlap shall be discouraged introduces another empirical tuning parameter, $r_o$, in the practical implementation.
The optimal value of $r_o$ allows for a balanced consideration of fast completion and low overlap, resulting in the optimal performance of the training.
The optimal hyperparameter sets, including the one required to tune the physics-informed reward function, are obtained with grid searches.
    
\textbf{Hyperparameter tuning for CaH$^+$ }{\boldmath$J\le 4$}\textbf{ and }{\boldmath$J\le 6$}
\textbf{system} --- Hyperparameter grid scans are performed for $\tau$, $r_l$, and $r_o$ independently (scan over two magnitudes for $\tau$ and $r_l$, and one magnitude for $r_o$) and the optimal combination of parameters are reported in the caption of Fig.~\ref{fig:ex_CaH_J4_6}. 
Furthermore, we found that a $\varepsilon_{\rm end}$ of 0.025 in Eq.~\ref{eq:rl_epsilon} gives the best performance. 
We also tested the use of a neural network with four fully connected layers in the RL calculations for $J\le 6$ system (with the same, and longer training episodes). The state preparation episodes finish with longer durations.

\textbf{MDP and quantum MDP modeling: temporal difference update} --- The major difference between the MDP and quantum MDP (qMDP) modelings of the process in Fig.~1c is the temporal difference update to the $Q(s, a)$ value estimates. 
In the training with the MDP modeling, the value update at step $t$ is 
\begin{align} \label{cmdp_tdupdate}
\Delta Q^{\rm MDP}(S_t, A_t) \propto \max_{a} Q(S_{t+1}(S_t, A_t), a) + R_{t+1} - Q(S_t, A_t)
\end{align}
where $S_{t+1}(S_t, A_t)$ is stochastically determined from a random sampling according to Eq.~1. 
On the other hand, the qMDP value update is 
\begin{align} \label{qmdp_tdupdate}
\Delta Q^{\rm qMDP}(S_t, A_t) \propto 
\Big{[} &p_0 \max_{a} Q(S_{t+1}(S_t, A_t, k=0), a) \nonumber \\ 
+ &p_1 \max_{a} Q(S_{t+1}(S_t, A_t, k=1), a) \Big{]} + R_{t+1} - Q(S_t, A_t) 
\end{align}
with $k$ the measurement outcome, and $p_0$, $p_1$ defined as in the POVM, Eqs.~\ref{qevo_2}--\ref{qevo_3}.
Although the above equations are presented with the discount factor set to $\gamma = 1$, practical reinforcement learning calculations may benefit from using $\gamma < 1$ for numerical benefits.
Note that the time evolution of the state for a given measurement outcome $k$ is fully determined, and the probability evolution is explicitly coded in each temporal difference learning update.
Comparing the two modelings, Eqs.~\ref{cmdp_tdupdate} and~\ref{qmdp_tdupdate} yield the same training results at the limit of infinite training samples because the stochastic evolution in Eq.~\ref{cmdp_tdupdate} eventually reproduces the same probabilistic distribution for two measurement outcomes. However, since $p_1$ can be relatively small for molecules with a large Hilbert space, the qMDP modeling promises an improved learning efficiency. 
This expectation is confirmed in the control simulation of hydronium ion, as the explicit incorporation of the POVM in qMDP modeling is a necessity for exploring the immense state-action space.

\textbf{Hyperparameter tuning for H$_3$O$^+$} {\boldmath$J\le 2$}\textbf{ system} --- We preformed grid search of optimal hyperparameters ($\tau=0.0001$, $0.00025$, $0.0005$; $r_l=0.0001$, $0.00025$, $0.0005$, $0.001$, $r_o$=1, 2, 5, $\varepsilon_{\rm end} = 0.0125$, $0.025$, $0.005$).
The calculation procedure is described as follows,
(\emph{i}) During training, we calculated the mean number of steps (length) of the last 100 training episodes under different hyperparameter combinations.
(\emph{ii}) We selected half of the training trajectories in (\emph{i}) with lower length and saved models from the training trajectories per 100 training episodes.
(\emph{iii}) We then tested all these saved models and selected the one with the lowest mean as the final optimal model from this search.
The results in Fig.~4 is obtained with $\tau=0.0001$, $r_l=0.0005$, $r_o$=1, $\varepsilon_{\rm end} = 0.125$, $\gamma=0.9$.
%


\clearpage
\subsection{Supplementary Figures and Tables}

\begin{table}
    \includegraphics[width=0.6\linewidth]{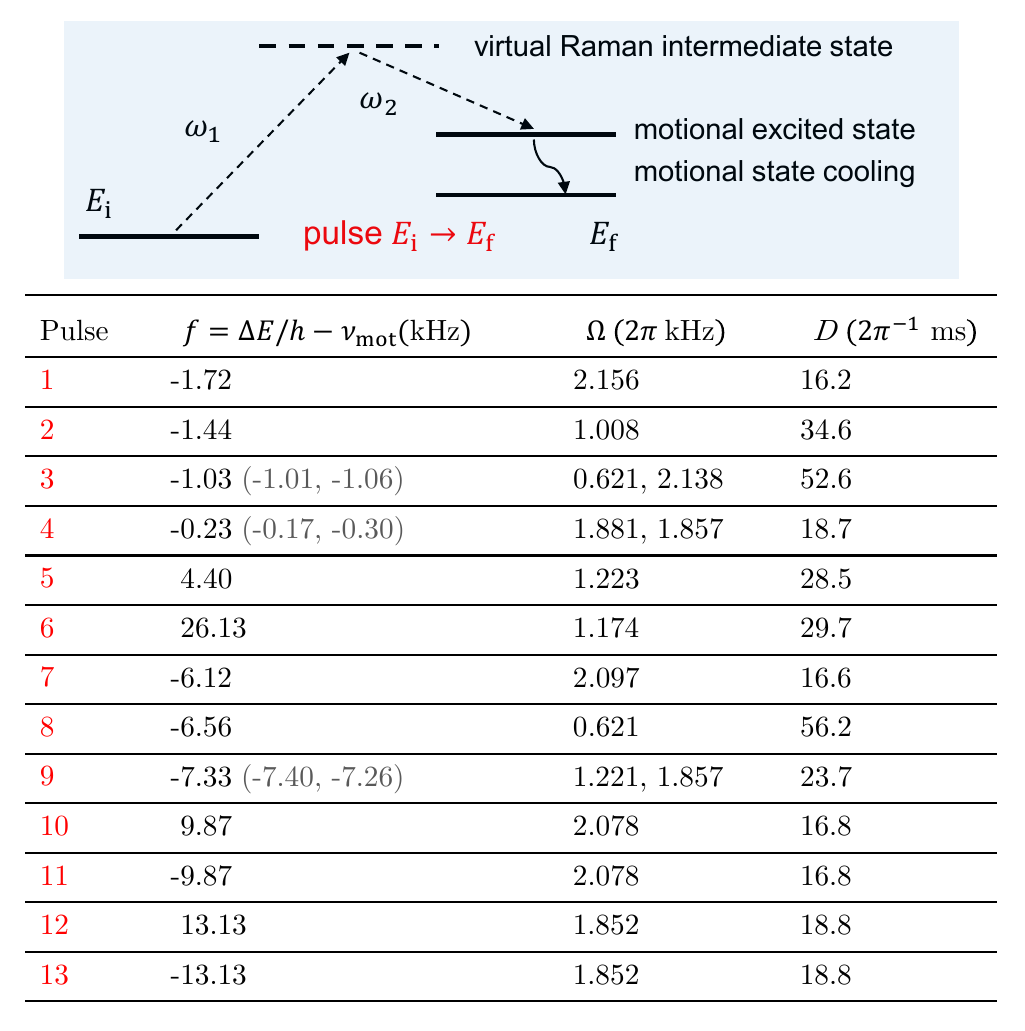}
    \caption{Optical pumping transitions in Fig.~2a consist of a two-photon stimulated Raman process, followed by motional state cooling. 
    Pulse sequence parameters, including transition frequencies $f$ (transition energies $\Delta E= E_{\rm f}-E_{\rm i}$, reported as difference from the motional mode frequency, $\nu_{\rm mot}=5.164$ MHz), Rabi rates $\Omega$, and the pulse duration $D$. 
    The duration is chosen such that the pulses are close to $\pi$-pulses, i.e. $D = \pi/ (\lambda_{\rm L-D} \Omega)$ with $\lambda_{\rm L-D}$ the Lamb-Dcke parameter.}
    \label{tbl:ex_pulse}
\end{table}

\clearpage

\begin{figure}[tb]
    \centering
    \includegraphics[width=0.65\textwidth]{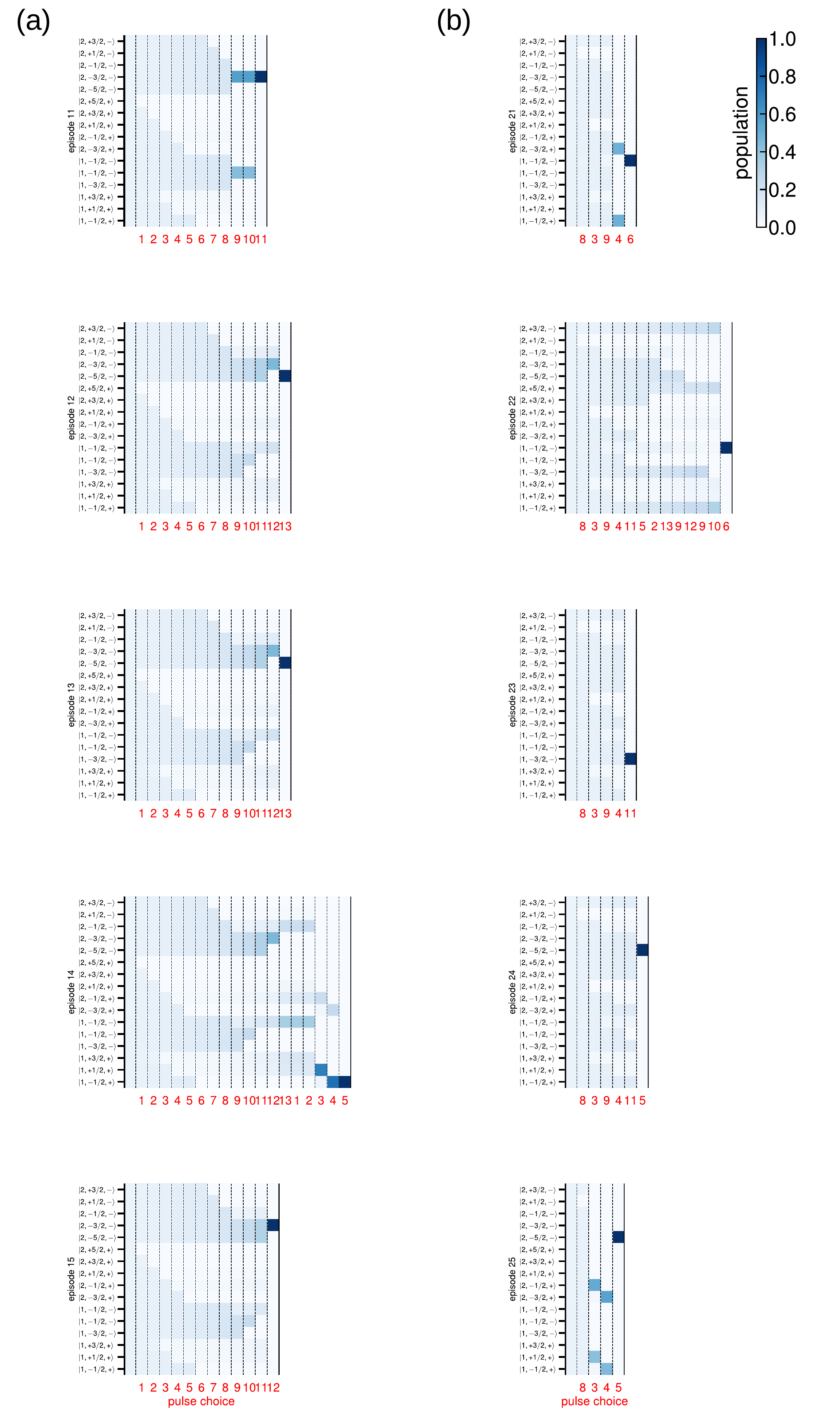}
    \caption{Dynamics of state populations, $P(\mathcal{J})= {\rm tr} (\rho \ket{\mathcal{J}} \bra{\mathcal{J}})$,  from typical state-action trajectories in the Markov decision processes with \textbf{(a)} sweeping and \textbf{(b)} the RL-designed protocol.
    }
    \label{fig:ex_fig3}
\end{figure} 
\clearpage

\begin{figure}[tb]
    \centering
    \includegraphics[width= 1 \textwidth]{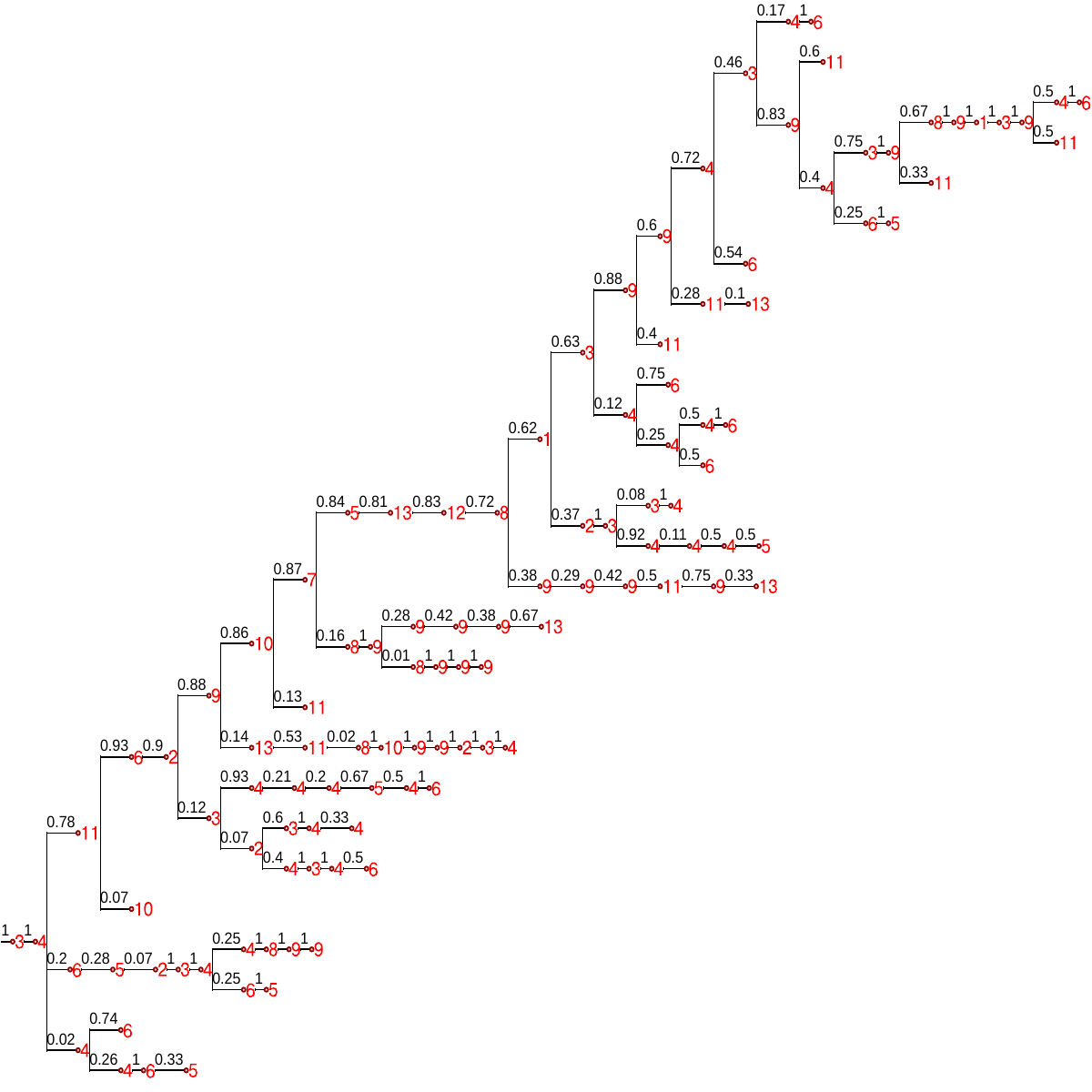}
    \caption{A complete version of the decision tree plotted in the main Fig.~2d. 
    The pulse choices are reported in red, and the branching probabilities are reported in black. 
    Most of the time the parent node has two offspring nodes, corresponding to the two, $k=0$ and $k=1$, measurement results, respectively. 
    However, we also observe nodes with more than two offspring because sometimes the optimal pulse choices are coincidentally the same for two different quantum states (c.f. Fig.~2d). 
    The branch that leads to episode termination (blue boxes in Fig.~2d) is omitted in the plot. 
    The tree plot is produced with the ETE library \cite{huerta2016ete}.}
    \label{fig:ex_fig4}
\end{figure} 

\begin{figure}[tb]
    \centering
    \includegraphics[width= 0.8\textwidth]
    {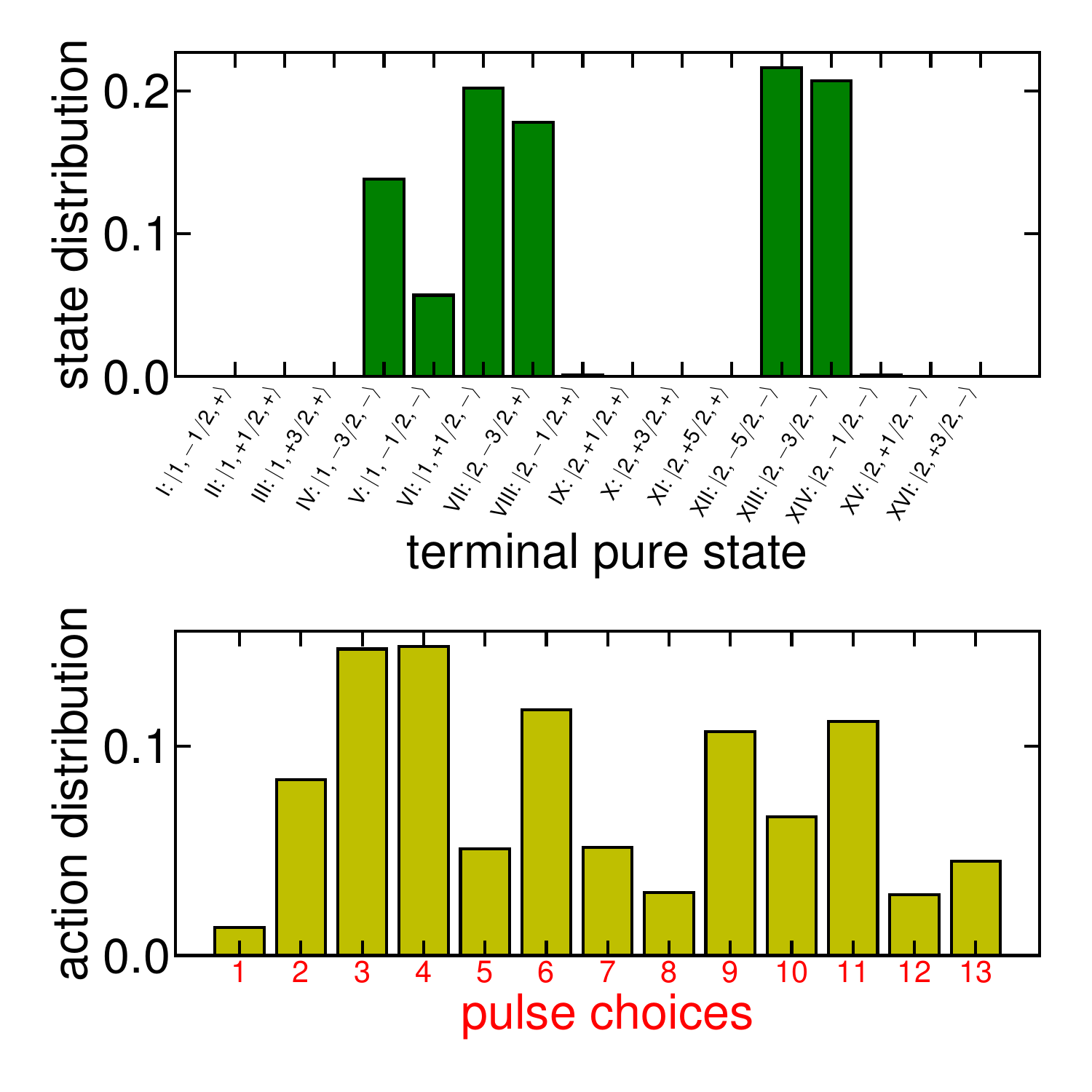}
    \caption{Statistical distribution of the termination single states (top), and of the pulse sequence choices (bottom) in the testing episodes for the results presented in Fig.~2.
    The model being tested is the ``model 600'' as in Fig.~2c. 
    Fractionally, 62\% of the episodes end on the $\ket{J, -J+1/2, -}$ or $\ket{J, -J-1/2, -}$ states.
    }
    \label{fig:ex_fig5}
\end{figure}

\begin{figure}[tb]
    \centering
    \includegraphics[width= 0.8\textwidth]
    {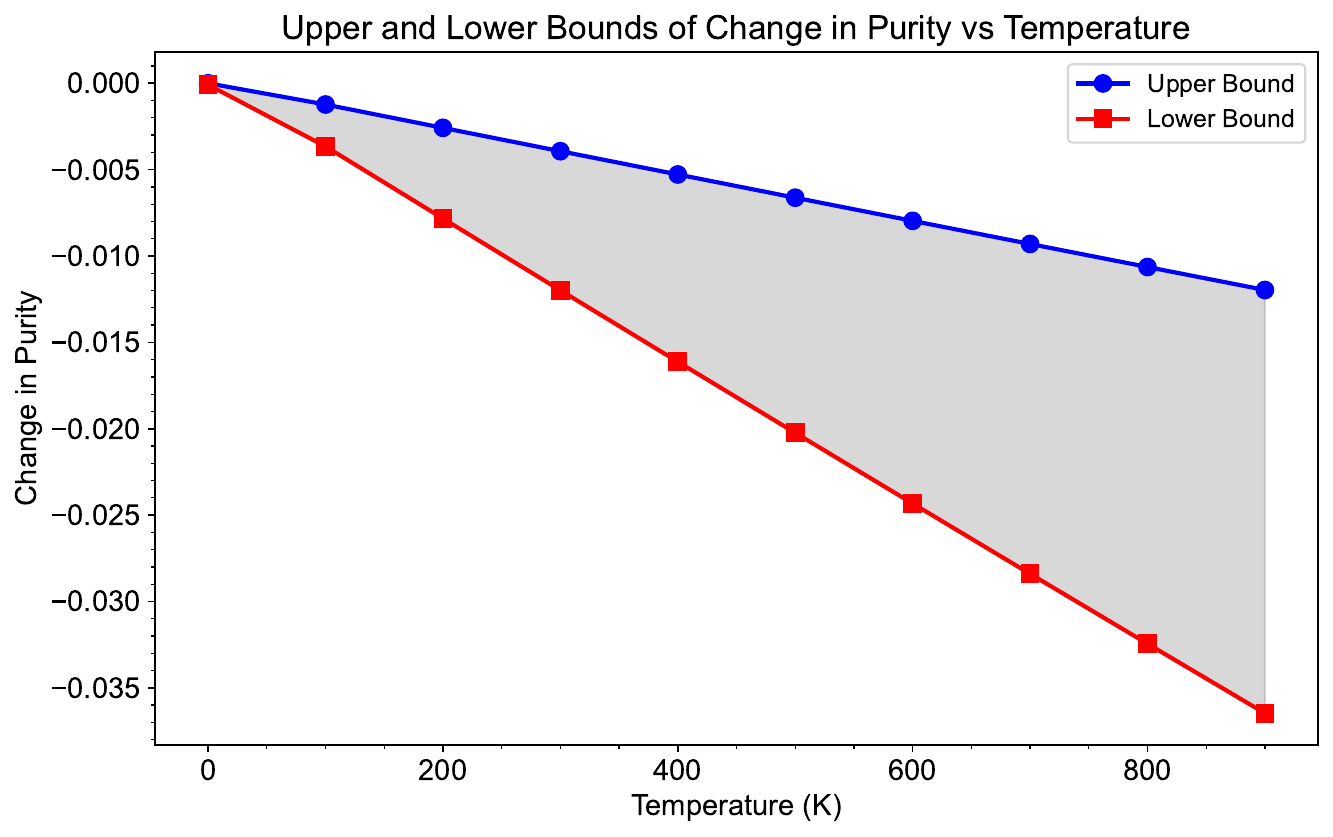}
    \caption{Lower and upper bounds of the degradation in purity due to thermal excitations during a single pulse/measurement step in the state space of manifolds $J=1,2$. The lower bound is calculated by evolving each pure rotational state under blackbody radiation (BBR) for the duration of the longest laser pulse in the pulse library. Each state's purity degrades differently based on the coupling strengths of BBR-driven transitions. The maximum purity degradation defines the lower bound, while the upper bound is similarly obtained using the shortest pulse duration. This analysis shows that BBR is responsible when the RL protocol is near termination but fails due to thermal excitations.}
    \label{fig:ex_thermalnoise}
\end{figure}

\begin{figure}[tb]
    \centering
    \includegraphics[width= 0.8\textwidth]
    {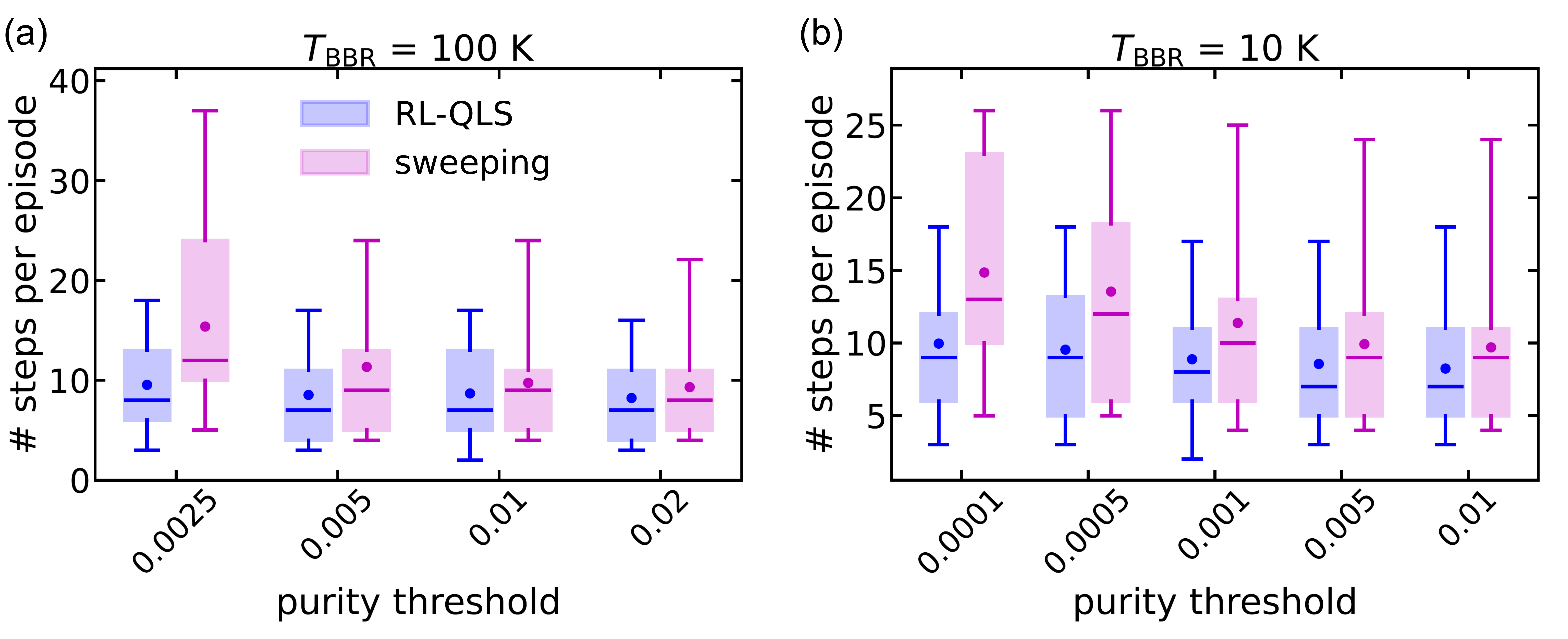}
    \caption{The distribution of the number of steps (i.e., episode lengths) against purity thresholds
    for the results reported in Fig.~3b. 
    Circular markers denote the mean numbers of steps.
    Short horizontal lines indicate the medians, and boxes denote the interquartile ranges (25\%–75\% of the distributions).
    The bars denote the 5\%–95\% range of the distributions.}
    \label{fig:ex_3b_distribution}
\end{figure}

\begin{figure}[!tb]
    \centering 
    \includegraphics[width= \textwidth]{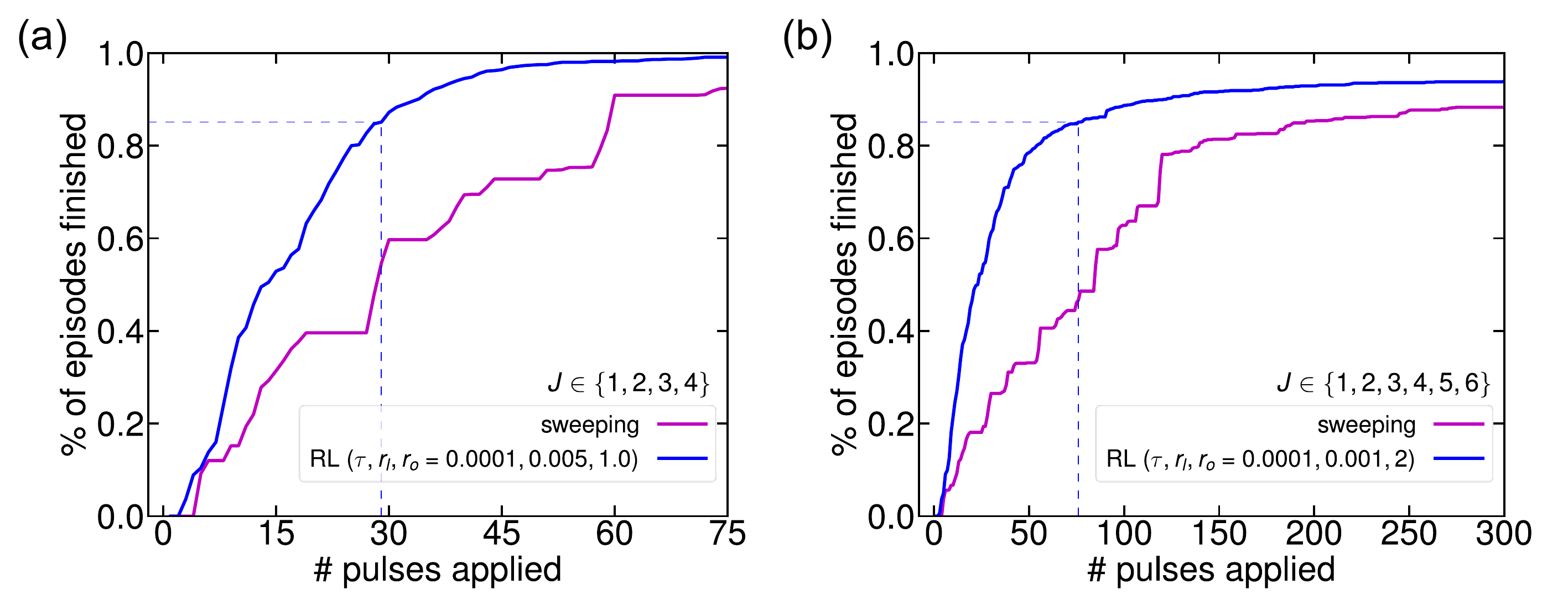}
    \caption{(\textbf{a}-\textbf{b}) Percentage of finished episodes as a function of the number of the pulses applied, 
    in the system that consists of rotational states from $J \in \{1,2,3,4\}$
    (\textbf{a}, with 48 states and a library of 68 pulses to choose from), and $J \in \{1,2,3,4,5,6\}$, (\textbf{b}, with 96 states and a library of 131 pulses). 
    Results are obtained with the sweeping protocol (solid purple) and the RL algorithm (solid blue).
    The optimal hyperparameters, including the one required to tune the physics-informed reward function, are reported in the legend and the dashed lines indicate the number of pulses required to achieve an 85\% success rate in RL-designed state preparation.
    Pulses libraries are reported at the end of this SM, Figs.~\ref{fig:ex_J6pulse}.}
    \label{fig:ex_CaH_J4_6}
\end{figure}

\begin{figure}[!tb]
    \centering
    \includegraphics[width=\textwidth]{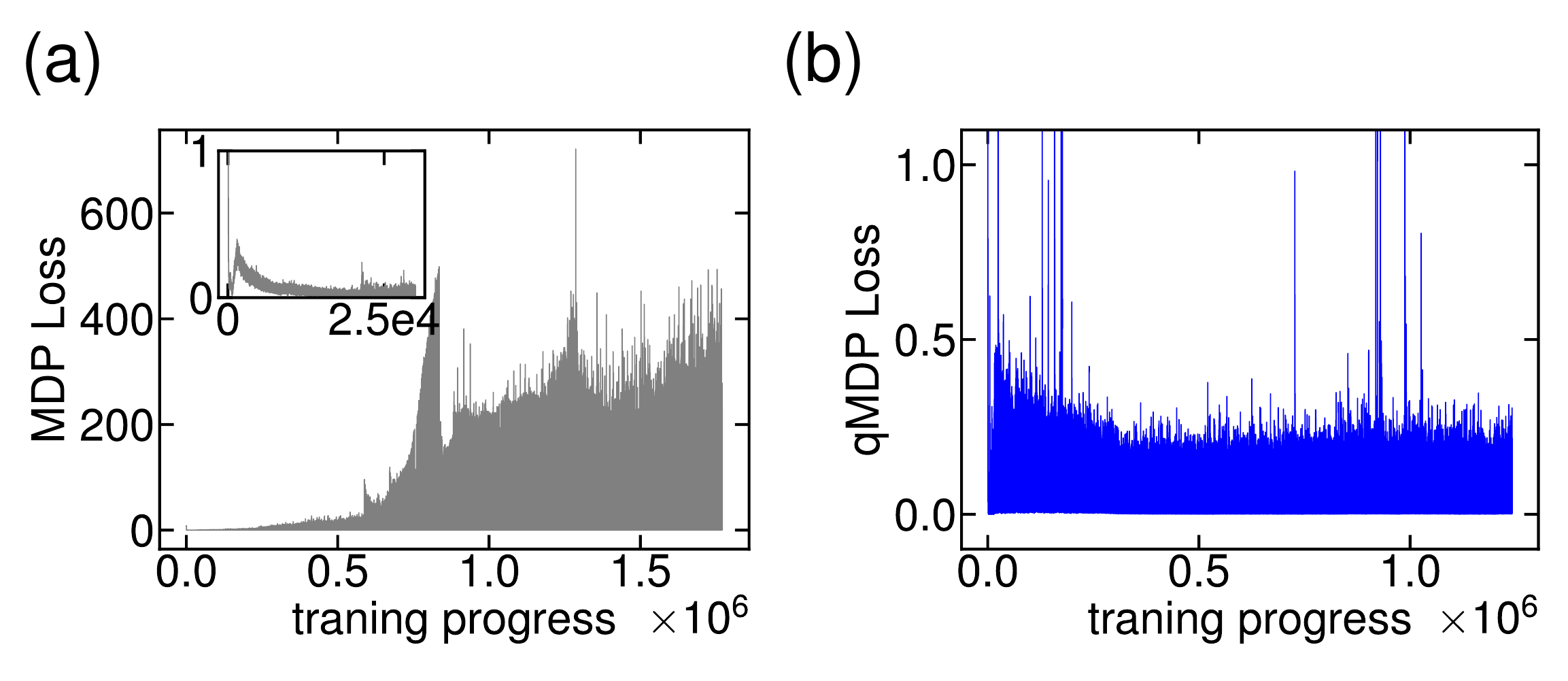}
    \caption{Loss functions of state-action values along typical training trajectories of H$_3$O$^+$ molecular ion control, with \textbf{(a)} MDP and \textbf{(b)} quantum MDP (qMDP) modeling approaches, respectively. With MDP modeling, the loss function decreases and remains at the same magnitude for $\sim$10$^5$ training steps (inset) and keeps accumulating to three magnitudes higher. As a sharp contrast, when the qMDP modeling is applied, the loss function remains at the same magnitude throughout the whole training process.
    The large loss in panel \textbf{(a)} is an artifact of averaging the state-action value update for different projective measurement outcomes (see Sec.~SD).
    Indeed, the inefficient training of MDP approach leads to training failure for more complex molecular control like H$_3$O$^+$ ion, even though the models obtained at a large loss could generate valid decision trees for simple systems (CaH$^+$, $J \in \{1,2\}$, $J \in \{1,2,3,4,5,6\}$, not presented here). 
    Both loss function results are obtained with 1000 training episodes and the same hyperparameter sets, $\tau=0.0001$, $r_l=0.001$, $r_o=2$, $\varepsilon_{\rm end}=0.025$.
    }
    \label{fig:ex_qmdp_loss}
\end{figure}

\clearpage
\begin{table}[t]
\raggedright
\caption{Energy levels of the H$_3$O$^+$ ion used in this study. 
The table labels the states with five quantum numbers; $J$ and $K$---the rotational quantum numbers, the parity of the inversion mode, $m_F$---the projection quantum number of total angular momentum $F$, and $\xi$. $\xi$ labels the state in the order of ascending energies when the other four quantum numbers are the same. $\mathfrak{J}$ labels a $\ket{J,K, \textrm{parity} }$ manifold.
}
\label{table:h3o_energy}
\begin{tabular}{
p{0.2cm}
>{\raggedleft\arraybackslash}p{1.2cm}
>{\raggedleft\arraybackslash}p{1.2cm}
>{\raggedleft\arraybackslash}p{1.2cm}
>{\raggedleft\arraybackslash}p{1.2cm}
>{\raggedleft\arraybackslash}p{1.2cm}
>{\raggedleft\arraybackslash}p{4cm}}
 $\mathfrak{J}$ & $J$ & $K$ & parity &  $m_F$ & $\xi$ & Energy, $E/h$ (kHz) \\
\hline
\hline
\multirow[t]{6}{*}{1}  
  &   1 & 1 &      + &  1.5 &   1 &           0 \\
  &   1 & 1 &      + &  0.5 &   1 &          5.441 \\
  &   1 & 1 &      + & -0.5 &   1 &         11.610 \\
  &   1 & 1 &      + & -1.5 &   1 &         19.383 \\
  &   1 & 1 &      + & -0.5 &   2 &         42.893 \\
  &   1 & 1 &      + &  0.5 &   2 &         45.007 \\
  \hline
\multirow[t]{12}{*}{2}  
  &   1 & 0 &      + &  2.5 &   1 &   142,813,552.326 \\
  &   1 & 0 &      + &  1.5 &   1 &   142,813,561.506 \\
  &   1 & 0 &      + &  0.5 &   1 &   142,813,571.109 \\
  &   1 & 0 &      + & -0.5 &   1 &   142,813,581.133 \\
  &   1 & 0 &      + & -1.5 &   1 &   142,813,591.529 \\
  &   1 & 0 &      + & -2.5 &   1 &   142,813,602.245 \\
  &   1 & 0 &      + &  1.5 &   2 &   142,813,615.597 \\
  &   1 & 0 &      + &  0.5 &   2 &   142,813,624.377 \\
  &   1 & 0 &      + & -0.5 &   2 &   142,813,637.053 \\
  &   1 & 0 &      + & -1.5 &   2 &   142,813,650.820 \\
  &   1 & 0 &      + &  0.5 &   3 &   142,813,656.782 \\
  &   1 & 0 &      + & -0.5 &   3 &   142,813,680.065 \\
  \hline
\multirow[t]{10}{*}{3}
  &   2 & 2 &      + &  2.5 &   1 &   906,028,510.578 \\
  &   2 & 2 &      + &  1.5 &   1 &   906,028,514.838 \\
  &   2 & 2 &      + &  0.5 &   1 &   906,028,519.262 \\
  &   2 & 2 &      + & -0.5 &   1 &   906,028,523.894 \\
  &   2 & 2 &      + & -1.5 &   1 &   906,028,528.801 \\
  &   2 & 2 &      + & -2.5 &   1 &   906,028,534.096 \\
  &   2 & 2 &      + & -1.5 &   2 &   906,028,577.169 \\
  &   2 & 2 &      + & -0.5 &   2 &   906,028,577.982 \\
  &   2 & 2 &      + &  0.5 &   2 &   906,028,578.519 \\
  &   2 & 2 &      + &  1.5 &   2 &   906,028,578.849 \\
  \hline
\multirow[t]{4}{*}{4}  
  &   0 & 0 &      - &  1.5 &   1 & 1,134,822,941.064 \\
  &   0 & 0 &      - &  0.5 &   1 & 1,134,822,956.392 \\
  &   0 & 0 &      - & -0.5 &   1 & 1,134,822,971.720 \\
  &   0 & 0 &      - & -1.5 &   1 & 1,134,822,987.048 \\
  \hline
\multirow[t]{10}{*}{5}  
  &   2 & 1 &      + &  2.5 &   1 & 1,334,121,279.666 \\
  &   2 & 1 &      + &  1.5 &   1 & 1,334,121,283.874 \\
  &   2 & 1 &      + &  0.5 &   1 & 1,334,121,288.248 \\
  &   2 & 1 &      + & -0.5 &   1 & 1,334,121,292.831 \\
  &   2 & 1 &      + & -1.5 &   1 & 1,334,121,297.692 \\
  &   2 & 1 &      + & -2.5 &   1 & 1,334,121,302.944 \\
  &   2 & 1 &      + & -1.5 &   2 & 1,334,121,346.035 \\
  &   2 & 1 &      + & -0.5 &   2 & 1,334,121,346.921 \\
  &   2 & 1 &      + &  0.5 &   2 & 1,334,121,347.530 \\
  &   2 & 1 &      + &  1.5 &   2 & 1,334,121,347.929 \\
  \hline
\multirow[t]{6}{*}{6}  
  &   1 & 1 &      - &  1.5 &   1 & 1,655,832,000.000 \\
  &   1 & 1 &      - &  0.5 &   1 & 1,655,832,005.441 \\
  &   1 & 1 &      - & -0.5 &   1 & 1,655,832,011.610 \\
  &   1 & 1 &      - & -1.5 &   1 & 1,655,832,019.383 \\
  &   1 & 1 &      - & -0.5 &   2 & 1,655,832,042.893 \\
  &   1 & 1 &      - &  0.5 &   2 & 1,655,832,045.007 \\
  \hline
\end{tabular}
\end{table}

\begin{table}[t]
\raggedright
\caption*{(Continued) 
}
\begin{tabular}{
p{0.2cm}
>{\raggedleft\arraybackslash}p{1.2cm}
>{\raggedleft\arraybackslash}p{1.2cm}
>{\raggedleft\arraybackslash}p{1.2cm}
>{\raggedleft\arraybackslash}p{1.2cm}
>{\raggedleft\arraybackslash}p{1.2cm}
>{\raggedleft\arraybackslash}p{4cm}}
 $\mathfrak{J}$ & $J$ & $K$ & parity &  $m_F$ & $\xi$ & Energy, $E/h$ (kHz) \\
\hline
\hline
\multirow[t]{28}{*}{7}  
  &   3 & 3 &      + &  4.5 &   1 & 2,193,342,406.643 \\
  &   3 & 3 &      + &  3.5 &   1 & 2,193,342,412.748 \\
  &   3 & 3 &      + &  2.5 &   1 & 2,193,342,418.938 \\
  &   3 & 3 &      + &  1.5 &   1 & 2,193,342,425.216 \\
  &   3 & 3 &      + &  0.5 &   1 & 2,193,342,431.588 \\
  &   3 & 3 &      + & -0.5 &   1 & 2,193,342,438.056 \\
  &   3 & 3 &      + & -1.5 &   1 & 2,193,342,444.625 \\
  &   3 & 3 &      + & -2.5 &   1 & 2,193,342,451.298 \\
  &   3 & 3 &      + & -3.5 &   1 & 2,193,342,458.079 \\
  &   3 & 3 &      + & -4.5 &   1 & 2,193,342,464.971 \\
  &   3 & 3 &      + &  3.5 &   2 & 2,193,342,517.202 \\
  &   3 & 3 &      + &  2.5 &   2 & 2,193,342,521.756 \\
  &   3 & 3 &      + &  1.5 &   2 & 2,193,342,526.482 \\
  &   3 & 3 &      + &  0.5 &   2 & 2,193,342,531.406 \\
  &   3 & 3 &      + & -0.5 &   2 & 2,193,342,536.559 \\
  &   3 & 3 &      + & -1.5 &   2 & 2,193,342,541.978 \\
  &   3 & 3 &      + & -2.5 &   2 & 2,193,342,547.700 \\
  &   3 & 3 &      + & -3.5 &   2 & 2,193,342,553.756 \\
  &   3 & 3 &      + &  2.5 &   3 & 2,193,342,607.166 \\
  &   3 & 3 &      + &  1.5 &   3 & 2,193,342,608.648 \\
  &   3 & 3 &      + &  0.5 &   3 & 2,193,342,610.361 \\
  &   3 & 3 &      + & -0.5 &   3 & 2,193,342,612.453 \\
  &   3 & 3 &      + & -1.5 &   3 & 2,193,342,615.235 \\
  &   3 & 3 &      + & -2.5 &   3 & 2,193,342,619.533 \\
  &   3 & 3 &      + & -1.5 &   4 & 2,193,342,661.159 \\
  &   3 & 3 &      + & -0.5 &   4 & 2,193,342,667.699 \\
  &   3 & 3 &      + &  0.5 &   4 & 2,193,342,673.184 \\
  &   3 & 3 &      + &  1.5 &   4 & 2,193,342,677.963 \\
\hline
\multirow[t]{10}{*}{8}  
  &   2 & 2 &      - &  2.5 &   1 & 2,563,282,510.578 \\
  &   2 & 2 &      - &  1.5 &   1 & 2,563,282,514.838 \\
  &   2 & 2 &      - &  0.5 &   1 & 2,563,282,519.262 \\
  &   2 & 2 &      - & -0.5 &   1 & 2,563,282,523.894 \\
  &   2 & 2 &      - & -1.5 &   1 & 2,563,282,528.801 \\
  &   2 & 2 &      - & -2.5 &   1 & 2,563,282,534.096 \\
  &   2 & 2 &      - & -1.5 &   2 & 2,563,282,577.169 \\
  &   2 & 2 &      - & -0.5 &   2 & 2,563,282,577.982 \\
  &   2 & 2 &      - &  0.5 &   2 & 2,563,282,578.520 \\
  &   2 & 2 &      - &  1.5 &   2 & 2,563,282,578.849 \\
\hline
\multirow[t]{14}{*}{9} 
  &   3 & 2 &      + &  3.5 &   1 & 2,906,370,436.587 \\
  &   3 & 2 &      + &  2.5 &   1 & 2,906,370,440.248 \\
  &   3 & 2 &      + &  1.5 &   1 & 2,906,370,443.973 \\
  &   3 & 2 &      + &  0.5 &   1 & 2,906,370,447.773 \\
  &   3 & 2 &      + & -0.5 &   1 & 2,906,370,451.656 \\
  &   3 & 2 &      + & -1.5 &   1 & 2,906,370,455.636 \\
  &   3 & 2 &      + & -2.5 &   1 & 2,906,370,459.729 \\
  &   3 & 2 &      + & -3.5 &   1 & 2,906,370,463.959 \\
  &   3 & 2 &      + &  2.5 &   2 & 2,906,370,527.479 \\
  &   3 & 2 &      + &  1.5 &   2 & 2,906,370,527.768 \\
  &   3 & 2 &      + &  0.5 &   2 & 2,906,370,527.983 \\
  &   3 & 2 &      + & -2.5 &   2 & 2,906,370,528.071 \\
  &   3 & 2 &      + & -0.5 &   2 & 2,906,370,528.115 \\
  &   3 & 2 &      + & -1.5 &   2 & 2,906,370,528.150 \\
  \hline
\end{tabular}
\end{table}

\begin{table}[t]
\raggedright
\caption*{(Continued)}
\begin{tabular}{
p{0.2cm}
>{\raggedleft\arraybackslash}p{1.2cm}
>{\raggedleft\arraybackslash}p{1.2cm}
>{\raggedleft\arraybackslash}p{1.2cm}
>{\raggedleft\arraybackslash}p{1.2cm}
>{\raggedleft\arraybackslash}p{1.2cm}
>{\raggedleft\arraybackslash}p{4cm}}
 $\mathfrak{J}$ & $J$ & $K$ & parity &  $m_F$ & $\xi$ & Energy, $E/h$ (kHz) \\
\hline
\hline
\multirow[t]{10}{*}{10} 
  &   2 & 1 &      - &  2.5 &   1 & 2,966,001,279.666 \\
  &   2 & 1 &      - &  1.5 &   1 & 2,966,001,283.874 \\
  &   2 & 1 &      - &  0.5 &   1 & 2,966,001,288.248 \\
  &   2 & 1 &      - & -0.5 &   1 & 2,966,001,292.831 \\
  &   2 & 1 &      - & -1.5 &   1 & 2,966,001,297.692 \\
  &   2 & 1 &      - & -2.5 &   1 & 2,966,001,302.944 \\
  &   2 & 1 &      - & -1.5 &   2 & 2,966,001,346.035 \\
  &   2 & 1 &      - & -0.5 &   2 & 2,966,001,346.921 \\
  &   2 & 1 &      - &  0.5 &   2 & 2,966,001,347.530 \\
  &   2 & 1 &      - &  1.5 &   2 & 2,966,001,347.929 \\
\hline
\multirow[t]{20}{*}{11} 
  &   2 & 0 &      - &  3.5 &   1 & 3,100,076,809.911 \\
  &   2 & 0 &      - &  2.5 &   1 & 3,100,076,817.055 \\
  &   2 & 0 &      - &  1.5 &   1 & 3,100,076,824.370 \\
  &   2 & 0 &      - &  0.5 &   1 & 3,100,076,831.864 \\
  &   2 & 0 &      - & -0.5 &   1 & 3,100,076,839.546 \\
  &   2 & 0 &      - & -1.5 &   1 & 3,100,076,847.422 \\
  &   2 & 0 &      - & -2.5 &   1 & 3,100,076,855.494 \\
  &   2 & 0 &      - & -3.5 &   1 & 3,100,076,863.764 \\
  &   2 & 0 &      - &  2.5 &   2 & 3,100,076,897.278 \\
  &   2 & 0 &      - &  1.5 &   2 & 3,100,076,902.913 \\
  &   2 & 0 &      - &  0.5 &   2 & 3,100,076,909.062 \\
  &   2 & 0 &      - & -0.5 &   2 & 3,100,076,915.876 \\
  &   2 & 0 &      - & -1.5 &   2 & 3,100,076,923.520 \\
  &   2 & 0 &      - & -2.5 &   2 & 3,100,076,931.955 \\
  &   2 & 0 &      - &  1.5 &   3 & 3,100,076,963.923 \\
  &   2 & 0 &      - &  0.5 &   3 & 3,100,076,965.459 \\
  &   2 & 0 &      - & -0.5 &   3 & 3,100,076,968.234 \\
  &   2 & 0 &      - & -1.5 &   3 & 3,100,076,978.053 \\
  &   2 & 0 &      - & -0.5 &   4 & 3,100,077,002.681 \\
  &   2 & 0 &      - &  0.5 &   4 & 3,100,077,012.083 \\
\hline
\end{tabular}
\end{table}

\clearpage
\begin{table}[t]
\raggedright
\caption{Rabi rates of the two-photon inversion-rotational transitions of H$_3$O$^+$ ion used in this molecular control study. Superscripts i and f indicate the initial and final states of the transition, respectively. The selection rules for the electric dipole-induced Raman transition is $\Delta J=0, \pm2$, $\Delta K=0$, ${\rm parity}^{\rm i} = {\rm parity}^{\rm f}$. 
Two Raman pulses are $\pi$-polarized and $\sigma^+/\sigma^-$-polarized, respectively, thus only $\Delta m_F = \pm 1$ transitions are allowed. The table only lists the transitions with $\Delta m_F = +1$ since the transitions i$\to$f and f$\to$i share the same Rabi rate.
The rates are obtained by assuming laser pulse amplitudes such that the Rabi rate of the * transition is $2.000 \times 2 \pi$ kHz.
}
\label{table:h3o_rabi_rate}
\begin{tabular}{
>{\raggedleft\arraybackslash}p{1cm}
>{\raggedleft\arraybackslash}p{1cm}
>{\raggedleft\arraybackslash}p{1cm}
>{\raggedleft\arraybackslash}p{0.75cm}
>{\raggedleft\arraybackslash}p{0.75cm}
>{\raggedleft\arraybackslash}p{1cm}
>{\raggedleft\arraybackslash}p{1cm}
>{\raggedleft\arraybackslash}p{1cm}
>{\raggedleft\arraybackslash}p{1cm}
>{\raggedleft\arraybackslash}p{1cm}|
>{\raggedleft\arraybackslash}p{2cm}}
$J^{\rm i}$ & $K^{\rm i}$ & ${\rm parity}^{\rm i}$ & $m_{F}^{\rm i}$ & $\xi^{\rm i}$ &
$J^{\rm f}$ & $K^{\rm f}$ & ${\rm parity}^{\rm f}$ & $m_{F}^{\rm f}$ & $\xi^{\rm f}$ & 
Rabi rate, \\ 
&&&&&&&&&&$ \Omega$ (2$\pi$ kHz) \\
\hline
\hline
      2 &      2 &   + &     2.5 &    1 &    3 &    2 &   + &      3.5 &     1 & 2.601 \\
      3 &      2 &   + &    -3.5 &    1 &    2 &    2 &   + &     -2.5 &     1 & 2.601 \\
      3 &      3 &   + &     3.5 &    1 &    3 &    3 &   + &      4.5 &     1 & 2.487 \\
      3 &      3 &   + &     2.5 &    1 &    3 &    3 &   + &      3.5 &     1 & 2.399 \\
      3 &      3 &   + &     0.5 &    4 &    3 &    3 &   + &      1.5 &     4 & 2.354 \\
      3 &      3 &   + &    -3.5 &    1 &    3 &    3 &   + &     -2.5 &     1 & 2.306 \\
      2 &      2 &   + &     1.5 &    2 &    3 &    2 &   + &      2.5 &     2 & 2.282 \\
      0 &      0 &   - &     1.5 &    1 &    2 &    0 &   - &      2.5 &     1 & 2.223 \\
      3 &      3 &   + &    -4.5 &    1 &    3 &    3 &   + &     -3.5 &     1 & 2.209 \\
      3 &      2 &   + &     0.5 &    1 &    2 &    2 &   + &      1.5 &     1 & 2.185 \\
      2 &      2 &   + &    -1.5 &    1 &    3 &    2 &   + &     -0.5 &     1 & 2.146 \\
      2 &      0 &   - &    -1.5 &    1 &    0 &    0 &   - &     -0.5 &     1 & 2.105 \\
      2 &      2 &   + &    -1.5 &    2 &    3 &    2 &   + &     -0.5 &     2 & 2.085 \\
      3 &      3 &   + &     2.5 &    2 &    3 &    3 &   + &      3.5 &     2 & 2.035 \\
      2 &      2 &   + &    -0.5 &    1 &    3 &    2 &   + &      0.5 &     1 & 2.027 \\
      3 &      2 &   + &    -0.5 &    2 &    2 &    2 &   + &      0.5 &     2 & 2.024 \\
      3 &      2 &   + &     0.5 &    2 &    2 &    2 &   + &      1.5 &     2 & 2.020 \\
      0 &      0 &   - &     0.5 &    1 &    2 &    0 &   - &      1.5 &     1 & 2.005 \\
      2 &      1 &   + &     0.5 &    1 &    1 &    1 &   + &      1.5 &     1 & 2.003 \\
      2 &      1 &   - &     0.5 &    1 &    1 &    1 &   - &      1.5 &     1 & 2.003 \\
      2 &      2 &   - &     1.5 &    1 &    2 &    2 &   - &      2.5 &     1 & *2.000 \\
      2 &      2 &   + &     1.5 &    1 &    2 &    2 &   + &      2.5 &     1 & 2.000 \\
      1 &      1 &   + &    -0.5 &    1 &    2 &    1 &   + &      0.5 &     1 & 1.977 \\
      1 &      1 &   - &    -0.5 &    1 &    2 &    1 &   - &      0.5 &     1 & 1.977 \\
      2 &      0 &   - &    -2.5 &    2 &    0 &    0 &   - &     -1.5 &     1 & 1.976 \\
      3 &      3 &   + &    -3.5 &    2 &    3 &    3 &   + &     -2.5 &     3 & 1.975 \\
      2 &      1 &   + &    -0.5 &    2 &    1 &    1 &   + &      0.5 &     2 & 1.967 \\
      2 &      1 &   - &    -0.5 &    2 &    1 &    1 &   - &      0.5 &     2 & 1.967 \\
      3 &      2 &   + &    -2.5 &    2 &    2 &    2 &   + &     -1.5 &     2 & 1.961 \\
      1 &      1 &   + &     1.5 &    1 &    2 &    1 &   + &      2.5 &     1 & 1.946 \\
      1 &      1 &   - &     1.5 &    1 &    2 &    1 &   - &      2.5 &     1 & 1.946 \\
      2 &      1 &   + &    -2.5 &    1 &    1 &    1 &   + &     -1.5 &     1 & 1.946 \\
      2 &      1 &   - &    -2.5 &    1 &    1 &    1 &   - &     -1.5 &     1 & 1.946 \\
      2 &      0 &   - &     0.5 &    2 &    0 &    0 &   - &      1.5 &     1 & 1.941 \\
      2 &      0 &   - &    -2.5 &    1 &    0 &    0 &   - &     -1.5 &     1 & 1.917 \\
      2 &      2 &   + &    -0.5 &    2 &    3 &    2 &   + &      0.5 &     2 & 1.909 \\
      3 &      3 &   + &     1.5 &    3 &    3 &    3 &   + &      2.5 &     3 & 1.905 \\
      3 &      2 &   + &    -0.5 &    1 &    2 &    2 &   + &      0.5 &     1 & 1.871 \\
      3 &      3 &   + &    -2.5 &    1 &    3 &    3 &   + &     -1.5 &     1 & 1.861 \\
      2 &      2 &   - &     0.5 &    2 &    2 &    2 &   - &      1.5 &     2 & 1.857 \\
      2 &      2 &   + &     0.5 &    2 &    2 &    2 &   + &      1.5 &     2 & 1.857 \\
      1 &      0 &   + &    -1.5 &    1 &    1 &    0 &   + &     -0.5 &     3 & 1.843 \\
      1 &      0 &   + &     0.5 &    1 &    1 &    0 &   + &      1.5 &     2 & 1.838 \\
      1 &      0 &   + &    -0.5 &    1 &    1 &    0 &   + &      0.5 &     3 & 1.822 \\
      2 &      2 &   - &    -2.5 &    1 &    2 &    2 &   - &     -1.5 &     1 & 1.808 \\
      2 &      2 &   + &    -2.5 &    1 &    2 &    2 &   + &     -1.5 &     1 & 1.808 \\
      2 &      0 &   - &    -0.5 &    1 &    0 &    0 &   - &      0.5 &     1 & 1.797 \\
      1 &      0 &   + &    -2.5 &    1 &    1 &    0 &   + &     -1.5 &     2 & 1.791 \\
      3 &      3 &   + &    -2.5 &    3 &    3 &    3 &   + &     -1.5 &     4 & 1.790 \\
      3 &      3 &   + &    -4.5 &    1 &    3 &    3 &   + &     -3.5 &     2 & 1.790 \\
\end{tabular}
\end{table}

\begin{table}[t]
\raggedright
\caption*{(Continued.)}
\begin{tabular}{
>{\raggedleft\arraybackslash}p{1cm}
>{\raggedleft\arraybackslash}p{1cm}
>{\raggedleft\arraybackslash}p{1cm}
>{\raggedleft\arraybackslash}p{0.75cm}
>{\raggedleft\arraybackslash}p{0.75cm}
>{\raggedleft\arraybackslash}p{1cm}
>{\raggedleft\arraybackslash}p{1cm}
>{\raggedleft\arraybackslash}p{1cm}
>{\raggedleft\arraybackslash}p{1cm}
>{\raggedleft\arraybackslash}p{1cm}|
>{\raggedleft\arraybackslash}p{2cm}}
$J^{\rm i}$ & $K^{\rm i}$ & ${\rm parity}^{\rm i}$ & $m_{F}^{\rm i}$ & $\xi^{\rm i}$ &
$J^{\rm f}$ & $K^{\rm f}$ & ${\rm parity}^{\rm f}$ & $m_{F}^{\rm f}$ & $\xi^{\rm f}$ & 
Rabi rate, \\ 
&&&&&&&&&&$ \Omega$ (2$\pi$ kHz) \\
\hline
\hline
      3 &      2 &   + &     1.5 &    1 &    2 &    2 &   + &      2.5 &     1 & 1.776 \\
      2 &      0 &   - &    -1.5 &    3 &    0 &    0 &   - &     -0.5 &     1 & 1.772 \\
      0 &      0 &   - &    -0.5 &    1 &    2 &    0 &   - &      0.5 &     2 & 1.744 \\
      2 &      0 &   - &     2.5 &    1 &    2 &    0 &   - &      3.5 &     1 & 1.739 \\
      3 &      3 &   + &     1.5 &    1 &    3 &    3 &   + &      2.5 &     1 & 1.738 \\
      1 &      1 &   + &    -0.5 &    2 &    2 &    1 &   + &      0.5 &     2 & 1.728 \\
      1 &      1 &   - &    -0.5 &    2 &    2 &    1 &   - &      0.5 &     2 & 1.728 \\
      1 &      1 &   + &    -1.5 &    1 &    2 &    1 &   + &     -0.5 &     2 & 1.723 \\
      1 &      1 &   - &    -1.5 &    1 &    2 &    1 &   - &     -0.5 &     2 & 1.723 \\
      2 &      1 &   + &    -1.5 &    2 &    1 &    1 &   + &     -0.5 &     1 & 1.707 \\
      2 &      1 &   - &    -1.5 &    2 &    1 &    1 &   - &     -0.5 &     1 & 1.707 \\
      1 &      0 &   + &    -0.5 &    2 &    1 &    0 &   + &      0.5 &     1 & 1.686 \\
      3 &      3 &   + &     2.5 &    3 &    3 &    3 &   + &      3.5 &     2 & 1.679 \\
      3 &      3 &   + &    -1.5 &    4 &    3 &    3 &   + &     -0.5 &     3 & 1.672 \\
      1 &      1 &   + &    -1.5 &    1 &    2 &    1 &   + &     -0.5 &     1 & 1.650 \\
      1 &      1 &   - &    -1.5 &    1 &    2 &    1 &   - &     -0.5 &     1 & 1.650 \\
      3 &      3 &   + &     1.5 &    2 &    3 &    3 &   + &      2.5 &     2 & 1.641 \\
      1 &      0 &   + &     0.5 &    3 &    1 &    0 &   + &      1.5 &     1 & 1.628 \\
      0 &      0 &   - &     1.5 &    1 &    2 &    0 &   - &      2.5 &     2 & 1.624 \\
      0 &      0 &   - &     0.5 &    1 &    2 &    0 &   - &      1.5 &     3 & 1.619 \\
      2 &      1 &   + &    -0.5 &    1 &    1 &    1 &   + &      0.5 &     1 & 1.619 \\
      2 &      1 &   - &    -0.5 &    1 &    1 &    1 &   - &      0.5 &     1 & 1.619 \\
      0 &      0 &   - &    -1.5 &    1 &    2 &    0 &   - &     -0.5 &     4 & 1.616 \\
      2 &      2 &   + &    -2.5 &    1 &    3 &    2 &   + &     -1.5 &     1 & 1.607 \\
      2 &      2 &   - &    -1.5 &    2 &    2 &    2 &   - &     -0.5 &     2 & 1.597 \\
      2 &      2 &   + &    -1.5 &    2 &    2 &    2 &   + &     -0.5 &     2 & 1.597 \\
      2 &      1 &   + &    -1.5 &    1 &    1 &    1 &   + &     -0.5 &     2 & 1.567 \\
      2 &      1 &   - &    -1.5 &    1 &    1 &    1 &   - &     -0.5 &     2 & 1.567 \\
      0 &      0 &   - &    -1.5 &    1 &    2 &    0 &   - &     -0.5 &     3 & 1.560 \\
      3 &      3 &   + &    -1.5 &    3 &    3 &    3 &   + &     -0.5 &     2 & 1.555 \\
      2 &      0 &   - &    -3.5 &    1 &    2 &    0 &   - &     -2.5 &     2 & 1.546 \\
      3 &      3 &   + &     3.5 &    2 &    3 &    3 &   + &      4.5 &     1 & 1.544 \\
      1 &      0 &   + &     1.5 &    1 &    1 &    0 &   + &      2.5 &     1 & 1.540 \\
      2 &      0 &   - &     0.5 &    3 &    0 &    0 &   - &      1.5 &     1 & 1.532 \\
      2 &      0 &   - &    -3.5 &    1 &    2 &    0 &   - &     -2.5 &     1 & 1.500 \\
      3 &      3 &   + &    -3.5 &    2 &    3 &    3 &   + &     -2.5 &     2 & 1.481 \\
      3 &      3 &   + &    -2.5 &    2 &    3 &    3 &   + &     -1.5 &     2 & 1.479 \\
      3 &      3 &   + &     0.5 &    2 &    3 &    3 &   + &      1.5 &     3 & 1.476 \\
      1 &      0 &   + &     1.5 &    2 &    1 &    0 &   + &      2.5 &     1 & 1.475 \\
      3 &      3 &   + &    -0.5 &    3 &    3 &    3 &   + &      0.5 &     4 & 1.474 \\
      2 &      0 &   - &    -2.5 &    1 &    2 &    0 &   - &     -1.5 &     1 & 1.461 \\
      3 &      3 &   + &     1.5 &    4 &    3 &    3 &   + &      2.5 &     3 & 1.444 \\
      1 &      0 &   + &    -1.5 &    2 &    1 &    0 &   + &     -0.5 &     2 & 1.441 \\
      3 &      3 &   + &    -0.5 &    3 &    3 &    3 &   + &      0.5 &     2 & 1.415 \\
      0 &      0 &   - &    -1.5 &    1 &    2 &    0 &   - &     -0.5 &     2 & 1.411 \\
      0 &      0 &   - &    -0.5 &    1 &    2 &    0 &   - &      0.5 &     1 & 1.405 \\
      1 &      1 &   + &     0.5 &    2 &    2 &    1 &   + &      1.5 &     2 & 1.401 \\
      1 &      1 &   - &     0.5 &    2 &    2 &    1 &   - &      1.5 &     2 & 1.401 \\
      2 &      0 &   - &    -0.5 &    3 &    2 &    0 &   - &      0.5 &     2 & 1.386 \\
      0 &      0 &   - &    -0.5 &    1 &    2 &    0 &   - &      0.5 &     4 & 1.380 \\
      3 &      3 &   + &    -1.5 &    4 &    3 &    3 &   + &     -0.5 &     4 & 1.378 \\
      2 &      0 &   - &     1.5 &    1 &    2 &    0 &   - &      2.5 &     1 & 1.360 \\
      2 &      2 &   + &    -1.5 &    1 &    2 &    2 &   + &     -0.5 &     1 & 1.356 \\
      2 &      2 &   - &    -1.5 &    1 &    2 &    2 &   - &     -0.5 &     1 & 1.356 \\
      2 &      0 &   - &    -0.5 &    3 &    2 &    0 &   - &      0.5 &     4 & 1.348 \\
      2 &      0 &   - &    -0.5 &    2 &    0 &    0 &   - &      0.5 &     1 & 1.342 \\
      2 &      0 &   - &    -2.5 &    2 &    2 &    0 &   - &     -1.5 &     3 & 1.330 \\
      3 &      3 &   + &    -0.5 &    2 &    3 &    3 &   + &      0.5 &     3 & 1.321 \\
      3 &      3 &   + &    -0.5 &    4 &    3 &    3 &   + &      0.5 &     3 & 1.312 \\
      2 &      1 &   + &     0.5 &    2 &    1 &    1 &   + &      1.5 &     1 & 1.292 \\
      2 &      1 &   - &     0.5 &    2 &    1 &    1 &   - &      1.5 &     1 & 1.292 \\
      3 &      3 &   + &     0.5 &    3 &    3 &    3 &   + &      1.5 &     4 & 1.288 \\
      3 &      3 &   + &    -1.5 &    2 &    3 &    3 &   + &     -0.5 &     1 & 1.280 \\
\end{tabular}

\end{table}

\begin{table}[t]
\raggedright
\caption*{(Continued.)}
\begin{tabular}{
>{\raggedleft\arraybackslash}p{1cm}
>{\raggedleft\arraybackslash}p{1cm}
>{\raggedleft\arraybackslash}p{1cm}
>{\raggedleft\arraybackslash}p{0.75cm}
>{\raggedleft\arraybackslash}p{0.75cm}
>{\raggedleft\arraybackslash}p{1cm}
>{\raggedleft\arraybackslash}p{1cm}
>{\raggedleft\arraybackslash}p{1cm}
>{\raggedleft\arraybackslash}p{1cm}
>{\raggedleft\arraybackslash}p{1cm}|
>{\raggedleft\arraybackslash}p{2cm}}
$J^{\rm i}$ & $K^{\rm i}$ & ${\rm parity}^{\rm i}$ & $m_{F}^{\rm i}$ & $\xi^{\rm i}$ &
$J^{\rm f}$ & $K^{\rm f}$ & ${\rm parity}^{\rm f}$ & $m_{F}^{\rm f}$ & $\xi^{\rm f}$ & 
Rabi rate, \\ 
&&&&&&&&&&$ \Omega$ (2$\pi$ kHz) \\
\hline
\hline
      2 &      0 &   - &    -0.5 &    4 &    0 &    0 &   - &      0.5 &     1 & 1.277 \\
      1 &      0 &   + &     0.5 &    2 &    1 &    0 &   + &      1.5 &     1 & 1.272 \\
      2 &      0 &   - &     2.5 &    2 &    2 &    0 &   - &      3.5 &     1 & 1.271 \\
      3 &      2 &   + &    -2.5 &    2 &    2 &    2 &   + &     -1.5 &     1 & 1.269 \\
      2 &      0 &   - &     0.5 &    2 &    2 &    0 &   - &      1.5 &     3 & 1.259 \\
      1 &      0 &   + &    -1.5 &    2 &    1 &    0 &   + &     -0.5 &     1 & 1.254 \\
      1 &      1 &   + &     0.5 &    1 &    2 &    1 &   + &      1.5 &     2 & 1.251 \\
      1 &      1 &   - &     0.5 &    1 &    2 &    1 &   - &      1.5 &     2 & 1.251 \\
      3 &      3 &   + &     0.5 &    1 &    3 &    3 &   + &      1.5 &     2 & 1.238 \\
      2 &      2 &   + &     0.5 &    1 &    3 &    2 &   + &      1.5 &     1 & 1.218 \\
      2 &      2 &   + &    -2.5 &    1 &    3 &    2 &   + &     -1.5 &     2 & 1.215 \\
      1 &      1 &   + &     0.5 &    2 &    2 &    1 &   + &      1.5 &     1 & 1.209 \\
      1 &      1 &   - &     0.5 &    2 &    2 &    1 &   - &      1.5 &     1 & 1.209 \\
      2 &      0 &   - &     1.5 &    3 &    2 &    0 &   - &      2.5 &     2 & 1.193 \\
      3 &      3 &   + &    -2.5 &    3 &    3 &    3 &   + &     -1.5 &     2 & 1.186 \\
      2 &      0 &   - &    -1.5 &    2 &    2 &    0 &   - &     -0.5 &     1 & 1.172 \\
      2 &      2 &   + &    -2.5 &    1 &    2 &    2 &   + &     -1.5 &     2 & 1.170 \\
      2 &      2 &   - &    -2.5 &    1 &    2 &    2 &   - &     -1.5 &     2 & 1.170 \\
      3 &      3 &   + &    -0.5 &    2 &    3 &    3 &   + &      0.5 &     1 & 1.167 \\
      3 &      3 &   + &    -2.5 &    2 &    3 &    3 &   + &     -1.5 &     1 & 1.165 \\
      1 &      0 &   + &    -2.5 &    1 &    1 &    0 &   + &     -1.5 &     1 & 1.157 \\
      3 &      3 &   + &     1.5 &    1 &    3 &    3 &   + &      2.5 &     2 & 1.155 \\
      2 &      0 &   - &     0.5 &    1 &    2 &    0 &   - &      1.5 &     2 & 1.152 \\
      2 &      0 &   - &     0.5 &    1 &    0 &    0 &   - &      1.5 &     1 & 1.148 \\
      3 &      3 &   + &     1.5 &    2 &    3 &    3 &   + &      2.5 &     3 & 1.147 \\
      2 &      0 &   - &    -0.5 &    2 &    2 &    0 &   - &      0.5 &     1 & 1.146 \\
      2 &      0 &   - &    -0.5 &    4 &    2 &    0 &   - &      0.5 &     3 & 1.126 \\
      2 &      2 &   + &     0.5 &    1 &    2 &    2 &   + &      1.5 &     1 & 1.120 \\
      2 &      2 &   - &     0.5 &    1 &    2 &    2 &   - &      1.5 &     1 & 1.120 \\
      2 &      0 &   - &    -1.5 &    2 &    2 &    0 &   - &     -0.5 &     4 & 1.102 \\
      3 &      3 &   + &    -1.5 &    1 &    3 &    3 &   + &     -0.5 &     1 & 1.102 \\
      3 &      3 &   + &    -0.5 &    1 &    3 &    3 &   + &      0.5 &     2 & 1.083 \\
      3 &      3 &   + &     0.5 &    3 &    3 &    3 &   + &      1.5 &     3 & 1.058 \\
      2 &      0 &   - &    -0.5 &    2 &    2 &    0 &   - &      0.5 &     3 & 1.058 \\
      1 &      0 &   + &     0.5 &    2 &    1 &    0 &   + &      1.5 &     2 & 1.056 \\
      2 &      0 &   - &    -0.5 &    4 &    2 &    0 &   - &      0.5 &     4 & 1.050 \\
      3 &      3 &   + &    -1.5 &    3 &    3 &    3 &   + &     -0.5 &     3 & 1.050 \\
      2 &      1 &   + &    -0.5 &    1 &    1 &    1 &   + &      0.5 &     2 & 1.048 \\
      2 &      1 &   - &    -0.5 &    1 &    1 &    1 &   - &      0.5 &     2 & 1.048 \\
      2 &      0 &   - &    -1.5 &    3 &    2 &    0 &   - &     -0.5 &     2 & 1.043 \\
      1 &      0 &   + &    -1.5 &    1 &    1 &    0 &   + &     -0.5 &     1 & 1.022 \\
      2 &      1 &   + &     1.5 &    1 &    2 &    1 &   + &      2.5 &     1 & 1.000 \\
      2 &      1 &   - &     1.5 &    1 &    2 &    1 &   - &      2.5 &     1 & 1.000 \\
      1 &      1 &   + &    -0.5 &    1 &    1 &    1 &   + &      0.5 &     2 & 0.995 \\
      1 &      1 &   + &    -0.5 &    2 &    1 &    1 &   + &      0.5 &     1 & 0.995 \\
      1 &      1 &   - &    -0.5 &    2 &    1 &    1 &   - &      0.5 &     1 & 0.995 \\
      1 &      1 &   - &    -0.5 &    1 &    1 &    1 &   - &      0.5 &     2 & 0.995 \\
      2 &      0 &   - &    -1.5 &    1 &    2 &    0 &   - &     -0.5 &     1 & 0.986 \\
      2 &      0 &   - &     1.5 &    2 &    2 &    0 &   - &      2.5 &     2 & 0.985 \\
      3 &      3 &   + &    -1.5 &    2 &    3 &    3 &   + &     -0.5 &     2 & 0.985 \\
      2 &      2 &   - &    -1.5 &    2 &    2 &    2 &   - &     -0.5 &     1 & 0.974 \\
      2 &      2 &   + &    -1.5 &    2 &    2 &    2 &   + &     -0.5 &     1 & 0.974 \\
      0 &      0 &   - &     0.5 &    1 &    2 &    0 &   - &      1.5 &     2 & 0.968 \\
      3 &      2 &   + &    -2.5 &    1 &    2 &    2 &   + &     -1.5 &     2 & 0.962 \\
      1 &      1 &   - &     0.5 &    1 &    1 &    1 &   - &      1.5 &     1 & 0.958 \\
      1 &      1 &   + &     0.5 &    1 &    1 &    1 &   + &      1.5 &     1 & 0.958 \\
      2 &      0 &   - &    -0.5 &    3 &    0 &    0 &   - &      0.5 &     1 & 0.957 \\
      3 &      2 &   + &     1.5 &    2 &    2 &    2 &   + &      2.5 &     1 & 0.952 \\
      2 &      0 &   - &     1.5 &    1 &    2 &    0 &   - &      2.5 &     2 & 0.949 \\
      1 &      0 &   + &    -1.5 &    2 &    1 &    0 &   + &     -0.5 &     3 & 0.947 \\
      3 &      2 &   + &    -1.5 &    1 &    2 &    2 &   + &     -0.5 &     2 & 0.947 \\
      2 &      0 &   - &    -0.5 &    1 &    2 &    0 &   - &      0.5 &     2 & 0.945 \\
\end{tabular}

\end{table}

\begin{table}[t]
\raggedright
\caption*{(Continued.)}
\begin{tabular}{
>{\raggedleft\arraybackslash}p{1cm}
>{\raggedleft\arraybackslash}p{1cm}
>{\raggedleft\arraybackslash}p{1cm}
>{\raggedleft\arraybackslash}p{0.75cm}
>{\raggedleft\arraybackslash}p{0.75cm}
>{\raggedleft\arraybackslash}p{1cm}
>{\raggedleft\arraybackslash}p{1cm}
>{\raggedleft\arraybackslash}p{1cm}
>{\raggedleft\arraybackslash}p{1cm}
>{\raggedleft\arraybackslash}p{1cm}|
>{\raggedleft\arraybackslash}p{2cm}}
$J^{\rm i}$ & $K^{\rm i}$ & ${\rm parity}^{\rm i}$ & $m_{F}^{\rm i}$ & $\xi^{\rm i}$ &
$J^{\rm f}$ & $K^{\rm f}$ & ${\rm parity}^{\rm f}$ & $m_{F}^{\rm f}$ & $\xi^{\rm f}$ & 
Rabi rate, \\ 
&&&&&&&&&&$ \Omega$ (2$\pi$ kHz) \\
\hline
\hline
      2 &      1 &   - &     0.5 &    2 &    2 &    1 &   - &      1.5 &     2 & 0.929 \\
      2 &      1 &   + &     0.5 &    2 &    2 &    1 &   + &      1.5 &     2 & 0.929 \\
      2 &      2 &   + &     1.5 &    1 &    3 &    2 &   + &      2.5 &     2 & 0.912 \\
      3 &      2 &   + &    -1.5 &    1 &    2 &    2 &   + &     -0.5 &     1 & 0.908 \\
      2 &      0 &   - &     0.5 &    4 &    2 &    0 &   - &      1.5 &     3 & 0.907 \\
      3 &      3 &   + &    -2.5 &    3 &    3 &    3 &   + &     -1.5 &     3 & 0.905 \\
      2 &      1 &   - &    -2.5 &    1 &    2 &    1 &   - &     -1.5 &     1 & 0.904 \\
      2 &      1 &   + &    -2.5 &    1 &    2 &    1 &   + &     -1.5 &     1 & 0.904 \\
      1 &      0 &   + &    -0.5 &    1 &    1 &    0 &   + &      0.5 &     2 & 0.895 \\
      2 &      0 &   - &    -2.5 &    2 &    2 &    0 &   - &     -1.5 &     1 & 0.859 \\
      2 &      2 &   - &    -0.5 &    1 &    2 &    2 &   - &      0.5 &     2 & 0.856 \\
      2 &      2 &   - &    -0.5 &    2 &    2 &    2 &   - &      0.5 &     1 & 0.856 \\
      2 &      2 &   + &    -0.5 &    1 &    2 &    2 &   + &      0.5 &     2 & 0.856 \\
      2 &      2 &   + &    -0.5 &    2 &    2 &    2 &   + &      0.5 &     1 & 0.856 \\
      1 &      0 &   + &    -0.5 &    3 &    1 &    0 &   + &      0.5 &     1 & 0.856 \\
      1 &      1 &   + &     0.5 &    1 &    2 &    1 &   + &      1.5 &     1 & 0.830 \\
      1 &      1 &   - &     0.5 &    1 &    2 &    1 &   - &      1.5 &     1 & 0.830 \\
      3 &      3 &   + &     2.5 &    1 &    3 &    3 &   + &      3.5 &     2 & 0.827 \\
      3 &      3 &   + &     0.5 &    2 &    3 &    3 &   + &      1.5 &     1 & 0.826 \\
      3 &      3 &   + &     0.5 &    1 &    3 &    3 &   + &      1.5 &     1 & 0.824 \\
      3 &      3 &   + &    -3.5 &    2 &    3 &    3 &   + &     -2.5 &     1 & 0.818 \\
      1 &      1 &   + &    -0.5 &    2 &    2 &    1 &   + &      0.5 &     1 & 0.815 \\
      1 &      1 &   - &    -0.5 &    2 &    2 &    1 &   - &      0.5 &     1 & 0.815 \\
      0 &      0 &   - &    -0.5 &    1 &    2 &    0 &   - &      0.5 &     3 & 0.812 \\
      2 &      2 &   - &     0.5 &    1 &    2 &    2 &   - &      1.5 &     2 & 0.812 \\
      2 &      2 &   + &     0.5 &    1 &    2 &    2 &   + &      1.5 &     2 & 0.812 \\
      3 &      3 &   + &     0.5 &    3 &    3 &    3 &   + &      1.5 &     2 & 0.802 \\
      2 &      2 &   + &     1.5 &    2 &    2 &    2 &   + &      2.5 &     1 & 0.799 \\
      2 &      2 &   - &     1.5 &    2 &    2 &    2 &   - &      2.5 &     1 & 0.799 \\
      2 &      1 &   - &    -1.5 &    2 &    2 &    1 &   - &     -0.5 &     2 & 0.798 \\
      2 &      1 &   + &    -1.5 &    2 &    2 &    1 &   + &     -0.5 &     2 & 0.798 \\
      2 &      2 &   + &     1.5 &    2 &    3 &    2 &   + &      2.5 &     1 & 0.792 \\
      2 &      2 &   + &     0.5 &    2 &    3 &    2 &   + &      1.5 &     1 & 0.787 \\
      1 &      1 &   - &    -1.5 &    1 &    1 &    1 &   - &     -0.5 &     2 & 0.781 \\
      1 &      1 &   + &    -1.5 &    1 &    1 &    1 &   + &     -0.5 &     2 & 0.781 \\
      3 &      3 &   + &     0.5 &    2 &    3 &    3 &   + &      1.5 &     2 & 0.780 \\
      3 &      3 &   + &    -3.5 &    1 &    3 &    3 &   + &     -2.5 &     2 & 0.753 \\
      1 &      0 &   + &    -0.5 &    2 &    1 &    0 &   + &      0.5 &     2 & 0.749 \\
      2 &      0 &   - &     0.5 &    2 &    2 &    0 &   - &      1.5 &     1 & 0.741 \\
      0 &      0 &   - &    -1.5 &    1 &    2 &    0 &   - &     -0.5 &     1 & 0.736 \\
      1 &      1 &   + &    -1.5 &    1 &    1 &    1 &   + &     -0.5 &     1 & 0.726 \\
      1 &      1 &   - &    -1.5 &    1 &    1 &    1 &   - &     -0.5 &     1 & 0.726 \\
      3 &      2 &   + &    -0.5 &    1 &    2 &    2 &   + &      0.5 &     2 & 0.720 \\
      3 &      3 &   + &    -0.5 &    4 &    3 &    3 &   + &      0.5 &     4 & 0.717 \\
      3 &      2 &   + &    -1.5 &    2 &    2 &    2 &   + &     -0.5 &     2 & 0.716 \\
      3 &      3 &   + &    -1.5 &    3 &    3 &    3 &   + &     -0.5 &     4 & 0.706 \\
      3 &      3 &   + &    -1.5 &    1 &    3 &    3 &   + &     -0.5 &     2 & 0.695 \\
      3 &      3 &   + &    -2.5 &    2 &    3 &    3 &   + &     -1.5 &     4 & 0.693 \\
      3 &      2 &   + &    -1.5 &    2 &    2 &    2 &   + &     -0.5 &     1 & 0.687 \\
      1 &      0 &   + &    -0.5 &    3 &    1 &    0 &   + &      0.5 &     2 & 0.681 \\
      2 &      1 &   + &    -1.5 &    1 &    2 &    1 &   + &     -0.5 &     1 & 0.678 \\
      2 &      1 &   - &    -1.5 &    1 &    2 &    1 &   - &     -0.5 &     1 & 0.678 \\
      3 &      2 &   + &     0.5 &    2 &    2 &    2 &   + &      1.5 &     1 & 0.671 \\
      3 &      3 &   + &    -1.5 &    2 &    3 &    3 &   + &     -0.5 &     3 & 0.666 \\
      3 &      3 &   + &    -1.5 &    2 &    3 &    3 &   + &     -0.5 &     4 & 0.660 \\
      2 &      0 &   - &    -1.5 &    1 &    2 &    0 &   - &     -0.5 &     3 & 0.659 \\
      2 &      2 &   + &     0.5 &    1 &    3 &    2 &   + &      1.5 &     2 & 0.652 \\
      2 &      0 &   - &    -0.5 &    2 &    2 &    0 &   - &      0.5 &     4 & 0.650 \\
      2 &      0 &   - &     0.5 &    4 &    2 &    0 &   - &      1.5 &     2 & 0.646 \\
      2 &      0 &   - &    -2.5 &    1 &    2 &    0 &   - &     -1.5 &     3 & 0.641 \\
      2 &      2 &   + &    -1.5 &    1 &    3 &    2 &   + &     -0.5 &     2 & 0.639 \\
      3 &      2 &   + &    -2.5 &    1 &    2 &    2 &   + &     -1.5 &     1 & 0.622 \\
      2 &      0 &   - &     0.5 &    3 &    2 &    0 &   - &      1.5 &     3 & 0.619 \\
\end{tabular}

\end{table}

\begin{table}[t]
\raggedright
\caption*{(Continued.)}
\begin{tabular}{
>{\raggedleft\arraybackslash}p{1cm}
>{\raggedleft\arraybackslash}p{1cm}
>{\raggedleft\arraybackslash}p{1cm}
>{\raggedleft\arraybackslash}p{0.75cm}
>{\raggedleft\arraybackslash}p{0.75cm}
>{\raggedleft\arraybackslash}p{1cm}
>{\raggedleft\arraybackslash}p{1cm}
>{\raggedleft\arraybackslash}p{1cm}
>{\raggedleft\arraybackslash}p{1cm}
>{\raggedleft\arraybackslash}p{1cm}|
>{\raggedleft\arraybackslash}p{2cm}}
$J^{\rm i}$ & $K^{\rm i}$ & ${\rm parity}^{\rm i}$ & $m_{F}^{\rm i}$ & $\xi^{\rm i}$ &
$J^{\rm f}$ & $K^{\rm f}$ & ${\rm parity}^{\rm f}$ & $m_{F}^{\rm f}$ & $\xi^{\rm f}$ & 
Rabi rate, \\ 
&&&&&&&&&&$ \Omega$ (2$\pi$ kHz) \\
\hline
\hline
      2 &      2 &   + &    -0.5 &    2 &    3 &    2 &   + &      0.5 &     1 & 0.600 \\
      3 &      2 &   + &     1.5 &    2 &    3 &    2 &   + &      2.5 &     2 & 0.598 \\
      2 &      0 &   - &     0.5 &    1 &    2 &    0 &   - &      1.5 &     1 & 0.589 \\
      3 &      2 &   + &     2.5 &    1 &    3 &    2 &   + &      3.5 &     1 & 0.587 \\
      2 &      0 &   - &    -1.5 &    3 &    2 &    0 &   - &     -0.5 &     3 & 0.587 \\
      2 &      1 &   + &    -2.5 &    1 &    2 &    1 &   + &     -1.5 &     2 & 0.586 \\
      2 &      1 &   - &    -2.5 &    1 &    2 &    1 &   - &     -1.5 &     2 & 0.586 \\
      2 &      1 &   + &    -1.5 &    2 &    1 &    1 &   + &     -0.5 &     2 & 0.566 \\
      2 &      1 &   - &    -1.5 &    2 &    1 &    1 &   - &     -0.5 &     2 & 0.566 \\
      2 &      1 &   - &     0.5 &    1 &    2 &    1 &   - &      1.5 &     1 & 0.560 \\
      2 &      1 &   + &     0.5 &    1 &    2 &    1 &   + &      1.5 &     1 & 0.560 \\
      2 &      0 &   - &     0.5 &    3 &    2 &    0 &   - &      1.5 &     2 & 0.547 \\
      3 &      3 &   + &     2.5 &    2 &    3 &    3 &   + &      3.5 &     1 & 0.534 \\
      2 &      0 &   - &    -0.5 &    1 &    2 &    0 &   - &      0.5 &     3 & 0.532 \\
      3 &      2 &   + &     1.5 &    1 &    3 &    2 &   + &      2.5 &     1 & 0.529 \\
      1 &      0 &   + &     0.5 &    1 &    1 &    0 &   + &      1.5 &     1 & 0.527 \\
      2 &      0 &   - &    -1.5 &    3 &    2 &    0 &   - &     -0.5 &     4 & 0.525 \\
      2 &      0 &   - &     1.5 &    3 &    2 &    0 &   - &      2.5 &     1 & 0.514 \\
      2 &      0 &   - &    -0.5 &    3 &    2 &    0 &   - &      0.5 &     3 & 0.496 \\
      3 &      3 &   + &    -2.5 &    1 &    3 &    3 &   + &     -1.5 &     3 & 0.492 \\
      2 &      0 &   - &     0.5 &    3 &    2 &    0 &   - &      1.5 &     1 & 0.488 \\
      2 &      1 &   + &    -1.5 &    2 &    2 &    1 &   + &     -0.5 &     1 & 0.487 \\
      2 &      1 &   - &    -1.5 &    2 &    2 &    1 &   - &     -0.5 &     1 & 0.487 \\
      2 &      0 &   - &    -1.5 &    2 &    2 &    0 &   - &     -0.5 &     2 & 0.481 \\
      3 &      3 &   + &    -2.5 &    2 &    3 &    3 &   + &     -1.5 &     3 & 0.480 \\
      3 &      3 &   + &    -0.5 &    2 &    3 &    3 &   + &      0.5 &     4 & 0.477 \\
      2 &      0 &   - &    -2.5 &    1 &    2 &    0 &   - &     -1.5 &     2 & 0.470 \\
      1 &      1 &   + &     0.5 &    2 &    1 &    1 &   + &      1.5 &     1 & 0.468 \\
      1 &      1 &   - &     0.5 &    2 &    1 &    1 &   - &      1.5 &     1 & 0.468 \\
      3 &      3 &   + &    -1.5 &    1 &    3 &    3 &   + &     -0.5 &     3 & 0.460 \\
      3 &      3 &   + &     0.5 &    4 &    3 &    3 &   + &      1.5 &     2 & 0.452 \\
      3 &      3 &   + &     1.5 &    4 &    3 &    3 &   + &      2.5 &     2 & 0.450 \\
      3 &      2 &   + &     0.5 &    2 &    3 &    2 &   + &      1.5 &     2 & 0.441 \\
      3 &      3 &   + &    -3.5 &    1 &    3 &    3 &   + &     -2.5 &     3 & 0.435 \\
      3 &      2 &   + &    -2.5 &    2 &    3 &    2 &   + &     -1.5 &     2 & 0.430 \\
      3 &      2 &   + &    -3.5 &    1 &    3 &    2 &   + &     -2.5 &     1 & 0.429 \\
      2 &      1 &   + &    -0.5 &    2 &    2 &    1 &   + &      0.5 &     1 & 0.428 \\
      2 &      1 &   + &    -0.5 &    1 &    2 &    1 &   + &      0.5 &     2 & 0.428 \\
      2 &      1 &   - &    -0.5 &    2 &    2 &    1 &   - &      0.5 &     1 & 0.428 \\
      2 &      1 &   - &    -0.5 &    1 &    2 &    1 &   - &      0.5 &     2 & 0.428 \\
      2 &      2 &   + &     0.5 &    2 &    3 &    2 &   + &      1.5 &     2 & 0.422 \\
      1 &      0 &   + &    -0.5 &    1 &    1 &    0 &   + &      0.5 &     1 & 0.422 \\
      2 &      0 &   - &    -0.5 &    4 &    2 &    0 &   - &      0.5 &     2 & 0.407 \\
      3 &      2 &   + &     0.5 &    1 &    2 &    2 &   + &      1.5 &     2 & 0.407 \\
      2 &      1 &   + &     0.5 &    1 &    2 &    1 &   + &      1.5 &     2 & 0.406 \\
      2 &      1 &   - &     0.5 &    1 &    2 &    1 &   - &      1.5 &     2 & 0.406 \\
      2 &      1 &   - &     1.5 &    2 &    2 &    1 &   - &      2.5 &     1 & 0.399 \\
      2 &      1 &   + &     1.5 &    2 &    2 &    1 &   + &      2.5 &     1 & 0.399 \\
      3 &      3 &   + &     1.5 &    3 &    3 &    3 &   + &      2.5 &     1 & 0.390 \\
      1 &      1 &   - &    -0.5 &    1 &    1 &    1 &   - &      0.5 &     1 & 0.383 \\
      1 &      1 &   - &    -0.5 &    2 &    1 &    1 &   - &      0.5 &     2 & 0.383 \\
      1 &      1 &   + &    -0.5 &    1 &    1 &    1 &   + &      0.5 &     1 & 0.383 \\
      1 &      1 &   + &    -0.5 &    2 &    1 &    1 &   + &      0.5 &     2 & 0.383 \\
      2 &      0 &   - &     0.5 &    4 &    0 &    0 &   - &      1.5 &     1 & 0.382 \\
      2 &      0 &   - &    -1.5 &    1 &    2 &    0 &   - &     -0.5 &     2 & 0.380 \\
      3 &      3 &   + &    -0.5 &    1 &    3 &    3 &   + &      0.5 &     3 & 0.377 \\
      2 &      0 &   - &    -2.5 &    2 &    2 &    0 &   - &     -1.5 &     2 & 0.373 \\
      3 &      2 &   + &    -2.5 &    1 &    3 &    2 &   + &     -1.5 &     1 & 0.365 \\
      1 &      0 &   + &    -0.5 &    3 &    1 &    0 &   + &      0.5 &     3 & 0.355 \\
      3 &      3 &   + &     2.5 &    3 &    3 &    3 &   + &      3.5 &     1 & 0.353 \\
      3 &      3 &   + &     0.5 &    3 &    3 &    3 &   + &      1.5 &     1 & 0.353 \\
      2 &      0 &   - &     0.5 &    2 &    2 &    0 &   - &      1.5 &     2 & 0.344 \\
      2 &      0 &   - &    -0.5 &    3 &    2 &    0 &   - &      0.5 &     1 & 0.341 \\
\end{tabular}
\end{table}

\begin{table}[t]
\raggedright
\caption*{(Continued.)}
\begin{tabular}{
>{\raggedleft\arraybackslash}p{1cm}
>{\raggedleft\arraybackslash}p{1cm}
>{\raggedleft\arraybackslash}p{1cm}
>{\raggedleft\arraybackslash}p{0.75cm}
>{\raggedleft\arraybackslash}p{0.75cm}
>{\raggedleft\arraybackslash}p{1cm}
>{\raggedleft\arraybackslash}p{1cm}
>{\raggedleft\arraybackslash}p{1cm}
>{\raggedleft\arraybackslash}p{1cm}
>{\raggedleft\arraybackslash}p{1cm}|
>{\raggedleft\arraybackslash}p{2cm}}
$J^{\rm i}$ & $K^{\rm i}$ & ${\rm parity}^{\rm i}$ & $m_{F}^{\rm i}$ & $\xi^{\rm i}$ &
$J^{\rm f}$ & $K^{\rm f}$ & ${\rm parity}^{\rm f}$ & $m_{F}^{\rm f}$ & $\xi^{\rm f}$ & 
Rabi rate, \\ 
&&&&&&&&&&$ \Omega$ (2$\pi$ kHz) \\
\hline
\hline
      3 &      3 &   + &    -0.5 &    4 &    3 &    3 &   + &      0.5 &     2 & 0.335 \\
      2 &      2 &   + &    -1.5 &    2 &    3 &    2 &   + &     -0.5 &     1 & 0.334 \\
      3 &      2 &   + &     0.5 &    1 &    3 &    2 &   + &      1.5 &     1 & 0.328 \\
      1 &      0 &   + &    -1.5 &    1 &    1 &    0 &   + &     -0.5 &     2 & 0.322 \\
      2 &      2 &   + &     1.5 &    1 &    3 &    2 &   + &      2.5 &     1 & 0.317 \\
      2 &      0 &   - &     0.5 &    1 &    2 &    0 &   - &      1.5 &     3 & 0.308 \\
      3 &      3 &   + &    -0.5 &    3 &    3 &    3 &   + &      0.5 &     1 & 0.279 \\
      2 &      2 &   + &    -1.5 &    1 &    2 &    2 &   + &     -0.5 &     2 & 0.277 \\
      2 &      2 &   - &    -1.5 &    1 &    2 &    2 &   - &     -0.5 &     2 & 0.277 \\
      2 &      0 &   - &    -0.5 &    2 &    2 &    0 &   - &      0.5 &     2 & 0.275 \\
      3 &      3 &   + &     0.5 &    1 &    3 &    3 &   + &      1.5 &     3 & 0.265 \\
      1 &      0 &   + &    -0.5 &    2 &    1 &    0 &   + &      0.5 &     3 & 0.262 \\
      3 &      3 &   + &     1.5 &    2 &    3 &    3 &   + &      2.5 &     1 & 0.258 \\
      3 &      3 &   + &     0.5 &    2 &    3 &    3 &   + &      1.5 &     4 & 0.256 \\
      2 &      0 &   - &    -0.5 &    1 &    2 &    0 &   - &      0.5 &     1 & 0.255 \\
      3 &      3 &   + &     1.5 &    3 &    3 &    3 &   + &      2.5 &     2 & 0.242 \\
      1 &      0 &   + &     0.5 &    3 &    1 &    0 &   + &      1.5 &     2 & 0.230 \\
      2 &      2 &   + &     0.5 &    2 &    2 &    2 &   + &      1.5 &     1 & 0.228 \\
      2 &      2 &   - &     0.5 &    2 &    2 &    2 &   - &      1.5 &     1 & 0.228 \\
      3 &      2 &   + &    -3.5 &    1 &    3 &    2 &   + &     -2.5 &     2 & 0.210 \\
      2 &      0 &   - &    -1.5 &    2 &    2 &    0 &   - &     -0.5 &     3 & 0.209 \\
      3 &      2 &   + &    -1.5 &    2 &    3 &    2 &   + &     -0.5 &     2 & 0.208 \\
      3 &      2 &   + &     2.5 &    2 &    3 &    2 &   + &      3.5 &     1 & 0.204 \\
      3 &      3 &   + &     0.5 &    4 &    3 &    3 &   + &      1.5 &     3 & 0.203 \\
      2 &      2 &   + &    -0.5 &    1 &    2 &    2 &   + &      0.5 &     1 & 0.202 \\
      2 &      2 &   + &    -0.5 &    2 &    2 &    2 &   + &      0.5 &     2 & 0.202 \\
      2 &      2 &   - &    -0.5 &    1 &    2 &    2 &   - &      0.5 &     1 & 0.202 \\
      2 &      2 &   - &    -0.5 &    2 &    2 &    2 &   - &      0.5 &     2 & 0.202 \\
      3 &      3 &   + &    -1.5 &    3 &    3 &    3 &   + &     -0.5 &     1 & 0.188 \\
      2 &      0 &   - &    -1.5 &    3 &    2 &    0 &   - &     -0.5 &     1 & 0.188 \\
      3 &      2 &   + &    -1.5 &    1 &    3 &    2 &   + &     -0.5 &     1 & 0.173 \\
      3 &      3 &   + &    -0.5 &    2 &    3 &    3 &   + &      0.5 &     2 & 0.172 \\
      3 &      2 &   + &     0.5 &    1 &    3 &    2 &   + &      1.5 &     2 & 0.172 \\
      3 &      2 &   + &    -1.5 &    2 &    3 &    2 &   + &     -0.5 &     1 & 0.168 \\
      3 &      2 &   + &    -0.5 &    1 &    3 &    2 &   + &      0.5 &     2 & 0.167 \\
      3 &      3 &   + &    -0.5 &    1 &    3 &    3 &   + &      0.5 &     1 & 0.167 \\
      2 &      0 &   - &    -1.5 &    1 &    2 &    0 &   - &     -0.5 &     4 & 0.163 \\
      3 &      3 &   + &    -1.5 &    4 &    3 &    3 &   + &     -0.5 &     2 & 0.160 \\
      1 &      1 &   + &    -0.5 &    1 &    2 &    1 &   + &      0.5 &     2 & 0.160 \\
      1 &      1 &   - &    -0.5 &    1 &    2 &    1 &   - &      0.5 &     2 & 0.160 \\
      3 &      3 &   + &    -0.5 &    3 &    3 &    3 &   + &      0.5 &     3 & 0.158 \\
      3 &      3 &   + &     1.5 &    1 &    3 &    3 &   + &      2.5 &     3 & 0.144 \\
      3 &      2 &   + &    -2.5 &    2 &    3 &    2 &   + &     -1.5 &     1 & 0.142 \\
      3 &      2 &   + &    -0.5 &    2 &    3 &    2 &   + &      0.5 &     1 & 0.139 \\
      2 &      1 &   - &    -1.5 &    1 &    2 &    1 &   - &     -0.5 &     2 & 0.139 \\
      2 &      1 &   + &    -1.5 &    1 &    2 &    1 &   + &     -0.5 &     2 & 0.139 \\
      3 &      2 &   + &     1.5 &    1 &    3 &    2 &   + &      2.5 &     2 & 0.127 \\
      3 &      2 &   + &    -0.5 &    2 &    3 &    2 &   + &      0.5 &     2 & 0.127 \\
      2 &      1 &   - &     0.5 &    2 &    2 &    1 &   - &      1.5 &     1 & 0.114 \\
      2 &      1 &   + &     0.5 &    2 &    2 &    1 &   + &      1.5 &     1 & 0.114 \\
      3 &      2 &   + &    -1.5 &    1 &    3 &    2 &   + &     -0.5 &     2 & 0.109 \\
      2 &      1 &   - &    -0.5 &    1 &    2 &    1 &   - &      0.5 &     1 & 0.102 \\
      2 &      1 &   - &    -0.5 &    2 &    2 &    1 &   - &      0.5 &     2 & 0.102 \\
      2 &      1 &   + &    -0.5 &    1 &    2 &    1 &   + &      0.5 &     1 & 0.102 \\
      2 &      1 &   + &    -0.5 &    2 &    2 &    1 &   + &      0.5 &     2 & 0.102 \\
      2 &      1 &   + &    -0.5 &    2 &    1 &    1 &   + &      0.5 &     1 & 0.101 \\
      2 &      1 &   - &    -0.5 &    2 &    1 &    1 &   - &      0.5 &     1 & 0.101 \\
      3 &      3 &   + &    -2.5 &    3 &    3 &    3 &   + &     -1.5 &     1 & 0.099 \\
      2 &      0 &   - &    -0.5 &    4 &    2 &    0 &   - &      0.5 &     1 & 0.098 \\
      3 &      3 &   + &    -2.5 &    1 &    3 &    3 &   + &     -1.5 &     2 & 0.080 \\
      3 &      2 &   + &    -0.5 &    1 &    3 &    2 &   + &      0.5 &     1 & 0.075 \\
      2 &      0 &   - &     0.5 &    4 &    2 &    0 &   - &      1.5 &     1 & 0.073 \\
\end{tabular}

\end{table}

\begin{table}[t]
\raggedright
\caption*{(Continued.)}
\begin{tabular}{
>{\raggedleft\arraybackslash}p{1cm}
>{\raggedleft\arraybackslash}p{1cm}
>{\raggedleft\arraybackslash}p{1cm}
>{\raggedleft\arraybackslash}p{0.75cm}
>{\raggedleft\arraybackslash}p{0.75cm}
>{\raggedleft\arraybackslash}p{1cm}
>{\raggedleft\arraybackslash}p{1cm}
>{\raggedleft\arraybackslash}p{1cm}
>{\raggedleft\arraybackslash}p{1cm}
>{\raggedleft\arraybackslash}p{1cm}|
>{\raggedleft\arraybackslash}p{2cm}}
$J^{\rm i}$ & $K^{\rm i}$ & ${\rm parity}^{\rm i}$ & $m_{F}^{\rm i}$ & $\xi^{\rm i}$ &
$J^{\rm f}$ & $K^{\rm f}$ & ${\rm parity}^{\rm f}$ & $m_{F}^{\rm f}$ & $\xi^{\rm f}$ & 
Rabi rate, \\ 
&&&&&&&&&&$ \Omega$ (2$\pi$ kHz) \\
\hline
\hline
      2 &      2 &   + &    -0.5 &    1 &    3 &    2 &   + &      0.5 &     2 & 0.071 \\
      2 &      0 &   - &    -0.5 &    1 &    2 &    0 &   - &      0.5 &     4 & 0.069 \\
      2 &      0 &   - &     1.5 &    2 &    2 &    0 &   - &      2.5 &     1 & 0.064 \\
      3 &      2 &   + &     0.5 &    2 &    3 &    2 &   + &      1.5 &     1 & 0.064 \\
      3 &      3 &   + &    -2.5 &    1 &    3 &    3 &   + &     -1.5 &     4 & 0.059 \\
      3 &      2 &   + &    -0.5 &    2 &    2 &    2 &   + &      0.5 &     1 & 0.055 \\
      2 &      1 &   + &    -1.5 &    1 &    1 &    1 &   + &     -0.5 &     1 & 0.011 \\
      2 &      1 &   - &    -1.5 &    1 &    1 &    1 &   - &     -0.5 &     1 & 0.011 \\
\hline
\end{tabular}
\end{table}

\clearpage

\begin{figure}[!htb]
    \iffullcompile
    \includegraphics[width=0.8 \textwidth]{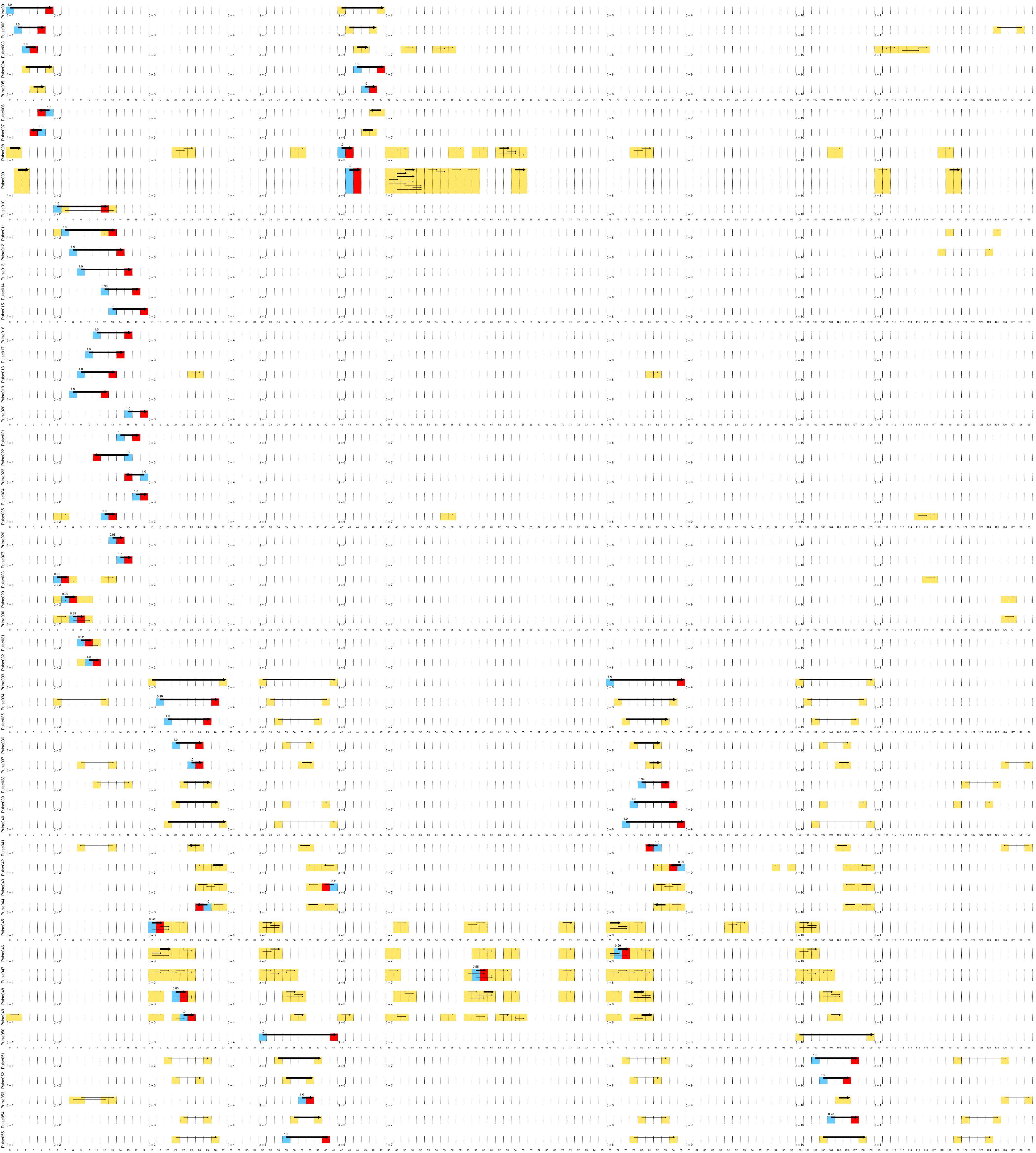}
    \fi
    \caption{Pulse library for the molecular control of H$_3$O$^+$ ion system in Fig.~4. The results are obtained with a simulation that includes 4 motional states. The main transition is color-coded as arrows from blue to red boxes, and the amount of the population transition is listed above the arrow. The width of the arrows indicates the amount of the population transition, and for each pulse, the most significant five transitions are plotted. (Part 1)}
    \label{fig:ex_H3O_pulse}
\end{figure}
\begin{figure}[!htb]
    \caption*{(Continued.)}
    \iffullcompile
    \includegraphics[width=0.85 \textwidth]{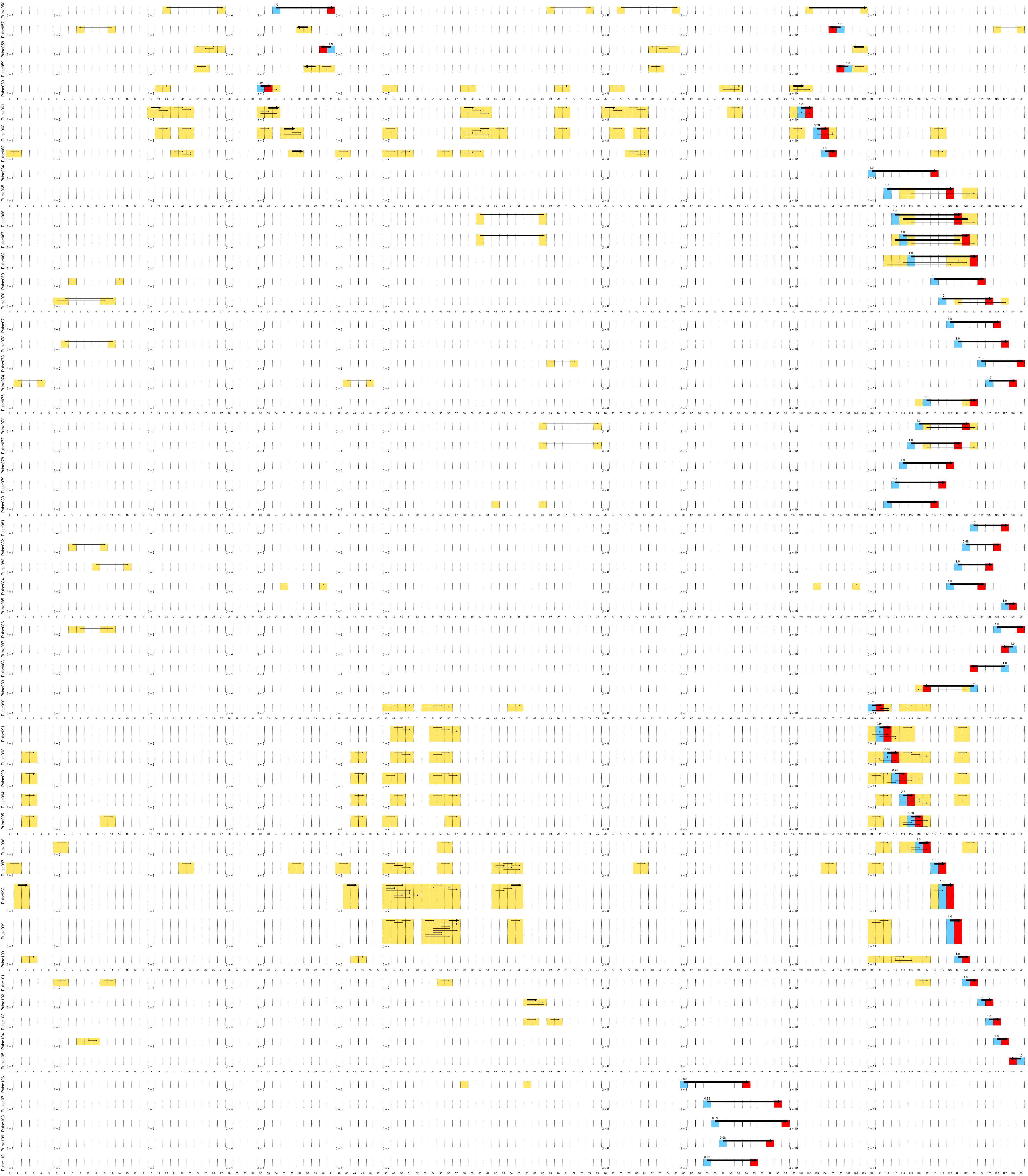}
    \fi
\end{figure} 
\begin{figure}[!htb]
    \caption*{(Continued.)}
    \iffullcompile
    \includegraphics[width=0.85 \textwidth]{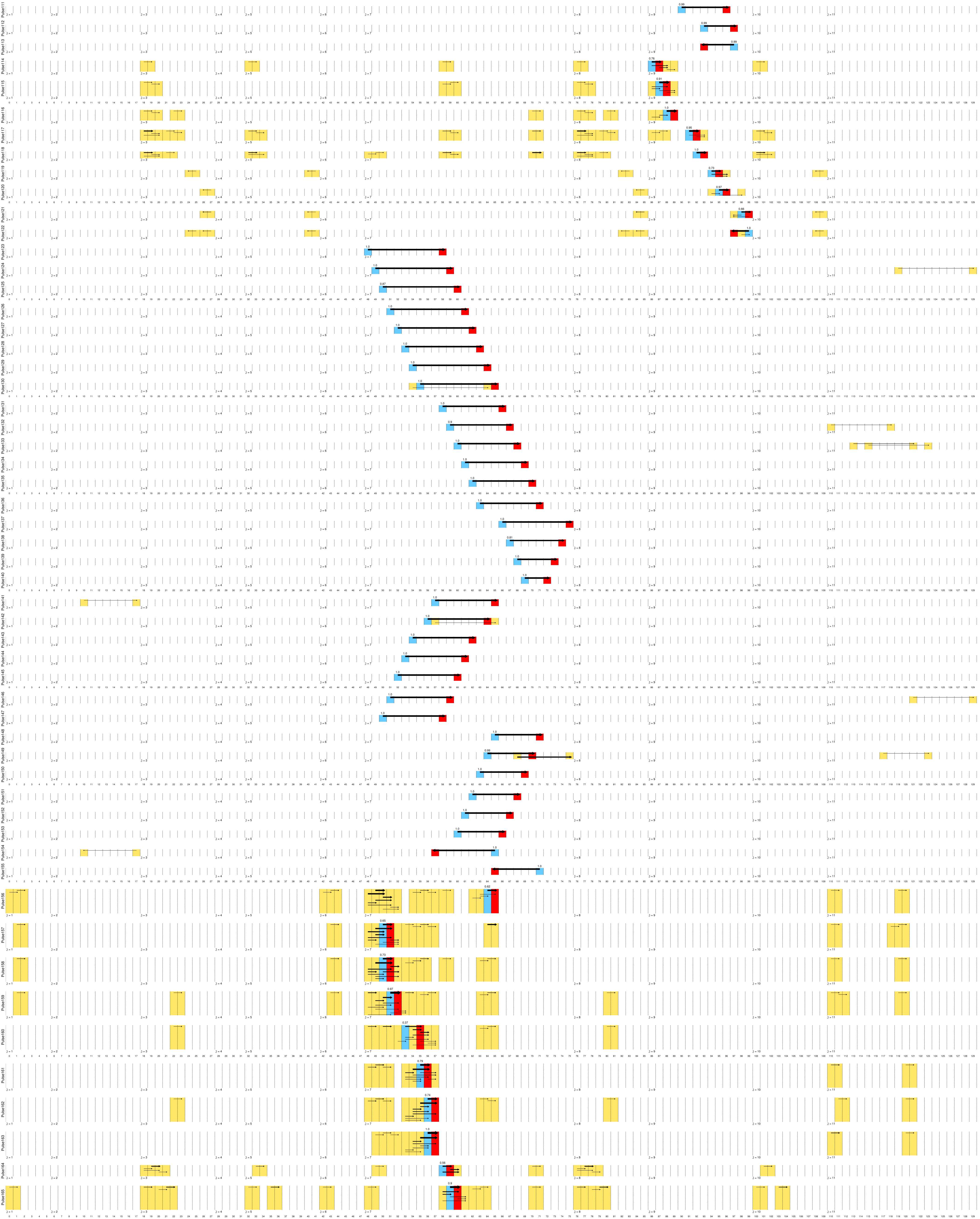}
    \fi
\end{figure} 
\begin{figure}[!htb]
    \caption*{(Continued.)}
    \iffullcompile
    \includegraphics[width=0.85 \textwidth]{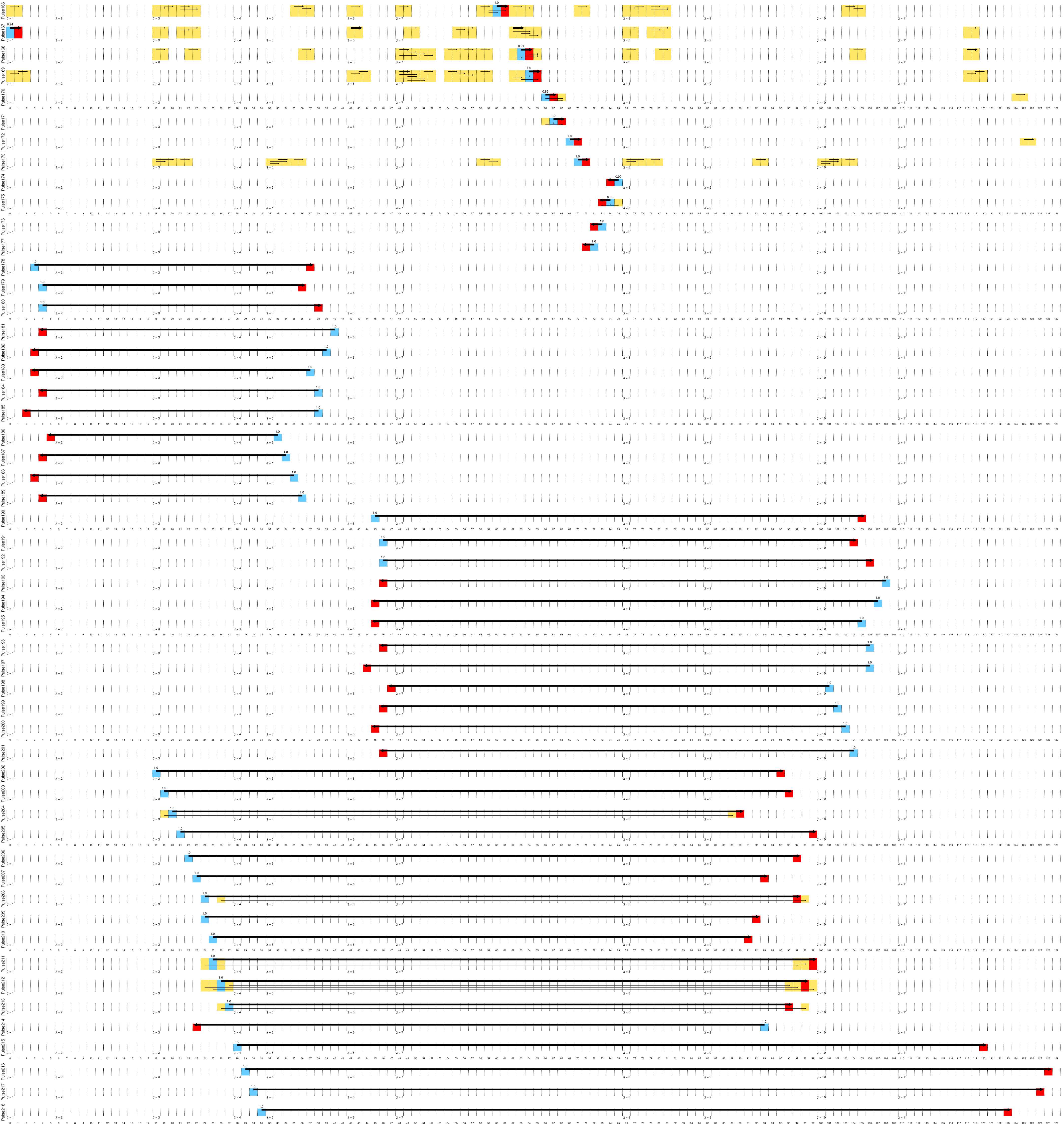}
    \fi
\end{figure} 

\begin{figure}[!htb]
    \centering
    \includegraphics[width=\textwidth]{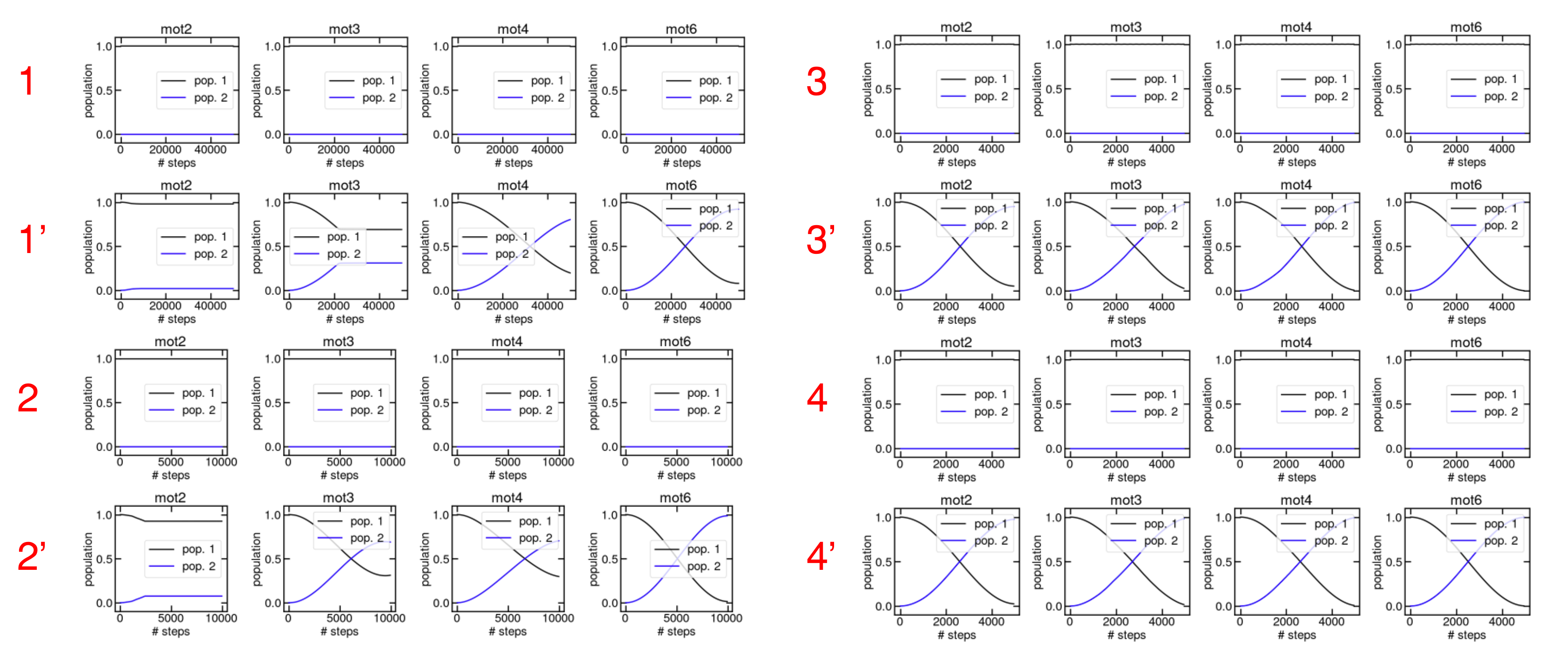}
    \caption{Time evolution of the populations under the influence of four selected pulses.
    The four pulses are ordered according to the strength of the Rabi oscillations, with the Rabi rates of 0.19, 0.53, 1.25, and 1.84,
    respectively, in the unit of $2\pi$ kHz.
    The frequencies of the pulses are set as the energy differences for the transitions
    (pulse 1: $\ket{2,5/2,+} \to \ket{2,3/2,-}$, 1': $\ket{2,3/2,-} \to \ket{2,5/2,+}$), 
    (pulse 2: $\ket{2,3/2,+} \to \ket{2,1/2,-}$, 2': $\ket{2,1/2,-} \to \ket{2,3/2,+}$),
    (pulse 3: $\ket{2,-3/2,+} \to \ket{2,-5/2,-}$, 3': $\ket{2,-5/2,-} \to \ket{2,-3/2,+}$), 
    (pulse 4: $\ket{2,-3/2,-} \to \ket{2,-5/2, -}$, 4': $\ket{2,-5/2,-} \to \ket{2,-3/2, -}$).
    The Raman pulses are with $\pi$ (abs.) and $\sigma^-$ (emit.) polarizations, thus only one direction of the population transfer (namely, those with $\Delta m=1$) can be driven. The number reported with `mot' indicates the number of motional manifolds included in the simulations. 
    }
    \label{fig:ex_weakpulse}
\end{figure} 

\begin{figure}[!htb]
    \iffullcompile
    \includegraphics[width=0.75 \textwidth]{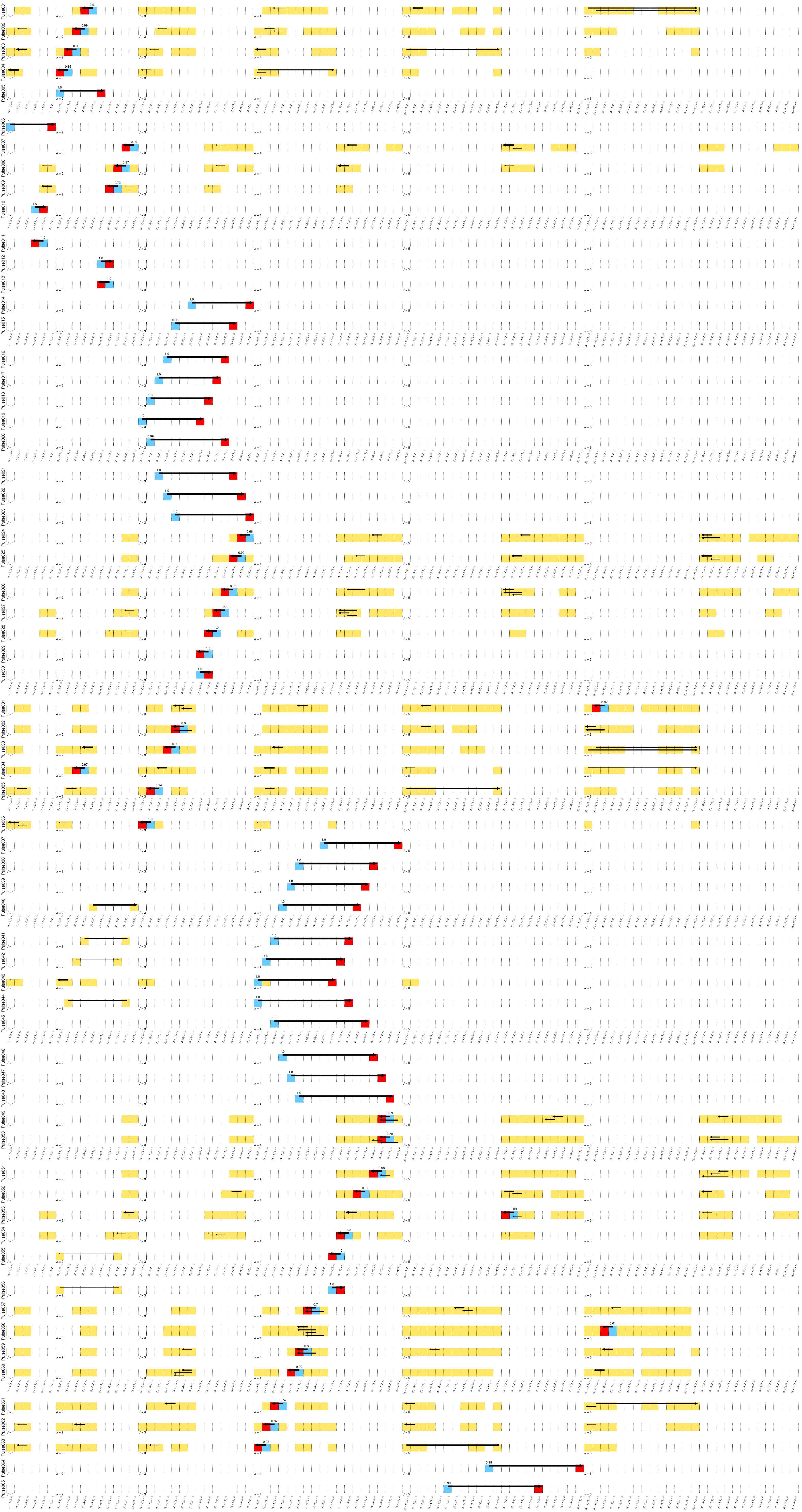}
    \fi
    \caption{Pulse library for CaH$^+$ $J \in \{1,2,3,4,5,6\}$ system learning in Fig.~\ref{fig:ex_CaH_J4_6}. The results are obtained with a simulation that includes 4 motional states. The main transition is color-coded as arrows from blue to red boxes, and the amount of the population transition is listed above the arrow. 
    The width of the arrows indicates the amount of the population transition, and for each pulse, the most significant five transitions are plotted. (Part 1)
    }
    \label{fig:ex_J6pulse}
\end{figure} 

\begin{figure}[!htb]
    \caption*{(Continued.)
    }
    \iffullcompile
    \includegraphics[width=0.75 \textwidth]{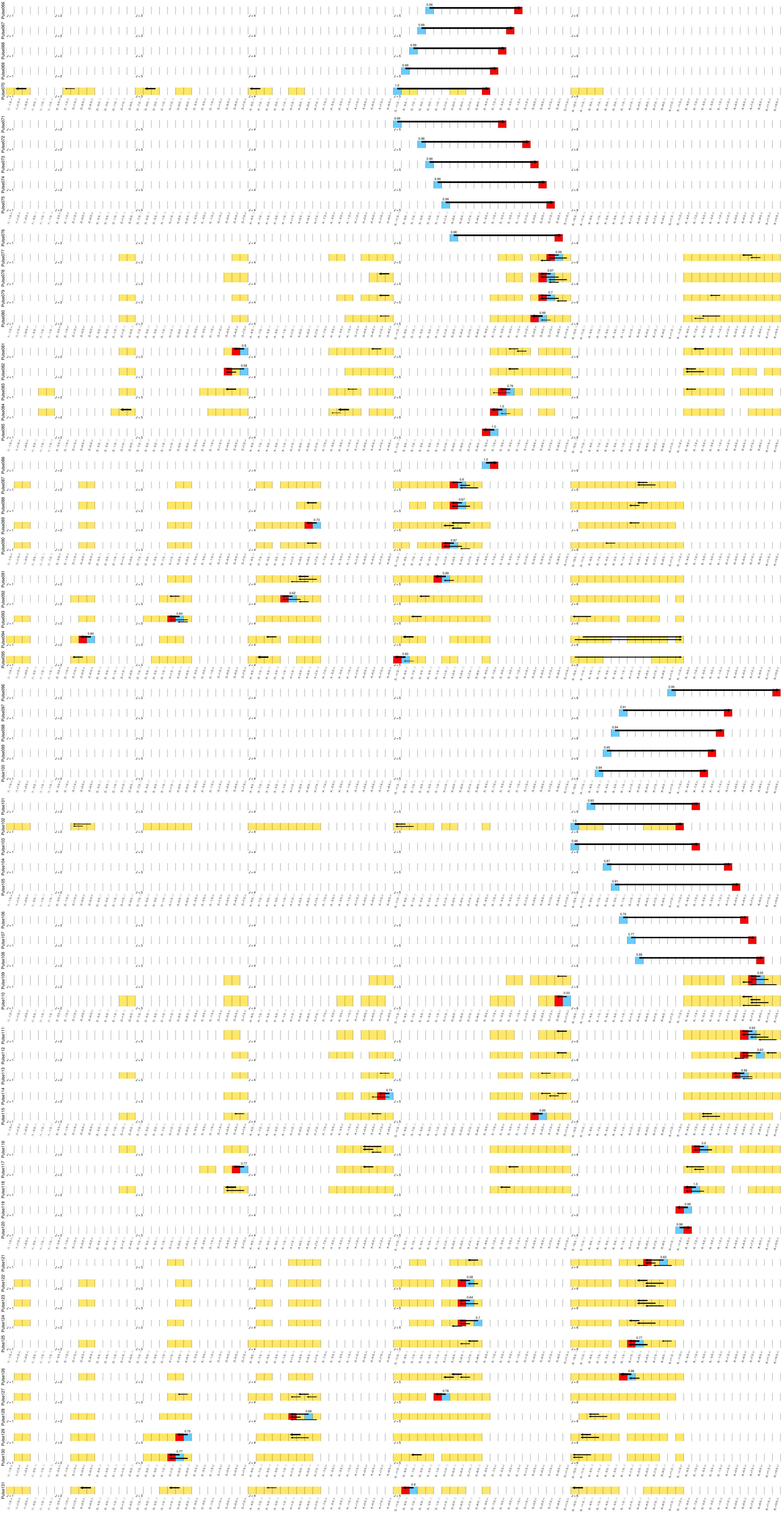}
    \fi
\end{figure}